# RADIO PULSARS

*Joeri van Leeuwen*

# RADIO PULSARS

(met een samenvatting in het Nederlands)

Proefschrift ter verkrijging van de graad van doctor aan de Universiteit Utrecht
op gezag van de Rector Magnificus, Prof. Dr. W.H. Gispen,
ingevolge het besluit van het College voor Promoties in het openbaar te verdedigen op
maandag 10 mei 2004 des middags te 2.30 uur
door

## Albert Gerardus Johannes van Leeuwen

geboren op 11 januari 1975, te Waardenburg

Promotor:     Prof. Dr. F.W.M. Verbunt
         Faculteit Natuur- en Sterrenkunde, Universiteit Utrecht

Promotiecommissie:  Prof. Dr. Ir. J.A.M. Bleeker
         Prof. Dr. J. Heise
         Prof. Dr. E.P.J. van den Heuvel
         Prof. Dr. M.B.M. van der Klis
         Prof. Dr. J. Kuijpers

# CONTENTS
# INHOUD

Radio pulsars blijven over als sterren aan het eind van hun leven ontploffen. Ze lijken nauwelijks nog op de sterren waar ze uit ontstaan: radio pulsars zijn wel even zwaar als sterren, maar bij de ontploffing zijn ze zo samengeperst dat ze 100.000 keer kleiner zijn geworden. Verder maken radio pulsars vooral radiostraling, terwijl sterren vooral licht maken.

De afgelopen jaren heb ik onderzocht hoe pulsars radiostraling maken, hoe ze veranderen als ze ouder worden en hoe je zoveel mogelijk nieuwe pulsars kunt ontdekken met verschillende radiotelescopen. Dat onderzoek wordt in de rest van dit proefschrift specifiek uitgebreid beschreven. Hier geef ik een wat algemenere inleiding en vat ik de belangrijkste resultaten samen.

Ik begin met de ontdekking van de eerste radio pulsar, veertig jaar geleden. Daarna zie we dat radio pulsars bij de ontploffing van bepaalde sterren ontstaan. Niet alle sterren worden radio pulsars. Ik vertel ook hoe de andere sterren eindigen: sommige worden 'witte dwergsterren', de resterende worden 'zwarte gaten'. Net als radio pulsars blijken die allebei te zijn gemaakt van bijna onvoorstelbaar zwaar materiaal. Ik ga wat dieper in op de bijzondere eigenschappen van radio pulsars en beschrijf ten slotte wat ik in mijn eigen onderzoek voor nieuws heb gevonden.

## Hoe onderzoek je sterren?

Bij onderzoek aan sterren is er één altijd terugkomend probleem: sterren staan zo ver weg dat je ze niet in detail kunt bekijken. Alleen uit het beetje licht dat ze onze kant op stralen, kunnen we leren hoe sterren werken. Dat begint heel eenvoudig: als je 's avond naar de sterrenhemel kijkt, zijn sommige sterren wat roder van kleur en andere wat witter. Dat kleurverschil vertelt je iets over de temperatuur van de ster: de rode zijn redelijk heet (roodgloeiend) maar de witte zijn heter (witheet). Zo kun je met je blote oog al iets te weten komen over een ster. Naast 'licht' (de straling die je kunt zien met je oog) zenden sterren ook nog andere soorten straling uit, zoals radiostraling, infraroodstraling, ultravioletstraling en röntgenstraling. Al deze soorten straling lijken erg veel op gewoon licht, alleen zijn onze ogen er niet gevoelig voor.

Om die straling toch op te kunnen vangen, gebruiken we apparaten als radioantennes, infraroodcamera's en röntgenfoto's. Wanneer je met deze apparaten bijvoorbeeld naar onze eigen zon (de dichtstbijzijnde ster) kijkt, krijg je vrijwel hetzelfde plaatje van de zon als met je oog.

In tegenstelling tot sterren maken radio pulsars eigenlijk alleen maar radiostraling en je kunt ze dus niet met het blote oog of een gewone telescoop zien. Daarom duurde het ook lang voor ze ontdekt werden.

## Hoe zijn radio pulsars ontdekt?

Eind jaren '60 werd er in Engeland een nieuwe radiotelescoop gebouwd, gewoon een groot veld vol radioantennes, allemaal met elkaar verbonden. Met die telescoop werd de hemel afgezocht naar sterren (alleen of in grote groepen ver weg) die radiostraling maken. Zoals wij sterren zien als lichtpunten aan de hemel, zo 'ziet' een radiotelescoop sterren als radiostralingspunten aan de hemel. Na een paar maanden waren er al veel radiostralingspunten gevonden. De meeste waren altijd even fel (dat bleken heel grote groepen sterren te zijn, heel ver weg), andere werden ieder paar weken feller en minder fel (dit waren sterren als de zon, niet zo ver weg). Maar er was ook één punt dat snel knipperde, ongeveer één keer per seconde. Dat is vreemd, want sterren kunnen helemaal niet zo snel veranderen, daar zijn ze veel te groot voor. De zon is bijvoorbeeld al 100 keer zo groot als de aarde.

Het was snel duidelijk dat er iets nieuws was ontdekt, al wist nog niemand precies wat. Vanwege de pulserende radiostraling werd het onbekende ding 'radio pulsar' gedoopt. De meest waarschijnlijke, maar even schokkende uitleg was dat er eindelijk signalen van buitenaardse wezens waren ontdekt, die probeerden contact met ons te maken. Een tijdje geleden vertelde de intussen 70-jarige ontdekker (die de Nobelprijs heeft gekregen voor deze vondst) me over de spannende tijd die aanbrak. Omdat het onderzoekers nog niet zeker waren van hun zaak, hielden ze het nieuws geheim. Ze begrepen wel dat de stormloop van leger, bevolking en pers hen het werk anders onmogelijk zou maken. In stilte onderzochten ze het knipperende ding verder. Als het signaal door buitenaardse wezens werd gemaakt, moest het vanaf een planeet rond een andere ster worden uitgezonden. Die planeet zou je dan, heel zwak, in het signaal moeten kunnen 'zien'. Na een paar weken continue meten was het duidelijk: het planeetsignaal zat er niet in. Toch geen planeet en toch geen buitenaards leven dus.

## Wat zijn radio pulsars dan wèl?

Toen eenmaal duidelijk was dat het geen buitenaardse wezens konden zijn, werd de ontdekking van de nieuwe 'stersoort' bekend gemaakt. De vondst was direct groot nieuws. Veel andere onderzoekers probeerden nu ook te bepalen wat de vreemde knipperster dan wèl was. Na een tijdje werd duidelijk dat radio pulsars veel kleiner zijn dan gewone sterren en dat ze, omdat ze relatief klein zijn, snel kunnen draaien.

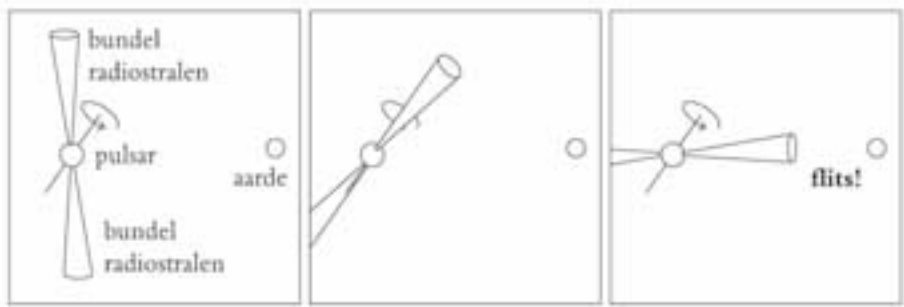

**Figuur 1.** Iedere keer als één van de bundels radiostraling over de aarde zwiept, zie je een flits.

Daarbij zwiepen ze twee felle bundels radiostraling rond. Iedere keer als één van die bundels radiostraling over de aarde zwiept zie je een radioflits, precies zoals een vuurtoren dat doet met gewoon licht (zie figuur 1).

Op dit moment zijn er ongeveer 2000 verschillende radio pulsars bekend. Sommige daarvan draaien een paar honderd keer per seconde rond. Ze draaien zo gelijkmatig dat je er een atoomklok op gelijk kunt zetten. Dat kan allemaal alleen maar omdat radio pulsars heel klein zijn, rond de 20 km doorsnede, maar toch meer wegen dan de zon (terwijl de zon zelf 1.400.000 km doorsnede is). Dat betekent dat zo'n radio pulsar 500 keer kleiner is dan de aarde, maar toch 500.000 keer zo zwaar. Radio pulsars zijn dan ook gemaakt van het zwaarste materiaal dat we kennen: een liter 'radio pulsar' zou op aarde maar liefst 1 biljard (miljoen miljard) kilo wegen. Dat is ongeveer net zoveel als de hele Mt. Everest. Dat zware materiaal kan alleen worden gemaakt wanneer een grote ster aan het eind van haar leven ontploft.

### Sterren ontploffen?

Sommige sterren ontploffen als hun brandstof op is en sommige van die ontploffende sterren worden dan radio pulsars. Ik beschrijf eerst kort hoe sterren hun licht maken en daarna wat er gebeurt als ze ermee ophouden: sommige sterren, de lichte, ontploffen niet maar worden 'witte dwergen'. Witte dwergen lijken al vrij veel op radio pulsars. Andere sterren, de zware, ontploffen wel en worden of een radio pulsar of een zwart gat.

Alle sterren die je overdag (de zon) en 's nachts (de andere sterren) aan de hemel ziet, werken hetzelfde. Ze maken hun straling (infraroodstraling, licht, ultravioletstraling, enz.) diep van binnen, waar ze lichte gassen ombouwen tot zwaardere. Grote, zware sterren zijn heter en zetten per jaar veel meer gas om dan kleine, lichte sterren. Daarom schijnen zware sterren feller. Dat betekent wel dat ze korter leven, want als alle lichte gassen op zijn, dooft de ster uit. De allerzwaarste sterren die we kennen (100 keer zwaarder dan de zon), leven 'maar' 10 miljoen jaar. De zon zelf is een vrij lichte ster. Toch weegt ze er al 300.000 keer zoveel als de aarde. De zon leeft ongeveer 1000 keer langer dan de zwaarste sterren en wordt dus 10 miljard jaar oud. Ze is nu ongeveer halverwege haar leven en gaat over grofweg 5 miljard jaar uit. Ter vergelijking, 5 miljoen jaar geleden (dus dat is 1000 keer korter) waren mensen en chimpansees nog hetzelfde; in vergelijking met mensenlevens doet de zon het dus nog vrijwel oneindig lang.

### Hoe eindigen lichte sterren?

Wanneer zware sterren geen licht meer maken, ontploffen ze. Daarna gaan ze soms als radio pulsar verder. Lichte sterren als de zon ontploffen niet en worden geen radio pulsar. Hoe ze wel aan hun einde komen beschrijf ik ter vergelijking hieronder, aan de hand van de zon.

Over een paar miljard jaar wordt de zon veel groter dan nu. De buitenste lagen drijven weg en vormen een nevel om de zon. De zon krimpt weer, tot er een vrij klein maar zwaar en fel sterretje overblijft, een 'witte dwergster'. Zo'n witte dwerg is ongeveer net zo groot als de aarde, maar ongeveer 300.000 keer zo zwaar (net zo zwaar als de zon). Sommige andere sterren die op de zon leken zijn al eerder tot witte dwerg verschrompeld. Met het blote oog zijn ze net niet te zien, maar met telescopen zijn er een paar duizend ontdekt (zie bijvoorbeeld figuur 3). Ooit waren dat allemaal sterren als de zon, waarschijnlijk zelfs met planeten zoals de aarde.

### Witte dwergen: kleine broertjes van radio pulsars

Witte dwergen zijn dus net zo zwaar als de zon, maar veel kleiner. Een liter 'witte dwerg' zou op aarde ongeveer 1 miljoen kilo wegen (ongeveer evenveel als een heel binnenvaartschip). Dat is veel meer alle materialen die we op aarde kunnen maken, maar nog steeds 1 miljard keer minder dan een liter 'radio pulsar'. Hoe kunnen witte dwergen en radio pulsars zoveel zwaarder zijn dan alles op aarde? Dat komt doordat alles wat wij uit het dagelijks leven kennen (waaronder dit boekje, en jijzelf) vooral bestaat uit lege ruimte. Al het materiaal op aarde is gemaakt uit atomen en op zijn beurt is ieder atoom weer opgebouwd uit protonen en neutronen in een atoomkern, met ver daarbuiten wat elektronen. Zo'n atoom is voor 99,9999999999999% leeg, en wijzelf en alle andere dingen op aarde dus ook (als in figuur 2, linker paneel). In een witte dwerg zijn al deze deeltjes onder grote druk en hitte op elkaar geperst (als in figuur 2, middenpaneel). Dat materiaal is veel zwaarder omdat de protonen en neutronen daarin veel dichter op elkaar zitten dan in jou.

Om radio pulsar materiaal te maken, moeten de deeltjes nog verder op elkaar worden geperst. Dat kan alleen bij onzettend hoge druk en temperatuur, en voor-

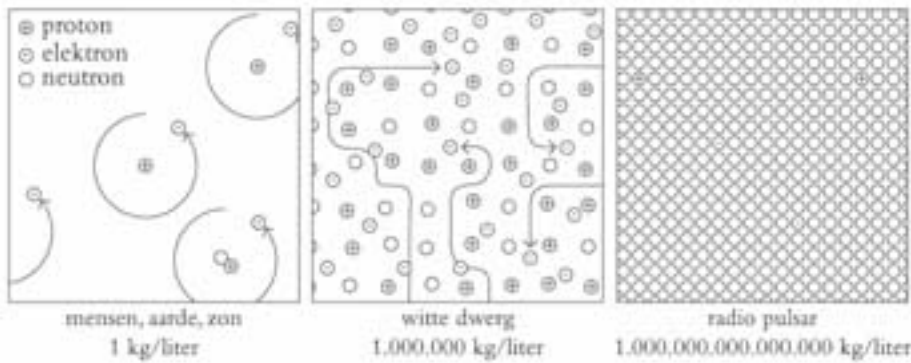

**Figuur 2.** Schema van de verschillen in de dichtheid van de zon (links), een witte dwergster (midden) en een neutronenster (rechts). In het echt zijn de verschillen in 'leegheid' veel groter.

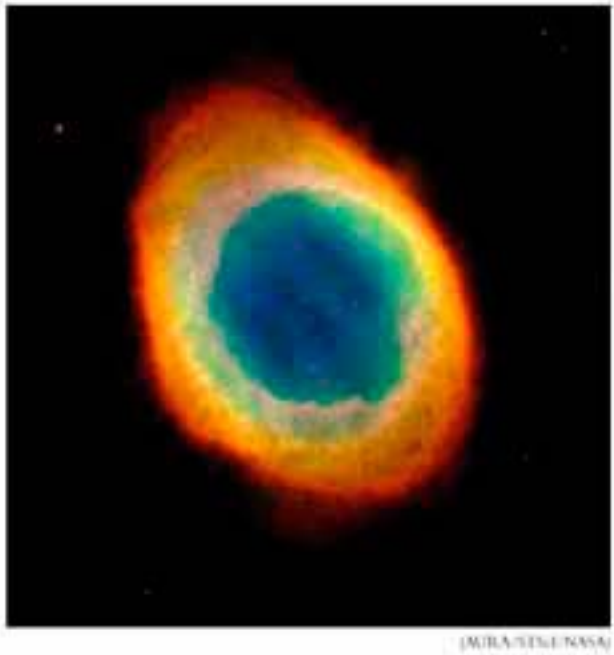

**Figuur 3.** Een wegdrijvende nevel rond een witte dwerg (het withete puntje in het midden).

[AURA/STScI/NASA]

Bij een supernova vervormt het binnenste van de ster tot een radio pulsar of een zwart gat. Ik leg kort uit wat zwarte gaten zijn en bespreek daarna de radio pulsars zelf.

*Zwarte gaten*

Bij de allergrootste sterren is de klap zo hard dat de hele ijzeren kern, zelf al meer dan twee keer zo zwaar als de zon, in elkaar wordt gedrukt tot ze kleiner is dan alles wat je kunt bedenken. De zwaartekracht om dit superzware punt is zo groot dat zelfs licht er niet weg kan komen: een zwart gat. Zwarte gaten zijn in principe vrijwel onzichtbaar, maar omdat ze een grote invloed om hun omgeving hebben, kun je ze toch vinden. De buitenste lagen van de oude ster vallen namelijk langzaam weer terug naar het zwarte gat, en draaien er met een grote kolk in, als in figuur 5. Dat maakt veel warmte en licht en dat is dan weer te zien met een telescoop. Op het ogenblik zijn er op die manier ongeveer 10 zwarte gaten gevonden.

zover we weten gebeurt het alleen maar middenin een 'supernova', de ontploffing van een zware ster.

*Waarom ontploffen zware sterren?*

Alle sterren geven licht omdat ze diep van binnen lichtere gassen omzetten in zwaardere gassen. Grote, zware sterren doen dat in verschillende rondes; eerst wordt al het lichtste gas omgezet in iets zwaarder gas. Als het lichtste gas op is, wordt al het iets zwaardere gas omgezet in nog zwaarder gas, enzovoort. Uiteindelijk is de binnenkant van de ster helemaal gemaakt van ijzergas. Dan is de koek op, want de ster kan het ijzer niet tot nog zwaarder gas omzetten. Voor het eerst in honderden miljoenen jaren kan de ster nu plotseling geen licht meer maken. Dat licht hield de ster in evenwicht en daarom stort ze nu het wegvalt opeens helemaal in. Maar als alles in het midden aankomt, kan het niet verder en met een gigantische klap stuitert een groot deel van de ster weer weg naar buiten. Deze ontploffing produceert een heel felle flits, die wel een paar weken kan duren (zie figuur 4). De meeste van de miljarden sterren in het heelal staan zo ver weg dat de flitsen van hun ontploffing niet zo opvallen, maar wanneer het een nabije ster is die ontploft, wordt ze een paar weken lang zo helder dat je haar zelfs overdag kunt zien. We weten uit oude Chinese geschriften dat dit ongeveer 1000 jaar geleden nog gebeurd is.

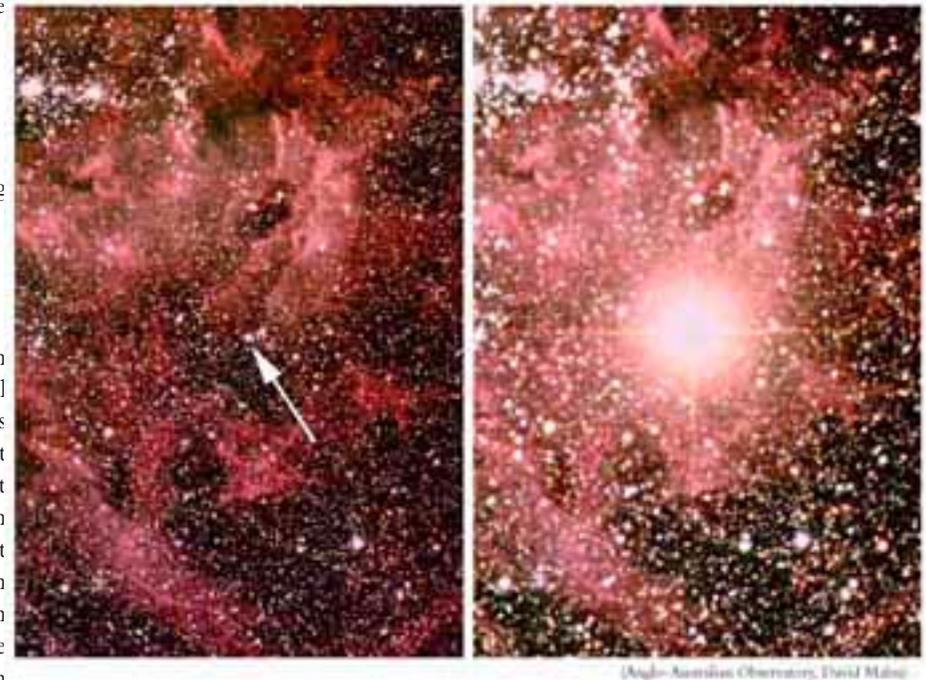

[Anglo-Australian Observatory, David Malin]

**Figuur 4.** In 1987 ging er één melkwegstelsel verderop een supernova af. Links een foto van de ster, twee jaar voor de ontploffing, rechts een foto uit de week erna.

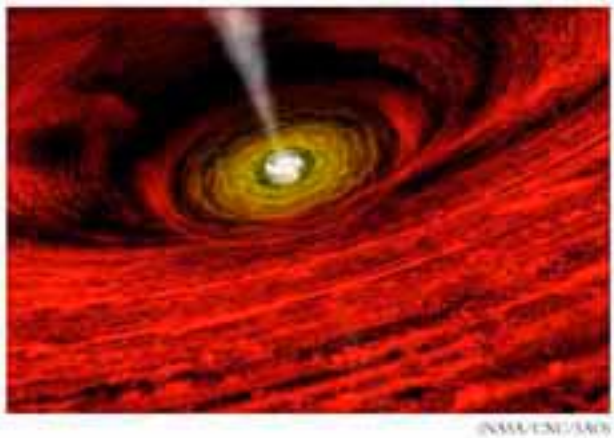

**Figuur 5.** Een impressie van de storm die rond veel zwarte gaten woedt.

(NASA/CXC/SAO)

*Een ontploffing met een verrassing*

Lichte sterren worden dus witte dwergen en de zwaarste sterren worden zwarte gaten. Bij middelzware sterren is er weliswaar een ontploffing, maar er blijft toch wat over: een radio pulsar. Veel sterren die wel zwaar genoeg zijn om te ontploffen, hebben namelijk een ijzerkern die net niet groot genoeg is om helemaal in te storten tot een zwart gat. In het laatste stadium voor de vorming van het zwarte gat, wanneer de oude ijzerkern al is samengeperst tot een grote bol van tegen elkaar aanzittende neutronen (als in het rechterpaneel van figuur 2), stopt de kern van de ster plotseling met instorten. Er blijft dan een bol over van maar 20 km diameter, maar zwaarder dan de zon en 500.000 keer zwaarder dan de aarde. Dat is de radio pulsar. Zo'n radio pulsar is ook gemaakt bij de laatste nabije supernova (die van 1000 jaar geleden, zie figuur 6).

Door de klap van het instorten wordt de radio pulsar weggeschoten uit de resten van de ster, met een snelheid van ongeveer 500.000 km per uur. De klap zorgt er ook voor dat de radio pulsar snel draait en een sterk magneetveld krijgt. Dit sterke magneetveld, ongeveer een miljoen keer sterker dan het sterkste veld dat we op het ogenblik in aardse laboratoria kunnen maken, produceert de twee felle bundels radiostraling die rondzwiepen door de ruimte. Er zijn waarschijnlijk miljoenen radio pulsars in het heelal, maar alleen als de aarde toevallig in het pad van één van hun bundels staat, zien wij ze als knipperende radiostralingspunten aan de hemel.

*Dit proefschrift*

Na 40 jaar weten we al veel over radio pulsars: waar ze ongeveer van gemaakt zijn, hoe ze geboren worden en hoeveel er ongeveer zijn. Er zijn ook nog hoop zaken onduidelijk: waar ze precies van gemaakt zijn bijvoorbeeld, maar ook hoe de felle radiostraling wordt gemaakt en waarom er wel veel jonge pulsars worden ontdekt, maar maar weinig oudere. Ik heb me vooral gericht op de laatste twee problemen, in wisselende samenwerkingsverbanden met andere onderzoekers. Ik heb onze resultaten gebundeld in dit proefschrift en vat ze hieronder samen.

In hoofdstuk 1 tot en met 6 proberen we beter te begrijpen hoe pulsars straling maken. We onderzoeken waarom pulsars op uiteenlopende manieren knipperen. Daarin zijn pulsars te vergelijken met vuurtorens op aarde, die ter herkenning vaak allemaal een verschillend aantal lichtbundels hebben: een bepaalde vuurtoren heeft bijvoorbeeld één enkele lichtbundel, terwijl de volgende er twee vlak naast elkaar heeft. De eerste vuurtoren produceert dan iedere keer een enkele flits, de tweede maakt steeds twee flitsen vlak na elkaar. Nu blijken sommige pulsars vrijwel hetzelfde te doen: in plaats van één lange puls per seconde, produceren ze iedere keer een paar korte pulsen vlak na elkaar. Blijkbaar hebben niet alle pulsars één brede radiobundel (als in figuren 1 en 7a), maar maken sommige pulsars meerdere smalle, opeenvolgende radiobundels. Volgens de meest gebruikte theorie worden die radiobundels gemaakt door grote vonken vlakbij het oppervlak van de radio pulsar (zoals in figuur 7b). Iedere vonk is waarschijnlijk enkele tientallen meters in doorsnede. De radiostraling van de vonken wordt door het magneetveld versterkt, tot de vonken zo helder zijn dat ze hier op aarde te zien zijn, een paar biljard kilometer verderop. Het vreemde is nu dat die vonken rond de magnetische pool lijken te draaien, als een soort draaimolen (zie figuur 7b).

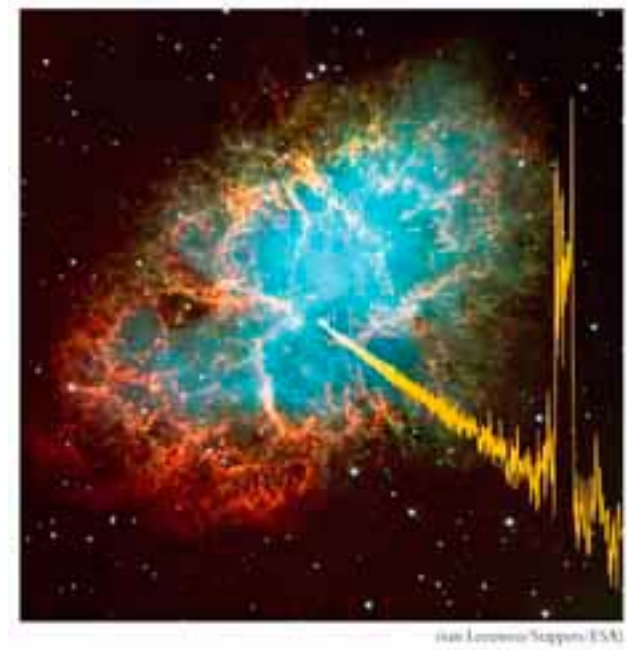

**Figuur 6.** De restanten van de laatste nabije supernova: een ring van gas en stof, ooit de buitenste lagen van de oude ster maar nu door de ontploffing naar buiten geslingerd. In het midden van deze nevel staat een radio pulsar. De pieken aan de rechterkant geven het knipperen van de radiostraling aan, gemeten met de Westerbork radiotelescoop.

(Van Lorenzo Stappers/ESA)

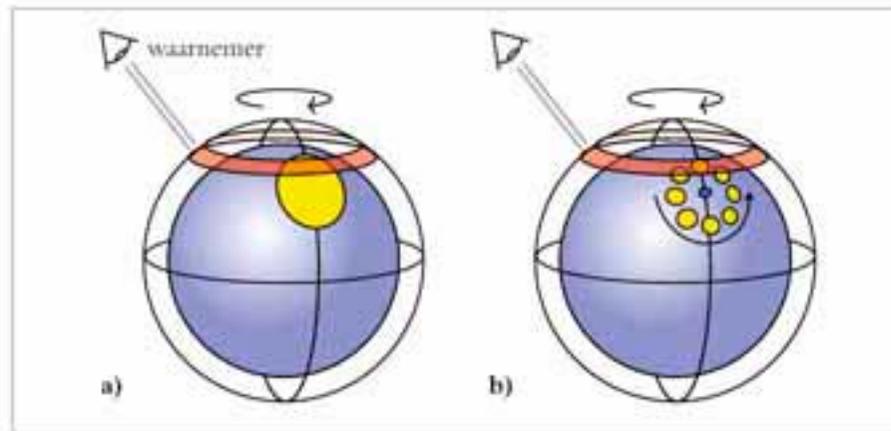

**Figuur 7.** Schets van twee verschillende opstellingen van de radiobundels. De bollen stellen radio pulsars voor. Alleen de lichte stippen maken straling. Terwijl de pulsar draait, ziet ons oog wat in de band ligt. Als een stip de band kruist, zien we een flits. **a)** één enkele bundel **b)** meerdere bundels. De vonken (lichte stippen) draaien om de magnetische pool. Ons oog volgt de rode band; ze gaat over de vonken, en ziet dan een tijd niets bijzonders. Ondertussen draaien de vonken door. Als ons oog weer bij de vonken is (de pulsar is dan eenmaal rond), zijn de vonken verschoven.

In de hoofdstukken 1 tot en met 3 beschrijven we ons onderzoek naar deze verschuivende vonken. Soms maken pulsars bijvoorbeeld even geen straling meer, maar wanneer ze na een paar seconden weer aangaan, bewegen de vonken anders. Uit die interactie proberen we meer te leren over de vonken; hoeveel er zijn en hoe snel ze ronddraaien bijvoorbeeld. Zo hebben we voor het eerst laten zien dat de vonken veel langzamer bewegen dan verwacht werd.

Nadat de vonken de straling hebben gemaakt, reist de straling door het gas en magneetveld rond de radio pulsar. Daarbij wordt de straling veranderd en afgebogen, vooral door het magneetveld. In hoofdstukken 4 tot en met 6 gebruiken we die veranderingen om eigenschappen van het gas en magneetveld te bepalen.

In hoofdstuk 7 verklaren we waarom er meer jonge pulsars gevonden worden dan oude. Omdat het energie kost om een sterk magneetveld door de ruimte te slepen, gaan pulsars steeds langzamer draaien. Pulsar leeftijden volgen daarom uit hun ronddraaisnelheden. De meeste pulsars blijken rond de 1 miljoen jaar oud te zijn. Er zijn er al een stuk minder van 10 miljoen jaar oud, en bijna geen van 100 miljoen jaar. Daarom is vroeger geopperd dat het magneetveld bij het ouder worden verzwakt. Daardoor zouden pulsars steeds minder helder worden, tot ze uitdoven. Eenmaal uitgedoofd zijn oude pulsars onvindbaar. Hoe magneetveld überhaubt uit zichzelf zwakker zou kunnen worden, bleef onduidelijk. In hoofdstuk 7 laten we zien dat het tekort aan oude pulsars eenvoudiger kan worden verklaard: er is bewijs dat jonge pulsars bredere stralingsbundels hebben dan oude, waardoor de kans dat de aarde toevallig in de radio bundel ligt voor jonge pulsars groter is dan voor oude. En omdat pulsars alleen maar kunnen worden ontdekt als hun radio bundel over de aarde zwiept, worden er meer jonge pulsars ontdekt dan oude, zelfs als er eigenlijk evenveel zijn.

Voor het onderzoek in hoofdstukken 1 tot en met 6 hebben we gebruik gemaakt van de Westerbork radiotelescoop. In Westerbork, Drente, staan veertien grote metalen spiegels die de straling van radio pulsars bundelen (figuur 8). Daarmee kan heel precies bepaald worden hoe een bepaald radiostralingspuntje aan de hemel helderder of zwakker wordt, net zoals een oog dat kan met een lichtpuntje. Over een paar jaar wordt er in Nederland een nieuwe radiotelescoop gebouwd. Deze telescoop, Lofar genaamd, wordt de meest geavanceerde telescoop ter wereld. In hoofdstuk 8 berekenen we hoe we met Lofar zoveel mogelijk nieuwe radio pulsars kunnen ontdekken. Lofar is ongeveer 20 keer zo groot als Westerbork en kan dus veel radiostraling bundelen. Daardoor kunnen we veel nieuwe, lichtzwakke pulsars vinden, wel 1500. Daarmee zal het aantal bekende pulsars in één klap worden verdubbeld.

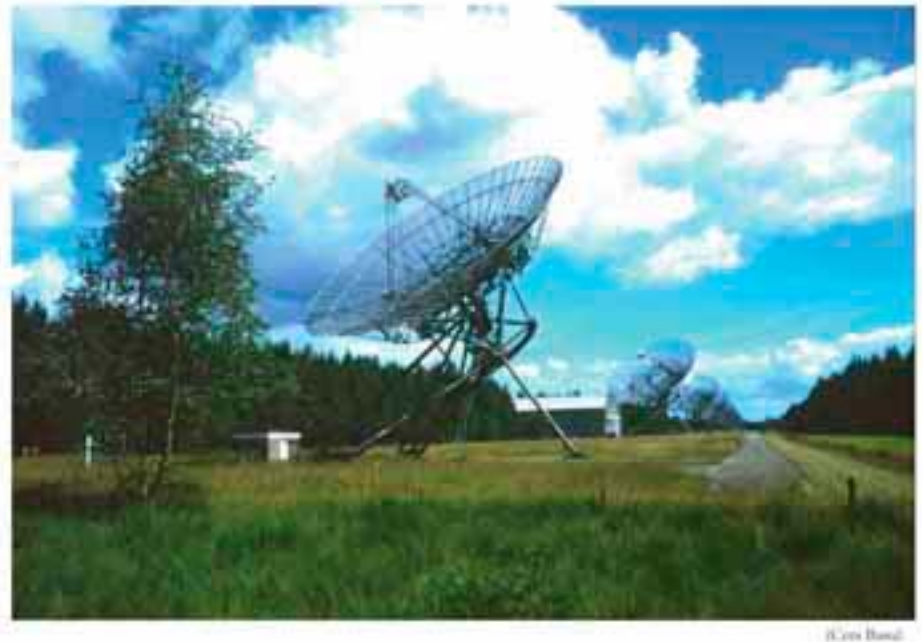

**Figuur 8.** In Westerbork, Drente, staat de radiotelescoop die we hebben gebruikt voor veel onderzoek in dit proefschrift.

### The discovery of radio pulsars

At a radio-pulsar conference in Crete, two years ago, Anthony Hewish (the man responsible for the discovery of the first four pulsars in 1967) related on how it all came about. He had a new radio telescope built, operating at a frequency of 80Mhz, with which he wanted to determine the angular sizes of a large number of galaxies from their scintillation properties. As this scintillation can be rapid, the telescope had a fast response time, and to survey large numbers of galaxies, all sources were tracked systematically. In hindsight, the new combination of these two features was responsible for recognising the first pulsars. Hewish' graduate student Jocelyn Bell, who followed the different sources on the sky, noticed that the intensity of one of the sources varied more strongly than would be expected just from scintillation. On a faster chart recorder, installed to further investigate the source type, the radiation appeared to be a series of short pulses, all only little more than a second apart. Such fast signals are usually man-made, but as the emission was linked to a fixed position on the sky, this now seemed unlikely. Hewish eliminated all terrestrial origins one after the other and it became more and more probable that the signal came from outside the solar system. The pulse period (roughly a second) was much shorter than the time scale on which normal stars can vary, and because the recurrence of the pulses was extremely regular, the most probable source appeared to be extra-terrestrial life signaling. Over the next few thrilling weeks, Hewish dared not make his discovery public; first for fear of finding his research taken over by the military; second for the uproar it would create in the world and for "all the madmen that would immediately start signalling back, while we had no clue on the intentions of the senders".

They did realise that, if the signal came from a planet around another star, it would have to be Doppler-shifted by the motion of that planet. After several weeks of timing pulse arrivals, the changing delays in these arrival times could be exactly accounted for by the motion of the earth alone: the source could not be on a planet, and was probably not artificial after all.

Still, for the source to vary on a time-scale of seconds, it had to be small. Only the already observed white dwarfs or the hypothetical neutron stars could oscillate with the 1-second period observed. Over the next weeks, Bell found three other pulsars. The shortest period found in these new pulsars, 0.25 seconds, was already unnervingly short for a white dwarf oscillation. [1]

### Rotating neutron stars

The oscillating white dwarf model remained popular at first (Hewish et al. 1968). White dwarfs were observable and understood, in contrast to neutron stars. The detection of a second periodicity in the original pulsar (the drifting subpulses we will discuss again later) supported the white dwarf model even more, as it could be a higher order of the oscillation (Drake & Craft 1968). But in late 1968, two new pulsars were discovered that would quickly change the situation: in the Vela supernova remnant a 0.089s pulsar was found (Large et al. 1968), and the Crab nebula turned out to host a 0.033s pulsar (Staelin & Reifenstein 1968). First of all, the rapidity of the pulsations ruled out oscillating white dwarfs; secondly pulsars were now linked clearly to supernovae. In 1934, Baade & Zwicky had already suggested that stars might end their lives as neutron stars. The discovery that the Crab pulsar slowed down then firmly established the rotating neutron-star model originally proposed by Gold (1968). The combination of the known age of 920 years and the measured slowdown rate could be well explained by a dipolar magnetic field, which had to be strong to power the Crab nebula (Pacini 1967).

[1] For Hewish' written account, see Hewish (1975); for historical overviews, see Manchester & Taylor (1977) and Lyne & Smith (1998)

### Acceleration

The outline of the radio-pulsar functioning was now complete. A radio pulsars is a rotating neutron star, progeny of a massive star, formed in a supernova, with a strong magnetic field.

Particles accelerated in this field had to be instrumental in forming emission higher up in the magnetosphere, that much was clear. Over the next ten years, the groundwork for two types of models for the acceleration of the particles was laid. In both, the combination of rotation and magnetic field creates an electric field parallel to the magnetic field, that accelerates the particles outward. If the pulsar magnetic field and rotation axes are aligned, the electric field they produce will extract electrons from the neutron star crust; if they are anti-parallel, the electric field will try to pull out ions in stead. In the 'polar gap' model, the ions in the crust are so tightly bound that the electric field cannot extract them, and when the particles previously present in the magnetosphere are expelled, a vacuum forms over both magnetic poles. As the electric field over the poles is now no longer screened out by the plasma, it increases with the vacuum buildup. When a particle pair is formed in this vacuum, the vacuum breaks down in a cascade of particles, half of which are accelerated outwards (Sturrock 1971; Ruderman & Sutherland 1975). The difference between this model and the 'steady space-charge-limited-flow' model is that in the latter model the binding energy of the ions in the crust is lower; ions as well as electrons can be continuously pulled out of the crust, independent of the alignment of the rotation and magnetic field axes. The pulsar rotation and the magnetic field create an electric field parallel to the field lines, that accelerates the particles outward (Arons & Scharlemann 1979).

The polar gap model qualitatively explains different phenomena (microstructure, drifting subpulses, plasma instabilities from the non-stationary flow) and can be used to make some quantitative predictions. Yet in the population of radio pulsars, magnetic field strengths range over 5 orders of magnitude, and rotation periods over more than 3; that the ions remain bound to the crust in all these pulsars seems problematic. Also, in polar gap models only half of the pulsars (those with opposite rotation and magnetic axes) emit, increasing the already uncomfortably high neutron-star birthrate needed to explain the number of pulsars observed.

The steady flow models appear more sound physically, but cannot begin to explain the phenomena mentioned above. Especially the so-called 'subpulse drift' is a problem; in many pulsars each single pulse is composed of several subpulses, and when comparing subsequent pulses, these subpulses move slowly through the pulse window (cf. figure 2, chapter 1). In the polar gap model, each of these subpulses is produced by a 'spark', a series of subsequent particle cascades at a fixed position over the polar cap. The magnetic field forces these sparks to move around the magnetic pole, which translates to a drift of subpulses through the pulse window. In chapters 1–3 we investigate observationally how the subpulse drift behaves after the 'nulls' (stops of all emission) that some pulsars show. We find that the drifting-nulling interaction tells much about the true motion of the sparks; the spread in the inferred velocities of the sparks is much larger than previously thought possible.

*Emission*

The radio emission probably originates from a height of several tens of neutron-star radii, quite far from the particle-acceleration region, probably by some kind of plasma instability or maser action (cf. Melrose 2003); the intensity of the radiation requires some kind of coherent emission process. The polarisation of the radiation follows the local magnetic field (Radhakrishnan & Cooke 1969). The radiation is refracted as it travels farther out through the pulsar magnetosphere. In chapters 4–6 we investigate the interaction of drifting subpulses, refraction and polarisation. We show how refraction changes subpulse patterns (chapter 4); we find that the large polarisation-angle changes observed within single subpulses can be explained in terms of orthogonal polarisation modes (chapter 5); and we demonstrate that, in contrast to the handedness, the amount of circular polarisation in pulse profiles is independent of frequency (chapter 6).

*Evolution*

If pulsars were neutron stars vibrating, the vibration amplitude might change in time, but the pulse period would remain constant. If caused by rotation, the pulsations should slow down; and indeed such a slowdown, first detected in the Crab pulsar, has been found to be a common feature of all radio pulsars.

As the magnetic field is the braking force in the pulsar rotation, one can estimate the field strength from the pulsar period and period derivative. Although more and more neutron stars with high magnetic fields are observed (mostly from their X-ray and gamma-ray emission) the field strength in most normal radio pulsars is roughly $10^{12}$ G. Ever since the discovery of pulsars the longevity of the magnetic field has been questioned (cf. Gunn & Ostriker 1970). In chapter 7 we show that field decay is not needed to explain the observed relative lack of long-period pulsars; furthermore, we find that the scarcity of high-field radio pulsars does not have to mean that high-field pulsars ('magnetars') do not emit in radio: even if many are born, their evolution is so fast that only few, if any, are visible at a given time.

*Detection*

All of chapters 1–6 use data taken with the Westerbork Synthesis Radio Telescope. Within several years, Westerbork will be superseded by Lofar, the new low-frequency radio telescope. For chapter 7 we already simulated several different telescopes and surveys; in chapter 8 we do the same for Lofar. We simulate and compare strategies for finding new pulsars and find that a 60-day all-sky survey can find 1500 new radio pulsars, doubling the currently known population.

# NULL-INDUCED MODE CHANGES IN PSR B0809+74

**with Marco Kouwenhoven, Ramachandran, Joanna Rankin and Ben Stappers**

We have found that there are two distinct emission modes in PSR B0809+74. Beside its normal and most common mode, the pulsar emits in a significantly different quasi-stable mode after most or possibly all nulls, occasionally for over 100 pulses. In this mode the pulsar is brighter and the subpulse separation is less. The subpulses drift more slowly and the pulse window is shifted towards earlier longitudes. We can now account for several previously unexplained phenomena associated with the nulling-drifting interaction: the unexpected brightness of the first active pulse and the low post-null driftrate. We put forward a new interpretation of the subpulse-position jump over the null, indicating that the speedup time scale of the post-null drifting is much shorter than previously thought. The speedup time scale we find is no longer discrepant with the time scales found for the subpulse-drift slowdown and the emission decay around the null.

## 1. Introduction

Lately, it has been quiet around bright PSR B0809+74, once the canonical example of regular subpulse drifting and the centre of lively discussions.

In the thirty years since their discovery by Drake & Craft (1968), drifting subpulses feature in many discussions about the nature of pulsars and the pulsar's emission mechanism. The subpulse-drifting phenomenon in itself is simple: when comparing adjacent individual pulses, the subpulses that comprise them are seen to shift regularly through the average pulse window. The left panel of Fig. 2, for example, shows a recent observation in which these subpulses form their driftbands.

In 1968, the interpretation of Drake & Craft that the subpulse drifting is the pulsation of the neutron star, is ground zero for much debate. Over the following years, the ever increasing quality and volume of observations of subpulses in different pulsars continue to limit possible models and require their refinement. It is in this process that observations of PSR B0809+74 take the lead.

Alexeev et al. (1969) and Vitkevich & Shitov (1970) are the first to detect the driftbands of PSR B0809+74. Cole (1970) then notes that the driftrate of the subpulses occasionally changes, without being able yet to identify the nulls as the trigger. One year later, Taylor & Huguenin (1971) do find that this occasional cessation of the pulsar's emission precedes a changing driftrate, lasting a few driftbands.

In his paper solely devoted to the drifting in PSR B0809+74, Page (1973) finds that the subpulse separation changes through the profile. He also introduces the 'subpulse phase', the difference between the actual position of the subpulse and its predicted position. While this makes it easier to plot long series of subpulse positions horizontally, it also introduces the notion that the subpulse position dramatically jumps in phase over a null. From the data however, it is clear that the subpulse position over the null does not change much at all – on the contrary, it is the preservation of phase that is the astonishing phenomenon.

Unwin et al. (1978) are the first to note this. They find that the position of the subpulses is identical before and after the null and conclude that, even though there is no emission from the pulsar, the emitting structures survive the null. Five years later, Lyne & Ashworth (1983) use a method devised by Ritchings (1976) to find the nulls of PSR B0809+74 and investigate the time scales associated with the change in drift pattern around a null. The first time scales they look into are those of the decay and rise of emission around a null. A finite decay and rise time would cause the pulses neighbouring a null to be less bright than the average. A relative dimness of the last pulse before the null is found indeed, pointing at a decay time shorter than 5% of the pulse period. Contrary to the expectations, however, the first pulses after the nulls are found to outshine the average ones.

The next time scales considered are those of the slowdown and speedup of the subpulse drift around the null. Lyne & Ashworth notice that, contrary to the findings of Unwin et al., there is a some change in subpulse positions over the null. They suggest that this change originates in subpulse drift that is caused by a gradual speedup of the drifting during the null. Being several tens of pulse periods, this speedup time scale is much larger than the time scales found for the emission decay and rise, and the subpulse drift slowdown.

In 1984, simultaneous observations at 102 and 1412 MHz by Davies et al. confirm the curvature of the driftbands found by Page (1973) and show that the subpulse width varies across the pulse profile.

Although work on PSR B0809+74 quiets down after this intense period, significant observational progress for the understanding of subpulse drifting in general is made on 0943+10. Deshpande & Rankin (1999, 2001) are able to detect periodicities in the subpulse strengths that lead to a determination of the recurrence time of individual subpulses.

On the theoretical front, the initial hypothesis that the subpulse drifting originates in pulsations of the neutron star (Drake & Craft 1968) is soon abandoned. The subpulses are now thought to be formed in the magnetosphere of the pulsar, and the topic of debate changes to the exact location of their formation. The observed periodicity is first thought to be caused by quasi-periodic fluctuations in the plasma near the co-rotation radius of the pulsar. The polarisation and beaming of the subpulses, however, later tie the emission mechanism to the surface of the pulsar.

In 1975, Ruderman & Sutherland build on the Goldreich & Julian (1969) and Sturrock (1971) models to propose a theory that incorporates the drifting subpulses and the nulling. They suggest that the pulsar emission is formed in discrete locations around the magnetic pole. These locations would rotate around the pole like a carousel. In one of many revisions of the Ruderman & Sutherland model, Filippenko & Radhakrishnan (1982) propose a model that explains the survival of the subbeam structure over the null. The work of Deshpande & Rankin (1999, 2001) visualises these models by mapping the observed subpulses onto their original positions on the carousel.

In this paper we investigate the behaviour of the pulsar emission around nulls. We will use observations of two extraordinary long sequences of post-null behaviour to explain the features of general subpulse drifting around nulls in a new framework.

## 2. Observations

*Data*

Over a period of about 18 months we have observed PSR B0809+74 with PuMa, the Pulsar Machine, at the Westerbork Synthesis Radio Telescope (WSRT) in the Netherlands. Most observations lasted several times the scintillation time scale of this pulsar. We have only used the data in which the pulsar was bright, amounting to a total of 13 hours. With a pulse period of 1.29 seconds, this comes down to $3.6 \times 10^4$ pulses.

The collecting area of WSRT and the low system noise made it possible for PuMa to record the data with a high timing resolution and a high signal-to-noise ratio (SNR). The WSRT consists of fourteen 25-meter dishes arranged in east-west direction. For our observations, signals from all the fourteen dishes were added after compensating for the relative geometrical delay between them, so that the array could be used like a single dish with an equivalent diameter of 94 meters. The pulsar was observed with PuMa (Kouwenhoven 2000; Voûte 2001; Voûte et al. 2002) at centre frequencies between 328 MHz and 382 MHz, with bandwidths of 10 MHz. Each 10-MHz band was split into 64 frequency channels, and the data was recorded, after being digitised to represent each sample by four bits (sixteen levels). After inspection for possible electrical interference, we corrected for the interstellar dispersion during our off-line analysis.

*Finding nulls*

Normally, the energy of individual pulses varies from about 0.5 to 2 times the average. When the pulsar is in the null state however, which it is for about 1.4% of the time, it is does not emit. In this case, very little energy is found where the pulse is expected. We have used this difference in on-pulse energies to identify the nulls. To do this, we compared the on-pulse energies with the energy found in off-pulse region, where no emission is expected.

To calculate the on-pulse and off-pulse energies, we split each pulse period $P_1$ (360° in longitude) into three sections. The first section, 0.15 $P_1$ wide, was centred around the peak of the average profile. The second section, equally wide, consisted of a part of the off-pulse region. A third section, 0.5 $P_1$ wide, contained a different, independent part of the off-pulse data and was used to estimate the noise and baseline of the signal. This baseline was subtracted from both the on-pulse and off-pulse data.

To calculate the energy in each pulse, we determined the central range in longitude that on average contained 90% of the power the pulsar emitted in the on-pulse section. For this region we summed the amplitudes of the signal to get the energy content. We did the same for an equally large piece of the off-pulse section. To correct for long term variations in the pulse energy due to scintillation, we subsequently scaled both energies with the average energy of the 100 surrounding pulses.

We defined that, in the null state, there is no observable radiation from the pulsar. This means that noise is all one observes at the time a new pulse is expected. The energy distribution of the null-state is then identical to that of the off-pulse region (high distribution in the background of Fig. 1). For our data this meant that all pulses with energies less than the highest energy found in the off-pulse distribution were nulls (small peak around zero energy in Fig. 1). Furthermore, the off-pulse distribution gives a better estimate of the expected spread in null energies than the null distribution itself, due to the larger number of pulses included.

This method, originally devised by Ritchings (1976), was first used on PSR B0809+74 by Lyne & Ashworth (1983). In their case, there was considerable overlap between the on-pulse and off-pulse distributions, hindering their identification of the nulls. We wanted to be certain that the set of nulls we found was genuine and complete; therefore we only used the 60% of our observations in which the on-pulse and off-pulse distribution were separated by more than 0.1 E/<E>.

With a 3-sigma point at 15 mJy, the low noise makes it possible to fully distinguish between pulses in the normal state on the one side, and those in the null state on the other, giving unprecedented insight into the nature of the null state.

*Fitting subpulses*

The major step in our data reduction was to extract the four basic elements of the individual pulses: the number of subpulses and their respective positions, widths and heights. By doing so we greatly increased the speed of subsequent operations, but at the same time kept a handle on the physics by retaining the parameters most important for visualising the data.

The high SNR of our observations showed a wealth of microstructure in the individual subpulses, which on average were Gaussian in shape. Plainly fitting Gaussians to the intricate, many featured subpulses did not immediately result in locating them reliably. Due to the changing amplitude of the signal throughout the window, the microstructure of the brightest subpulses was often more important to the goodness-of-fit than complete, weaker subpulses near the edge of the pulse window. In order to still find these weak subpulses, we devised a two-step approach of first finding the strong subpulses and then using their driftpath to suggest the position of weaker subpulses. This approach worked very well.

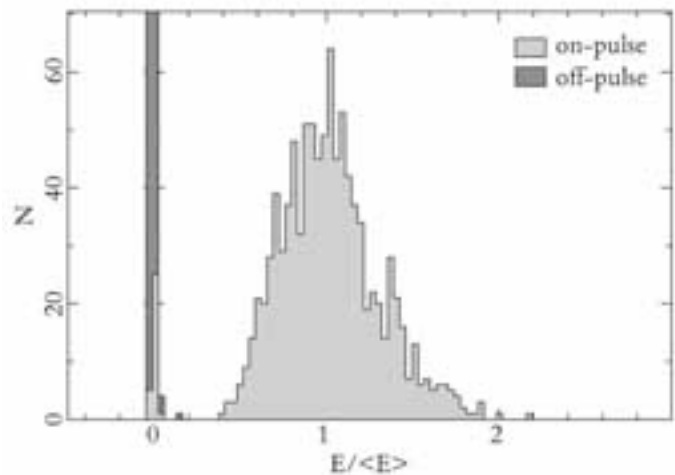

**Fig. 1.** Histogram of the energy in 1000 consecutive pulses, each scaled with the average energy of the 100 pulses that surround it. In the foreground (light gray) we show the distribution of the energies found in the on-pulse window. In the background (dark gray) the energies found in an equally large section of the off-pulse window are shown, peaking far off the scale around N=900.

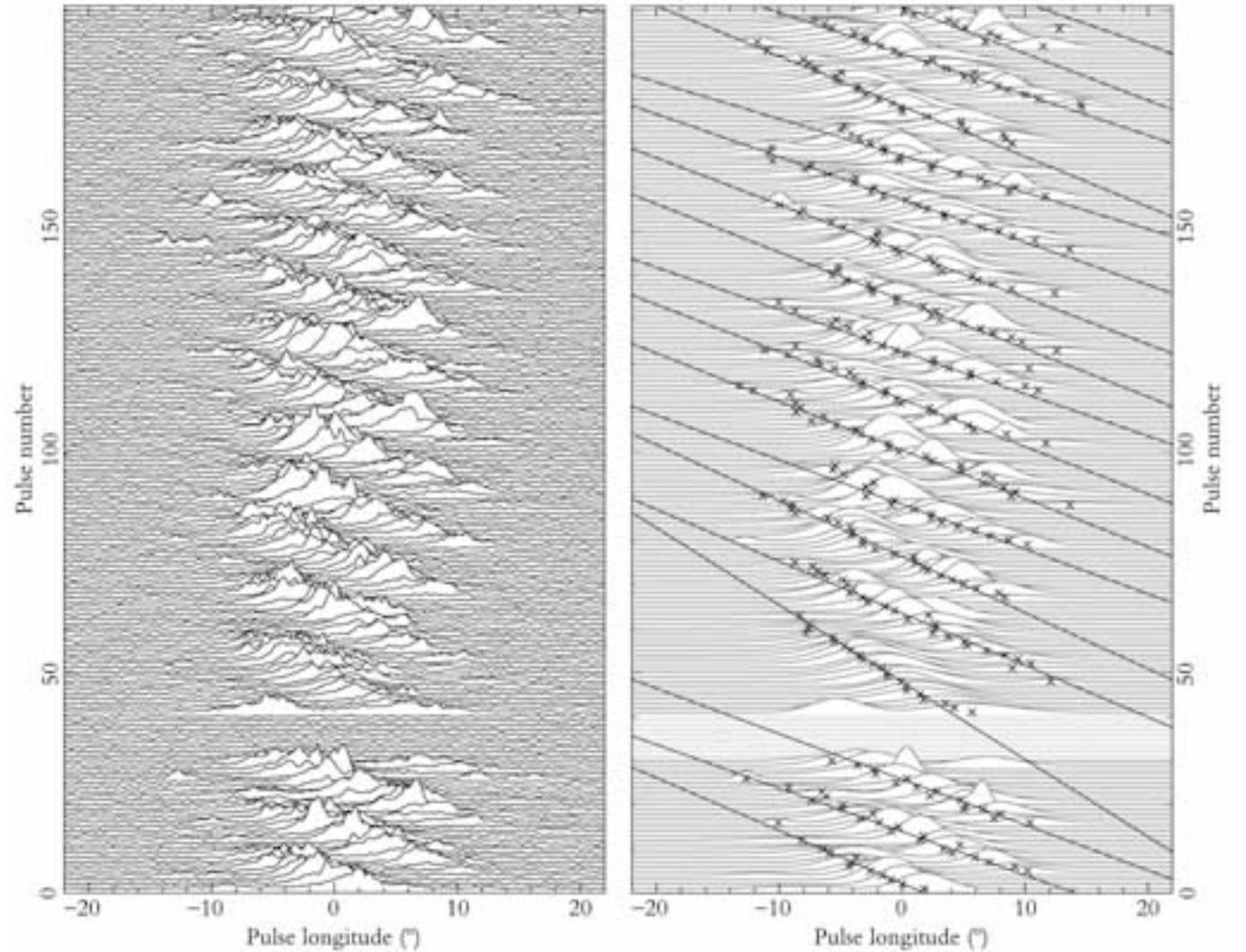

**Fig. 2.** Example of the fitting of the subpulses. Left panel: stacked pulses in the on-pulse region. Right panel: after finding nulls (lighter colour), and fitting subpulses (crosses at the base of the fitted Gaussians), we fit the driftbands (black lines). In this and all other figures, 0° longitude is defined as the centre of the best Gaussian fit to the average profile.

For our fitting, we assumed the subpulses to be Gaussian in shape, non overlapping, and to have a full width at half maximum (FWHM) less than 15°. We used a Levenberg-Marquadt method (Press et al. 1992) with multiple starting configurations to produce goodness-of-fit values for different numbers of subpulses. By comparing these $\chi^2$ values, we located the subpulses that had high significance levels.

Upon finding these normal to strong subpulses, we set out to detect the weaker ones over the noise. For this purpose we composed driftbands out of the individual subpulses already identified. On a one-by-one basis, a subpulse was either added to a path that had predicted its position to within $P_2/4$ ($P_2$ being the average separation between two subpulses within a single pulse, about 11° longitude), or taken to be the start of a new path. Paths ended if no subpulse fitted the path for more than 5 pulses, or upon reaching a null. On the first run, only paths longer than ten pulses were allowed to survive, so as to eliminate the interference of short run-away paths. This allowed a steady pattern of long paths to grow, incorporating about 90% of the subpulses found. On the second run, paths longer than 5 pulses were allowed to form: these shorter paths occur only around nulls, where the normal, long paths are interrupted.

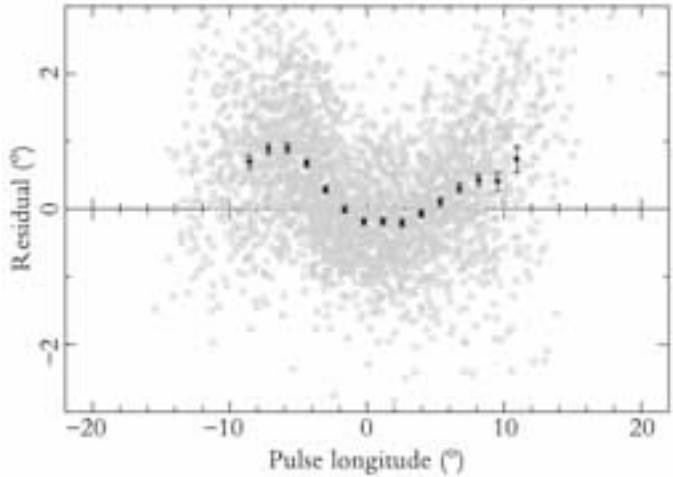

**Fig. 3.** Horizontal residuals to straight line fits to the driftbands, individual and binned, for 2000 pulses.

The drifting pattern thus formed predicted the locations of the weaker subpulses. Around each predicted position, we isolated a section of the data. This section was taken as large as possible without interfering with other, previously fitted subpulses. We checked the significance of fitting a single Gaussian to indicate the presence of a subpulse. In the same way as described above, the final driftpattern was then identified using the extended set of subpulses. To check the method, we compared the observed average profile with the one recreated from the Gaussians fits to the subpulses. From their match, we concluded that the fitting procedure was effective.

## 3. Results

### Driftband fitting

The driftbands are not straight, but show a systematic curvature (Davies et al. 1984). This is seen most strikingly when we plot the residuals to straight line fits: Figure 3 shows the longitudinal offset between the actual and predicted positions of the subpulses. Near the edges of the profile the subpulses arrive later than is to be expected based on a linear driftband, in the middle they arrive sooner.

This probably explains the results of Popov & Smirnova (1982). After fitting straight lines to all driftbands, they find that the driftrate of the last driftband before the null is 20% higher than that of a normal driftband. As seen in the curvature of the driftbands, however, a normal driftband consists of a fast drifting first half and a slow drifting second half. If a driftband is cut by a null, the part before the null consists only of the fast drifting first half, resulting in a higher average driftrate over this shorter driftband.

To make our results independent of the longitude, the non-linearity of the driftbands is taken into account in all the following driftrate calculations. In these cases, the positions of individual subpulses are corrected by subtracting the appropriate residual value.

### Shortest nulls

The main criterion used to decide whether or not to include certain datasets was the complete separation of the on-pulse and off-pulse energy distributions. Therefore, the set of null identified is genuine and complete. This allows us to investigate the underlying statistics of null occurrence and length, and estimate the influence of short nulls.

The chance of finding PSR B0809+74 in the null state is on average about 1.4%. If the null lengths were distributed in a Poissonian fashion, we would find a preponderance of one-pulse nulls and very few longer ones. However, in the null-length histogram (Fig. 4) we find a peak at nulls of length 2, a significant decrease towards shorter lengths and a considerable number of long nulls, showing that the occurrence of nulls is not governed by pure chance.

Comparing this histogram with the one previously found by Lyne & Ashworth (1983), we see that we identify about twice as many one-pulse nulls in the data. Still, many nulls shorter than one period must pass unnoticed, as they occur when the pulsar faces away from us. We know that nulls, especially long ones, have a distinct impact on the drifting pattern in their vicinity. Although the impact of shorter nulls is less, a large number of unnoticeable, short nulls ($<P_1$) might seriously influence the drifting pattern we are trying to understand.

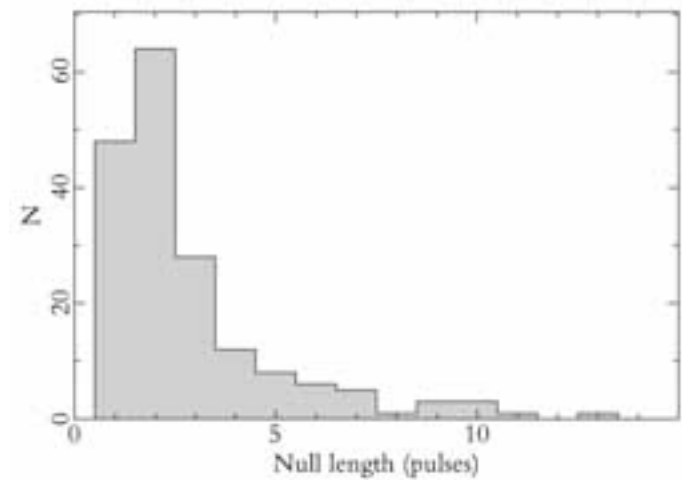

**Fig. 4.** Length histogram for the 180 nulls.

Using the null-length histogram we can estimate the number of these short nulls. It peaks at two-pulse nulls, and the distribution decreases towards shorter lengths. Assuming that the underlying distribution of null-lengths is continuous, this tendency of decreasing occurrence towards shorter nulls implies that there is a small number of nulls shorter than one pulse period. Extrapolating the decrease leads to an estimate of about 15 unnoticeable short nulls in the null-length interval from 0 to 0.5 pulses. The low frequency of their occurrence (0.04%) indicates the influence of short nulls on the drifting pattern is negligible.

### Null versus burst length

The next question we address involves the interval between adjacent nulls (the so-called burst) and the duration of the nulls. Does waiting longer for a null mean it will last longer, too? We have checked these relationships and have found that the lengths of neighbouring nulls and bursts are independent.

### Position jump over nulls

If the nulling mechanism is independent of the position of the subpulses, we expect that the distribution of subpulse positions is the same for the normal pulses and the pulses that immediately precede a null. We find no proof of differences in these distributions and conclude that there is no preference for a null to start at a certain subpulse position.

The positions of subpulses change over a null. We derived this shift of the subpulse pattern for each null in our sample. As the positions of the individual subpulses were already identified, this shift was simply extracted. Each subpulse before the null matched a subpulse after the null, if the latter fell within $-3P_2/4$ to $P_2/4$ of the former. This range is symmetric around the average expected jump over a null, so as to minimise the number of ambiguous cases. In the following analysis, we used the average of all individual subpulse shifts within one pulse.

Figure 5 shows this shift in the subpulse position over a null, corrected for the non-linear behaviour of the driftband. If the motion of the subpulses were independent of the emission, the subpulses would continue to drift invisibly throughout the null state, and reappear at a very different position. The associated shift in positions would then be spread around the diagonal dashed line in Fig. 5. If, on the other hand, the subpulse drift would cease abruptly and completely during the null, the jump in position would be between zero and two times the average shift in longitude between normal pulses. The uncertainty in this estimate arises from the fact that neither the start nor end of the null are known more accurately than to one pulse period. The mean jump in position over the null would then follow the horizontal dash-dotted line in Fig. 5.

After discarding the three ambiguous cases (points near the top edge of Fig. 5), we computed averages for each null length. These averages follow neither of the two cases outlined above. There is too much change in position over the null to be accounted for by just an abrupt stop of the drift, and there is no evidence for steady (albeit slower than regular) drift during the null. We do see that for nulls longer than one, the jump is constant, $1.47\pm0.16°$ above the offset value that we would expect in the case of no drift. This independence of null length and subpulse jump over the null is shown as the dotted horizontal line in Fig. 5.

### Driftrate around the null

We have computed the driftband slope around nulls, correcting for their general curvature.

For the driftrate before the null, we fitted straight lines to the last six subpulses of each driftband the ended with the null. We find that this driftrate just before a null does not deviate from the normal driftrate.

The driftrate after nulls is different from the normal driftrate, though. Although there is some spread, all the driftrates we find after longer nulls are lower than the normal average value (Fig. 6).

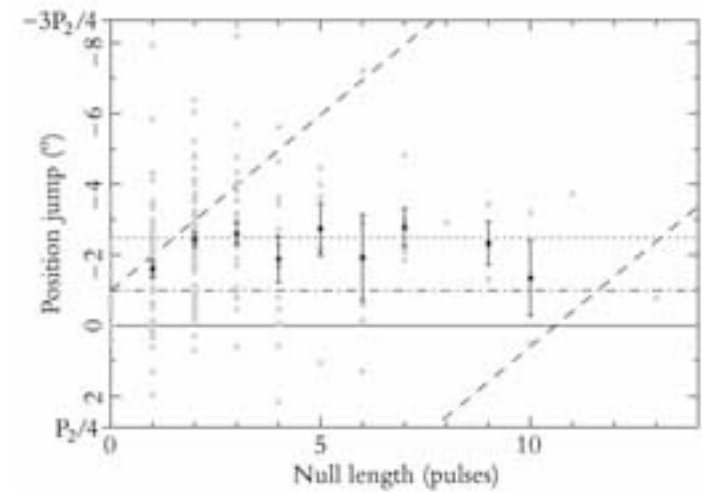

**Fig. 5.** Jump in subpulse position against null length. The bottom of the plot falls at $P_2/4$, the top at $-3P_2/4$. The diagonal dashed line is the predicted subpulse path if the drifting were independent of the nulls. The horizontal dash-dotted line is the predicted path for a sudden and complete stop of drifting during the null. The horizontal dotted line is the average jump found for the nulls longer than 1 period.

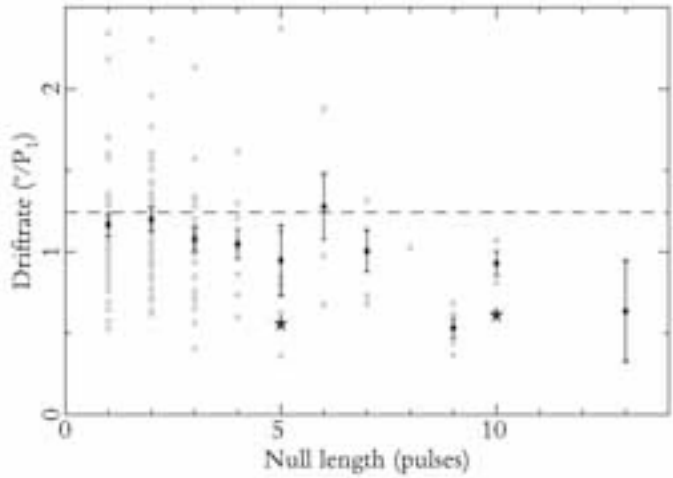

**Fig. 6.** Drift after the null versus the length of the null. We show the average driftrate of the first 6 pulses for each driftband after a null (gray points). The average per null length (black points) with its error is also plotted. The normal driftrate, corrected for the driftband curvature, is indicated by the horizontal dashed line. The stars denote the driftrate for the slow drifting mode sequences. One pulse often consists of more than one subpulse, so after many nulls we see several driftbands reappear. The number of driftrates plotted here is therefore larger than the total number of nulls.

*Average pulse profile around null*

To investigate whether the change from normal emission to the null state is sudden or more gradual, Lyne & Ashworth (1983) compared the energies of the pulses near a null. The finite chance that the emission from the pulsars drops or rises within the pulse window would influence the brightness of the pulses around the null. The last pulse before the null was indeed found to be less bright than a normal pulse. The first pulse after the null, however, was considerably more bright than a normal pulse.

We have not just looked at the energies of these neighbouring pulses, but also at their profiles. To this end, we have averaged all the last pulses before nulls longer than 1 period, and all the first pulses after these nulls. The results, shown in Fig. 7 and condensed in Table 1, are surprising. Not only do we find the expected differences in brightness, we also see a significant offset in the pulse position for the first pulses after nulls.

*Slow drifting mode*

Although we had set out to quantify the normally very regular drifting behaviour of PSR B0809+74, we unexpectedly found two occasions in which the pulsar clearly deviates from its normal drifting mode. We will refer to these sequences by the year in which they were observed, 1999 and 2000. In Fig. 8a we have plotted the derived longitude (Lyne & Ashworth 1983) of subsequent pulses around the mode changes. The derived longitude of the subpulses effectively converts the different short driftbands into one long one. Figure 9 shows a grayscale plot of the two slow drifting series.

The most striking difference, as the drifting pattern is concerned, is clearly the decrease of the driftrate by about 50%. We see that in both the observations the drifting mode changes during or immediately following a long null. After these nulls of length 10 and 5 respectively, almost all drifting properties reappear at new values, as laid out in Table 2. As some of these properties normally already show change over time, we compare the slow drifting sequences (labeled 'slow') to the 600 pulses that surround them (labeled 'normal').

In the first three columns of Table 2, we investigate the parameters of the individual subpulses in both modes. Immediately we see a very interesting change in the relative average position of the subpulses. Right after the null, there is a definite shift in position towards earlier arrival, that can already be seen in a plot like Fig. 9. Interestingly, this shift of the pulse window is similar to the average jump in subpulse position we see over a normal null.

In the slow drifting mode, the subpulses are slightly wider, but their heights remain the same. Next, we explore the driftband characteristics in the last three columns of Table 2. The average longitude separation of two adjacent driftbands, $P_2$, decreases significantly by about 15%. Hence, in the slow drifting mode the subpulses within each pulse are spaced closer together than normal.

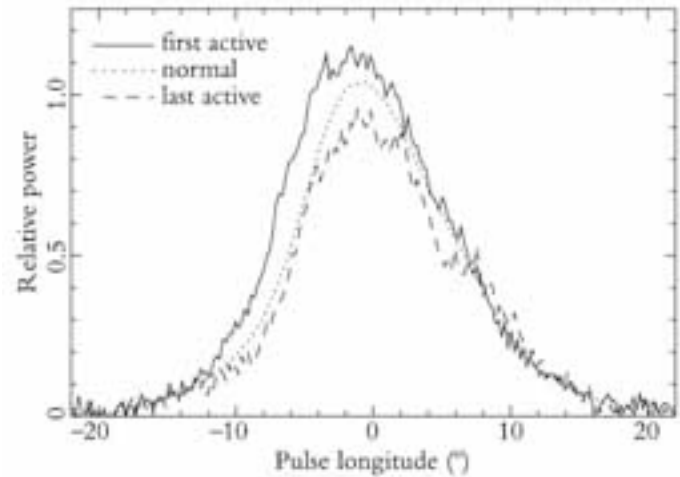

**Fig. 7.** Average profiles for the pulses adjacent to nulls, for all 131 nulls longer than 1 pulse. Shown are: the first active pulse after the null, the last active pulse before the null and, for reference, the normal average profile.

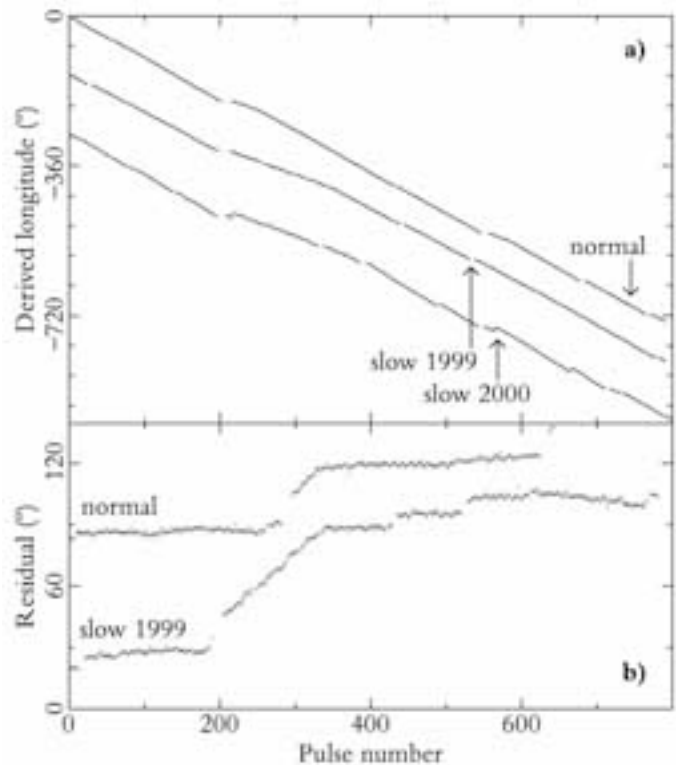

**Fig. 8. a)** Derived longitudes of three pulse sequences. The gaps in the curves are nulls. The sequences are aligned on the ends of the long nulls around pulse 200. The top line shows the normal drifting behaviour around nulls of lengths 13, 9, 3 and 2, respectively. The middle line shows the 1999 slow drifting sequence (offset −150°), the bottom line shows the 2000 sequence (offset −300°). **b)** Deviation of the derived longitude from normal drifting for the 1999 and the normal sequence shown in panel (a). The transitions from slow to normal drifting have been aligned.

When we compare the driftrates of the subpulses between the two modes, we find that the driftrate in the slow drifting mode is almost halved. This puts the new driftrate right in the range of driftrates we usually find after a longer null. We have indicated these driftrate values with black stars in Fig. 6. Being dependent on $P_2$ and the driftrate, the fractional change in $P_3$ (the recurrence time of a driftband) is comparably large.

In both cases the new drifting mode is stable for about 120 pulses and then changes back to normal. To illustrate this change, we have plotted the deviation of the derived longitude from normal drifting in Fig. 8b. For the 1999 observation, we see a normal driftrate up to the null at pulse 200. After this null, the driftrate is smaller, up to pulse 330. In about 20 pulses, the drift then speeds up back to its normal value.

For reference, we have also plotted the drifting behaviour after the longest null in our sample. Again, we see a speedup to the normal driftrate in about 20 pulses at pulse 330, very similar to the slow drifting speedup time scale.

The slow drifting interval in the 2000 observation (right panel of Fig. 9) is followed and stopped by a series of frequent, longer than average nulls, out of which the pulsar emerges in its normal drifting pattern.

These surprising changes in subpulse positions, widths and separations must influence the average pulse profile during the slow drifting mode. Noting the many similarities between slow drifting pulses and the pulses that follow nulls (halved driftrate, similar offsets, similar speedup) we are interested in the possible similarities between the average profiles of these pulses, especially since the average profile of the pulses after the null is singularly bright and shifted in longitude.

In Fig. 10 and Table 1 we compare the pulsar's normal average profile with the average profile in the slow drifting mode. In both observations, the slow drifting mode average profile is much brighter than the normal profile and offset towards earlier arrival − exactly the two peculiarities we found in the average pulse profile after a null as well.

## 4. Discussion

### Driftband fitting

We find that the number of invisible nulls is low. That means we can study of the drifting properties of PSR B0809+74 straightforwardly.

| | height | position (°) | fwhm (°) |
|---|---|---|---|
| normal | 1.000 ± 0.003 | 0.000 ± 0.016 | 13.33 ± 0.04 |
| last active | 0.875 ± 0.005 | 0.27 ± 0.12 | 13.96 ± 0.09 |
| first active | 1.112 ± 0.005 | −0.77 ± 0.12 | 13.84 ± 0.08 |
| slow 1999 | 1.272 ± 0.005 | −1.51 ± 0.3 | 11.99 ± 0.06 |
| slow 2000 | 1.311 ± 0.008 | −0.75 ± 0.3 | 10.52 ± 0.08 |

**Table 1.** Comparison of the pulsar's average emission profiles for different sets of pulses. We show the height, full width at half maximum (fwhm) and position of the Gaussians that fitted the profiles best. The 'normal' subset consist of all non-null pulses in the dataset. All heights and positions in this table are relative to the height and position of this 'normal' set. The set labeled 'last active' contains all 131 pulses that preceded a null that was longer than 1 pulse period. The characteristics of the pulses that followed these nulls are labeled 'first active'. The subsets 'slow 1999' and 'slow 2000' contain the 120 pulses that form the slow drifting sequences observed in 1999 and 2000, respectively.

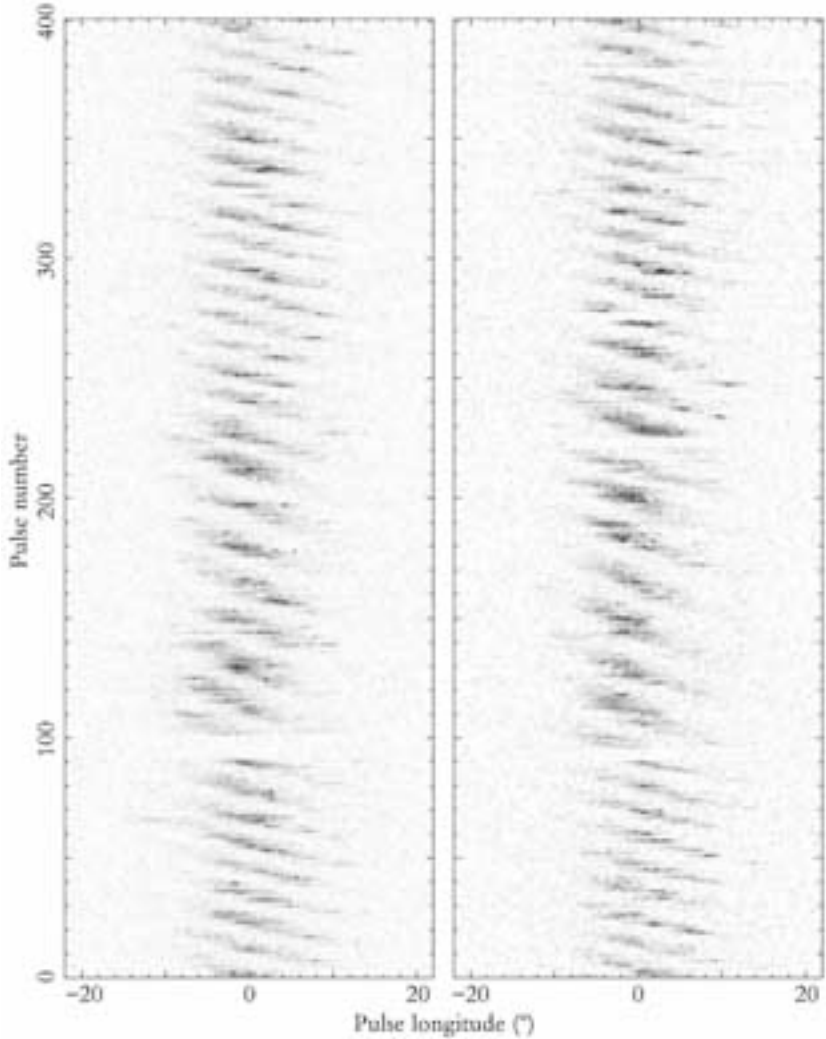

**Fig. 9.** Grayscale plots of the pulses composing the two slow drifting sequences. The grayscale linearly depicts the intensity of a sample. The left series was observed in 1999, the right series in 2000. After a null of length 10 (1999 series) and 5 (2000 series) around pulse number 100, the driftrate and driftband separation are steadily different for about 120 pulses. From pulse 240 on, the drifting changes back to normal.

The non-linearity of the driftbands was already noted by Page (1973), who integrated several hundreds of pulses to compare the shape of the driftbands in different observations. He found considerable differences between these driftband shapes. We compare the average drift path over several thousands of pulses however, and find that the curvature of the driftbands remains the same.

The curvature of the driftband can be a direct consequence of the curved geometry of the emission region (Krishnamohan 1980). In that case, the shape of the driftband is expected to be point-symmetric: the curve will then be unchanged after mirroring it in both axes. This point-symmetry would then return in the residuals to straight line fits (Fig. 3). The shape of the driftbands we find, however, is not point-symmetric but axisymmetric: the curve is unchanged after mirroring in the y-axis.

We could still tie the curvature of the driftbands to the emission-region geometry, by assuming that part of the pulse profile of this pulsar is missing. In that case, the driftband shape we find would represent only part of the total expected shape. This suggestion that part of the profile of PSR B0809+74 is missing fits in with multi-frequency observations (Bartel et al. 1981; Kuzmin et al. 1998) that imply 'absorption' of part of the profile at lower frequencies.

### Slow drifting mode and nulling

With our investigation of PSR B0809+74 still ongoing, we discuss here only the phenomena themselves and defer their interpretation to a subsequent paper.

The similarity between this pulsar's behaviour in the slow drifting mode and around nulls is striking:

- The slow drifting sequences start at nulls.
- The driftrate of the slow drifting mode is a lower limit to the driftrates found after all nulls.
- The speedup from slow drifting to normal is identical to the speedup after a null.
- The average profiles of both the post-null and slow drifting pulses are brighter than normal.
- These average profiles are both displaced to earlier arrival.
- This displacement of the average profile is caused by offsets in the individual subpulses.
- These offsets are identical to the jump of the subpulses over the null.

The natural conclusion is that the behaviour after a null and the slow drifting mode are the same, quasi-stable phenomenon. Normally the pulsar reappears from the null in the slow drifting mode. After a variable time, it quickly evolves to the normal mode. Therefore, right after a normal null we see either a short sequence of slow drifting, or the transition back to normal drifting. In the case of the long slow drifting mode sequences, the metamorphosis back to the normal mode is delayed.

This would explain all the similarities we found above. After a normal null, the pulsar is in the slow drifting mode or changing back to the normal mode. All the characteristics of the slow drifting mode can then be identified in the post-null behaviour, although they will be less pronounced; the transition to normal drifting may already be taking place.

| | averages from Gaussian fits to subpulses | | | averages from straight line fits to driftbands | | |
|---|---|---|---|---|---|---|
| | relative height | relative position (°) | fwhm (°) | $P_2$ (°) | driftrate (°/$P_1$) | $P_3$ ($P_1$) |
| normal 1999 | 1.00 ± 0.02 | 0.00 ± 0.11 | 5.39 ± 0.06 | 11.61 ± 0.14 | 1.086 ± 0.011 | 10.72 ± 0.13 |
| slow 1999 | 1.06 ± 0.05 | −1.2 ± 0.3 | 5.75 ± 0.11 | 9.9 ± 0.5 | 0.606 ± 0.019 | 16.5 ± 0.5 |
| normal 2000 | 1.00 ± 0.02 | 0.0 ± 0.2 | 5.22 ± 0.06 | 11.33 ± 0.14 | 1.043 ± 0.014 | 10.88 ± 0.13 |
| slow 2000 | 0.99 ± 0.04 | −1.8 ± 0.2 | 5.55 ± 0.11 | 9.63 ± 0.3 | 0.54 ± 0.02 | 17.6 ± 0.5 |

**Table 2.** Subpulse and driftband properties for the slow drifting mode sequences and the surrounding normal drifting pattern.

When the driftrate increases at the end of a slow drifting sequence, this speedup is quick and identical to the speedup seen after a normal null (see Fig. 8b).

The post-null driftrate values are found in between the slow drifting and normal driftrate value, depending on how soon the transition back to normal takes place (see Fig. 6). Although the return from the low to the normal driftrate will be quick, the time the driftrate is low may vary for different nulls of the same length. The average of these different sequences will then resemble the slow exponential decay found by Lyne & Ashworth (1983).

Comparing average pulse profiles, the increased brightness and the pulse offset of the post-null average profile are attenuated versions of similar deviations seen in the slow drifting mode profile (see Table 1, Figs. 7 and 10). The average profile offset we see in the slow drifting mode is caused by a shift of the window in which the subpulses appear (Tables 2 and 1).

The change in position of the post-null average pulse profile must then be caused by this shift of the pulse window as well. The magnitude of this shift is identical to the subpulse-longitude jump over normal nulls. This means that the subpulse-position jump over a null is caused by a displacement of the pulse window as a whole, like in the slow drifting mode.

Previously, this jump was thought to be the effect of the subpulse-drift speedup during the null. For this speedup to produce the observed jump in subpulse position, the time scale involved had to be long, contrasting the short time scales found for the slowdown of the subpulse drift and the rise and fall of the emission around a null.

With the displacement of the subpulses over the null accounted for, the estimated speedup time of the subpulse drifting is negligible: the preservation of the position of the subpulses over the null now only allows for a quick speedup of the subpulse drift, putting all time scales of emission and drift rise and decay around the nulls in the same range.

## 5. Conclusions

After many or all nulls, PSR B0809+74 emits in a mode different from the normal one. This mode is quasi-stable, normally changing back to normal in several pulses. This is seen as the normal behaviour after a null, of which the reduced driftrate is the most striking characteristic. Occasionally, the quasi-stable slow drifting configuration persists for over a hundred pulses before changing back to the normal mode.

The pulses in the slow drifting mode and, consequently, all pulses after the null, are brighter than normal pulses. In the slow drifting mode, the subpulses are closer together, they drift more slowly through the profile and the window in which they appear is offset towards earlier arrival.

This offset of the pulse window accounts for the displacement of subpulses over the null. When taking this shift of the window into account, we find that the longitude of the subpulses is perfectly conserved over a null. This indicates that the speedup time for the subpulse drift is short.

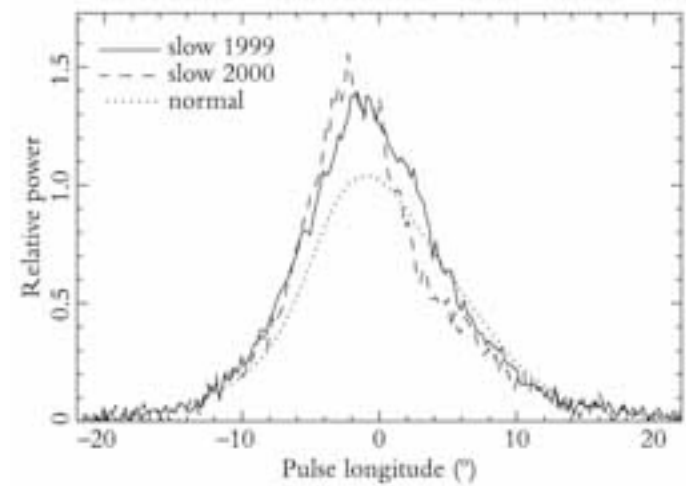

**Fig. 10.** Average profiles for the slow drifting mode sequences. The 'slow 1999' and 'slow 2000' profiles are averaged over the 120 pulses that composed the slow drifting sequences observed in 1999 and 2000, respectively. The normal average profile is plotted for comparison.

# PROBING DRIFTING AND NULLING MECHANISMS THROUGH THEIR INTERACTION IN PSR B0809+74

**with Ben Stappers, Ramachandran and Joanna Rankin**

Both nulling and subpulse drifting are poorly understood phenomena. We probe their mechanisms by investigating how they interact in PSR B0809+74. We find that the subpulse drift is not aliased but directly reflects the actual motion of the subbeams. The carousel-rotation time must then be over 200 seconds, which is much longer than theoretically predicted.

The drift pattern after nulls differs from the normal one, and using the absence of aliasing we determine the underlying changes in the subbeam-carousel geometry. We show that after nulls, the subbeam carousel is smaller, suggesting that we look deeper in the pulsar magnetosphere than we do normally. The many striking similarities with emission at higher frequencies, thought to be emitted lower too, confirm this. The emission-height change as well as the striking increase in carousel-rotation time can be explained by a post-null decrease in the polar gap height. This offers a glimpse of the circumstances needed to make the pulsar turn off so dramatically.

## 1. Introduction

In pulsars, the emission in individual pulses generally consists of one or more peaks ('subpulses'), that are much narrower than the average profile and the brightness, width, position and number of these subpulses often vary from pulse to pulse.

In contrast, the subpulses in PSR B0809+74 have remarkably steady widths and heights and form a regular pattern (see Fig. 1a). They appear to drift through the pulse window at a rate of $-0.09$ $P_2/P_1$, where $P_2$ is the average longitudinal separation of two subpulses within one rotational period $P_1$, which is 1.29 seconds. Figure 1a also shows how the pulsar occasionally stops emitting, during a so-called null.

In this paper, we will interpret the drifting subpulse phenomenon in the rotating carousel model (Ruderman & Sutherland 1975). In this model, the pulsar emission originates in discrete locations ('subbeams') positioned on a circle around the magnetic pole. The circle rotates as a whole, similar to a carousel, and is grazed by our line of sight. In between successive pulses, the carousel rotation moves the subbeams through this sight line, causing the subpulses to drift.

Generally, the average profiles of different pulsars evolve with frequency in a similar manner: the profile is narrow at high frequencies and broadens towards lower frequencies, occasionally splitting into a two-peaked profile (Kuzmin et al. 1998). This is usually interpreted in terms of 'radius to frequency mapping', where the high frequencies are emitted low in the pulsar magnetosphere. Lower frequencies originate higher, and as the dipolar magnetic field diverges the emission region grows, causing the average profile to widen.

The profile evolution seen in PSR B0809+74 is different. The movement of the trailing edge broadens the profile as expected, but the leading edge does the opposite. The profile as a whole decreases in width as we go to lower frequencies until about 400 MHz. Towards even lower frequencies the profile then broadens somewhat (Davies et al. 1984; Kuzmin et al. 1998). Our own recent observations of PSR B0809+74, simultaneously at 382, 1380 and 4880 MHz, confirm these results (Rankin et al. 2002). Why the leading part of the expected profile at 400 MHz is absent is not clear. While Bartel et al. (1981) suggest cyclotron absorption, Davies et al. (1984) conclude that the phenomenon is caused by a non-dipolar field configuration. We will refer to this non-standard profile evolution as 'absorption', but none of the arguments we present in this paper depends on the exact mechanism involved.

In Chapter 1 (van Leeuwen et al. 2002) we investigated the behaviour of the subpulse drift in general, with special attention to the effect of nulls. We found that after nulls the driftrate is less, the subpulses are wider but more closely spaced, and the average pulse profile moves towards earlier arrival. Occasionally this post-null drift pattern remains stable for more than 150 seconds.

For a more complete introduction to previous work on PSR B0809+74, as well as for information on the observational parameters and the reduction methods used, we refer the reader to Chapter 1. In this paper we will investigate the processes that underly the post-null pattern changes. We will quantify some of the timescales asso-ciated with the rotating carousel model and map the post-null changes in the drift pattern onto the emission region.

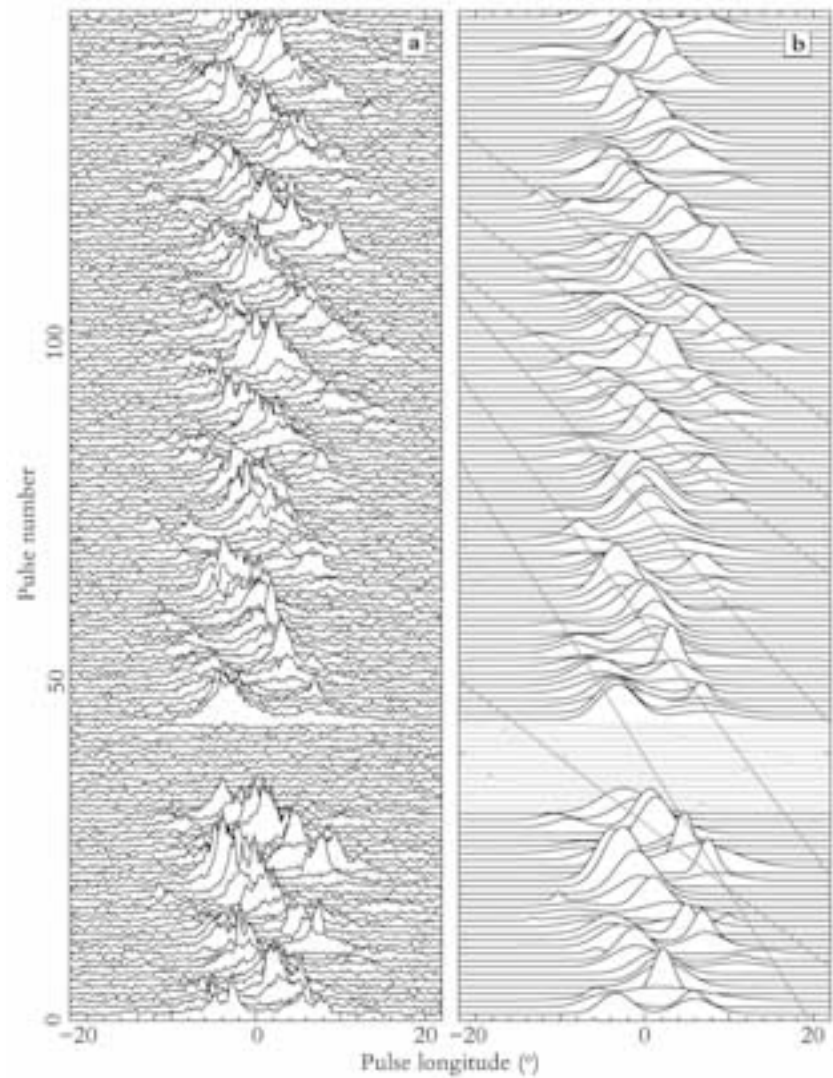

**Fig. 1.** Observed and fitted pulse sequences. A window on the pulsar emission is shown for 150 pulses. One pulse period is 360°. The centre of the Gaussian that fits the pulse profile best is at 0°. **a)** The observed pulse sequence, with a null after pulse 30. **b)** The Gaussian curves that fitted the subpulses best. Nulls are shown in lightest gray, driftbands fitted to the subpulse pattern are medium gray.

One of the interesting timescales is the time it takes one subbeam to complete a rotation around the magnetic pole. This carousel-rotation time is predicted to be of the order of several seconds in the Ruderman & Sutherland model. Only recently a carousel-rotation time was first measured: Deshpande & Rankin (1999) find a periodicity associated with a 41-second carousel-rotation time for PSR B0943+10.

The second goal is to determine the changes in the emission region that underly the different drift pattern we see after nulls. Mapping this emission region could increase our insight into what physically happens around nulls.

Achieving either goal requires solving the so-called aliasing problem: as the subpulses are indistinguishable and as we observe their positions only once every pulse period, we cannot determine their actual speed.

## 2. Solving the aliasing problem

The main obstacles in the aliasing problem are the under sampling of the subpulse motion and our inability to distinguish between subpulses. The pulsar rotation only permits an observation of the subpulse positions once every pulse period. Following them through subsequent pulses might still have led to a determination of their real speed, but unfortunately the subpulses are so much alike that a specific subpulse in one pulse cannot be identified in the next, making it impossible to learn its real speed.

In Fig. 2 we show a simulation of subpulse drifting, where we have marked all subpulses formed by a particular subbeam with a darker colour. We use these simulations to discuss how the driftrate, which is the observable motion of the subpulses through subsequent pulses, is related to the subbeam speed, which cannot be determined directly. In Fig. 2a the speed of the subbeams is low ($-0.09\ P_2/P_1$) and identical to the driftrate. In Fig. 2b the subbeam speed is higher ($0.91\ P_2/P_1$), but the driftrate is identical to the one seen in Fig. 2a. When the differences between subpulses formed by various subbeams are smaller than the fluctuations in subpulses from one single subbeam, these two patterns cannot be distinguished from one another. In that case the subpulses within one driftband, which seem to be formed by one subbeam, can actually be formed by a different subbeam each pulse period ('aliasing').

To solve the aliasing problem for PSR B0809+74 we follow driftrate changes after nulls to determine the subbeam speed. Nulls last between 1 and 15 pulse periods, and in Chapter 1 we have shown that for each null the positions of the subpulses before and after the null are identical if we correct for the shift of the pulse profile. So, as there is no apparent shift in subpulse position, either the subbeams have not moved at all, or their movement caused the new subpulses to appear exactly at the positions of the old ones.

As the lengths of the nulls are drawn from a continuous sample it is highly unlikely that the subbeam displacement is always an exact multiple of the subpulse separation: only a total stop of the subbeam carousel can explain why the subpulse positions are always unchanged over the null. At some point after the null, however, the subbeams have accelerated, and the drift pattern has returned to normal. In Fig. 1 we see how, after a null, the driftrate increases to its normal value in about 50 pulses.

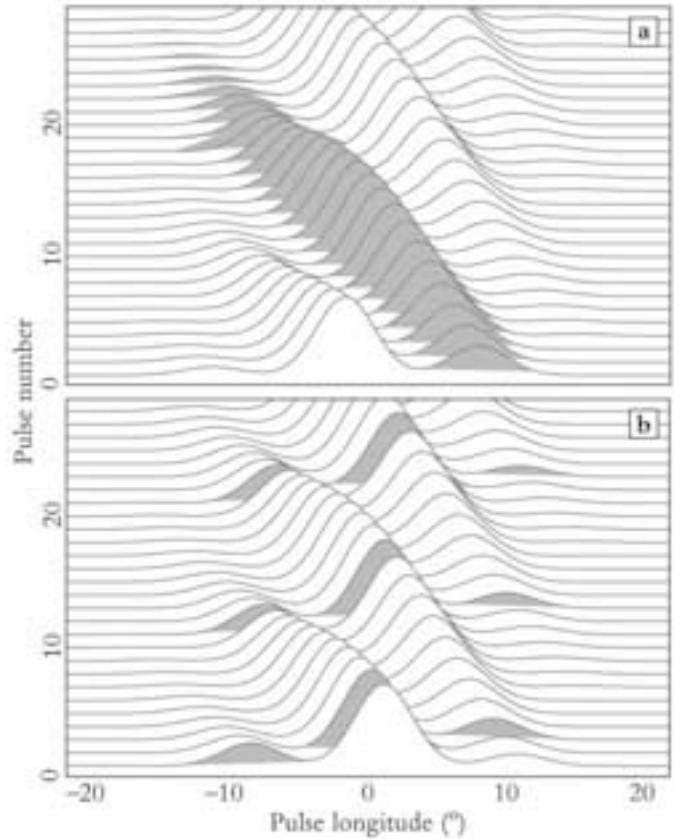

**Fig. 2.** Different alias orders illustrated. We show two series of stacked simulated drifting subpulses. We have marked the subpulses formed by one particular subbeam with a darker colour. **a)** At a low subbeam speed, a single subbeam traces an entire driftband by itself. The driftrate is identical to the subbeam speed: alias order 0. **b)** At alias order $-1$, the subbeam speed is higher than the driftrate and in opposite direction.

There are two scenarios for this subbeam acceleration. The first we will call gradual speedup. Here the changes in the subbeam speed occur on timescales larger than $P_1$. The second we will call instantaneous, as the entire acceleration happens within $1P_1$, effectively out of sight.

In Fig. 3 we show four simulated pulse sequences with different speedup parameters. In all cases, we simulate a drift pattern like that of PSR B0809+74. During a null, from pulses 30 to 45, there is no subbeam displacement. Immediately after the null the subbeams build up speed, and each pulse period we translate the subbeam displacement to a change in subpulse position. Although the final driftrate is the same for all scenarios ($-0.09\ P_2/P_1$), the subbeam speeds differ considerably. The bottom four graphs show these speeds for each scenario. In the top four diagrams we have marked the subpulses from one subbeam with a darker colour for clarification.

Let us look at the case of gradual speedup to alias order 0, where the subbeam speed is the same as the driftrate (Fig. 3a). In this case the driftrate will gradually increase and form a regular driftband pattern, much like the pattern found in the observations.

Next, we investigate a subbeam acceleration to a slightly higher speed. At 0.91 $P_2/P_1$, Fig. 3b shows alias mode $-1$, which is the simplest configuration in which the subbeams move opposite to the subpulse drift. Right after the null the drift consequently commences in this opposite direction. When the subbeam speed nears the first aliasing boundary $0.5\ P_2/P_1$, the subpulses seem to move erratically through the window. After the subbeam speed passes the first alias boundary, the subpulse drift resumes its normal direction, and as the subbeam speed approaches $0.91\ P_2/P_1$, the drift pattern returns to normal.

Also in the 'alias order 1' scenario (Fig. 3c), which is the simplest aliased mode in which the subbeams move in the same direction as the subpulses, the subpulses wander when the subbeam speed nears an alias boundary. Such a disturbance of the drift pattern turns out to be present in all simulations of non-zero alias orders. Because the observed pulse sequences always show smooth, non-wandering driftbands (like in Fig. 1a), we conclude that the subbeams cannot accelerate gradually to a high speed.

With instantaneous acceleration the subbeam speed switches suddenly. Most likely it will do so when the pulsar beam faces away from us. After we see pulse number $n$, the subbeams will move slowly for a certain time, quickly speed up, and then move fast until we see the subpulses of pulse $n+1$ appear. The actual speedup can occur any moment between seeing pulses $n$ and $n+1$. This means that the displacement of the subbeams can vary from very little (speedup just before pulse $n+1$) to a lot (speedup right after pulse $n$). The accompanying changes in subpulse position will then be evenly distributed between 0 and $P_2$; the subpulse positions after the speedup will not be related to those before.

If the subbeams accelerate instantaneously at the end of a null, before the first pulse can be observed, the subpulse phases are not preserved over the null. As this is opposite to what is observed, this possibility is ruled out.

If the subbeam acceleration occurs a few pulses after a null, we expect that the change in driftrate will nearly always be accompanied by a sudden change in the longitude of the driftband, as illustrated in Fig. 3d for a final subbeam speed of $-1.09$ $P_2/P_1$. The absence of a significant number of such sudden shifts in the data indicates that the subbeam carousel of PSR B0809+74 does not speed up instantaneously to a high speed.

As only the gradual speedup to alias order 0 can explain the observed drift patterns, the drift seen in the subpulses of PSR B0809+74 directly reflects the movement of the subbeams, without any aliasing.

## 3. Discussion

### Alias order of other pulsars

The subbeam speedup in PSR B0809+74 follows the simplest scenario possible: it is gradual, and the subpulse drift is not aliased. In other pulsars this might not be the case, and for those we predict drift-direction reversals or jumps in driftband longitude during the subpulse-drift speedup phase after nulls.

### Subbeam-carousel rotation time

Having resolved the aliasing of the subpulses' representation of the subbeams, we know that the subbeams move at $-0.09\ P_2/P_1$. Such a low rotation rate means it will take each subbeam $11P_1$ to move to the current position of its neighbour. The following estimate of the number of subbeams then leads directly to the carousel-rotation time.

Towards the edges of the profile the sight line and the carousel move away from each other, and a subbeam will cease to be visible when the sight line no longer crosses it. When the subbeams are small, or widely spaced compared to the curvature of the carousel, the sight line will cross only few. In that case there will not be many subpulses visible in one pulse. If, however, the subbeams are large and closely spaced compared to the carousel curvature, the sight line passes over more subbeams in one traverse, leading to many subpulses per pulse.

In most of the pulses of PSR B0809+74 we observe two subpulses, occasionally we discern three. By combining this with the ratio of the subpulse width and separation we find there must be more than 15 subbeams on the carousel. As one subbeam reaches the position of its neighbour in $11P_1$, a 15-subbeam carousel rotates in over 200 seconds.

Persistent differences in the properties of individual subbeams should introduce a long term periodicity in the pulse sequences, at the carousel-rotation frequency. Thus far, no such periodicity has been found, which is not surprising as the periodicity can only be measured if the lifetime of the subbeam characteristics is longer than the carousel-rotation time.

In PSR B0943+10, the only pulsar for which we know how the subbeams vary (Deshpande & Rankin 1999, 2001), the associated timescales are in the order of 100 seconds. If we assume their lifetimes are comparably long in PSR B0809+74, the subbeams will have lost most of their recognisable traits once they return into view after one carousel rotation. Having regular pulse sequences that contain several tens of rotation times might still show some periodicity at the carousel-rotation frequency.

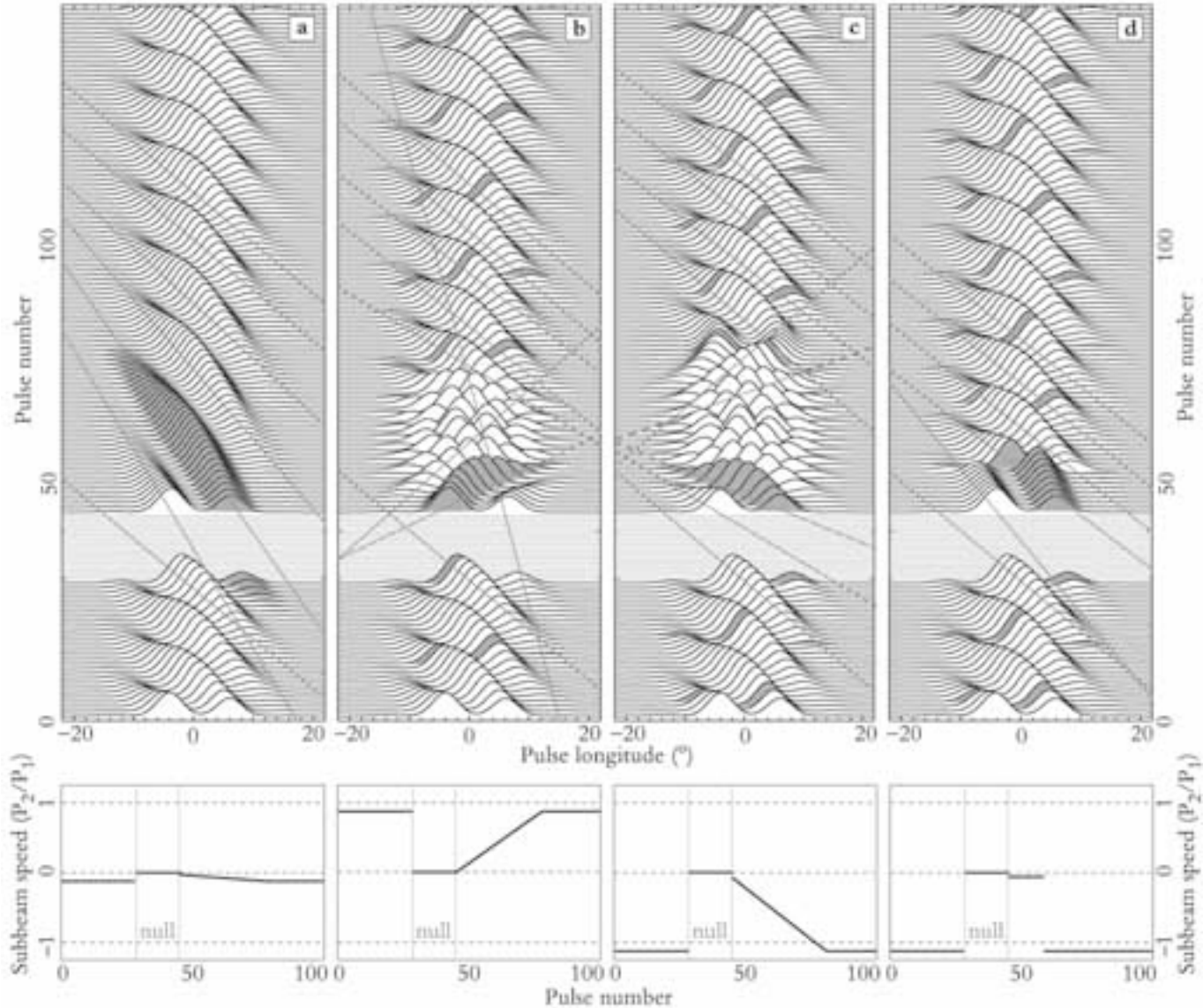

**Fig. 3.** Simulated scenarios for subbeam speedup. The top panels show the resulting drift patterns. Nulls are shown in lightest gray, driftbands fitted to the subpulse pattern are overdrawn in medium gray. We have marked the subpulses formed by one particular subbeam with a darker colour. The bottom panels show the subbeam speed versus the pulse number. **a)** Gradual acceleration to a subbeam speed equal to the driftrate (alias order 0). **b)** Gradual acceleration to a subbeam speed larger than the driftrate and in opposite direction (alias order –1) **c)** Gradual acceleration to a subbeam speed larger than the driftrate and in the same direction (alias order 1) **d)** Instantaneous acceleration, again to alias order 1.

Unfortunately nulls have a destructive influence on the driftband pattern and possibly on the subpulse characteristics. This has thus far made it impossible to observe bright sequences longer than 3 times our lower limit on the rotation time of 200 seconds.

After the prediction that the rotation time in most pulsars would be on the order of several seconds (Ruderman & Sutherland 1975), the observation that it is 41 seconds in PSR B0943+10 was somewhat surprising. The suggested dependence of the carousel-rotation time on the magnetic field strength and the pulse period predicts that the rotation time in PSR B0809+74 should be roughly 4 times smaller. The circulation time we observe, several hundreds of seconds, shows that the theoretical

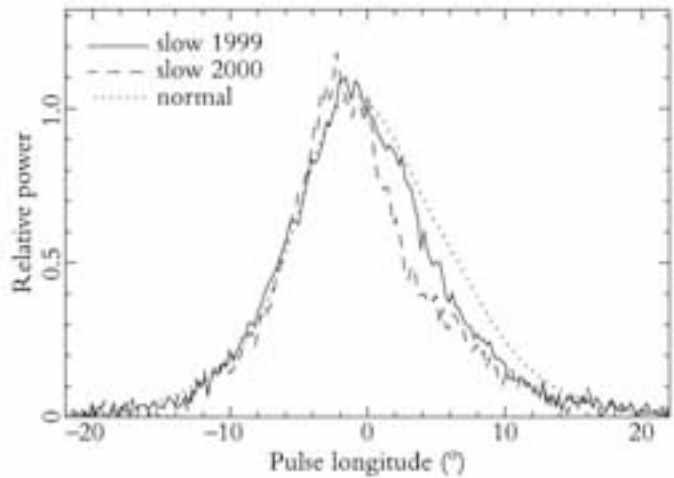

**Fig. 4.** Pulse profiles for the normal mode and the two longest slow drifting sequences after nulls described in Chapter 1. The broadening of the subpulses causes the slow–drifting profile to outshine the normal one. Here all three profiles have been scaled to the same height to show the change in the pulse location and shape.

predictions are incorrect in absolute numbers as well as in the scaling relations they propose.

*Emission region geometry*

In Chapter 1 we found many interesting changes in the driftband pattern after nulls. The subpulses reappear 5−10% broader and 15% closer together, which results in a brighter average profile. The individual subpulses move 1.5° towards earlier arrival, as does their envelope, the average profile, with 1.1° (Fig. 4). The subpulses drift 50% more slowly than usual. Occasionally, this slow drifting pattern remained stable for over 100 pulses.

Knowing, as we do now, that there is no aliasing involved, we can identify single subpulses with individual subbeams. This implies that we can interpret the drift pattern changes in terms of the subbeam-carousel geometry (Fig. 5). We will do so for each of the altered characteristics.

The first thing we note is that, relative to the shift of the average profile, the positions of the subpulses are unchanged over nulls. This means that the subbeam carousel has not rotated during the null.

After nulls, the subpulses are positioned about 15% closer together. In principle, this could be the result from a change in subbeam speed. At a certain moment, a particular subbeam will be pointing towards the observer. It takes some time before the pulsar has rotated the next subbeam into the observer's view, and during this time the subbeams themselves have moved, too. This motion translates directly to a change in the longitudinal subpulse separation. Yet, as we have shown that subbeam speed is low, this effect is negligible.

A more significant change in $P_2$ could be caused by an increase in the number of subbeams on a carousel of unchanged size. The other option is that a decrease in the carousel radius moves the subbeams closer together.

In the first scenario the number of subbeams would have to change during the null, causing the subbeam placement to change considerably. This would lead to subpulse-position jumps over the null. In that case, we would not expect the phases of the subpulses to be as unchanged over the nulls as is observed. Secondly, some time after the null the new configuration would have to return to normal. This means subbeams would have to appear or disappear. We see no evidence of this in any of the 200 nulls we observed. The change in the subbeam separation $P_2$ can therefore not be due to a changed number of subbeams.

In the second scenario the carousel radius decreases by 15%, but the number of subbeams remains the same. As the subbeams now share a reduced circumference, their separation also decreases.

The contraction of a carousel causes all subpulses to move towards the longitude of the magnetic axis. If we combine this contraction with the aforementioned absorption, we can immediately explain the observed shift to earlier arrival of the subpulses and their envelope, the average profile: with only the trailing part unabsorbed, all visible subpulses move towards earlier arrival (Figs. 4 and 5).

In general, the subpulses farthest from the longitude of the magnetic axis will move most, while those at it, if visible, will remain at their original positions. Enlargement of our sample of subpulse positions can therefore indicate where the magnetic axis of PSR B0809+74 is located. This would immediately indicate how much of the pulse profile is 'absorbed' and illuminate the thus far much debated alignment of the pulse profiles at different frequencies.

Normally, the edges of a pulse profile indicate where the overlap of the sight line and the subbeam carousel begins and ends. The contraction of a carousel will then move both edges inward. Yet when the first part of the profile is absorbed, the leading edge reflects the end of the absorption. If this edge is near the middle of the sight-line traverse over the carousel, it will not be affected by a reduction of the carousel size. In that case we will see a change in the position of the trailing edge but not in the leading one, which is exactly what we observe in PSR B0809+74 (Fig. 4).

The wider subpulses we see after a null translate directly to wider subbeams. We do note that a change in carousel radius leads to a new sight-line path, which may have an impact as well.

The significant chance that a null starts or ends within the pulse window of the pulses that surround it, would on average make these pulses dimmer than normal ones (Lyne & Ashworth 1983). Quite unexpectedly however, the pulses after nulls were found to be brighter than average. This is easily explained in our model: the change in the carousel geometry causes the subpulses to broaden and become more closely spaced, while their peak brightness remains the same. This leads to the post-null pulse–intensity increase that is observed.

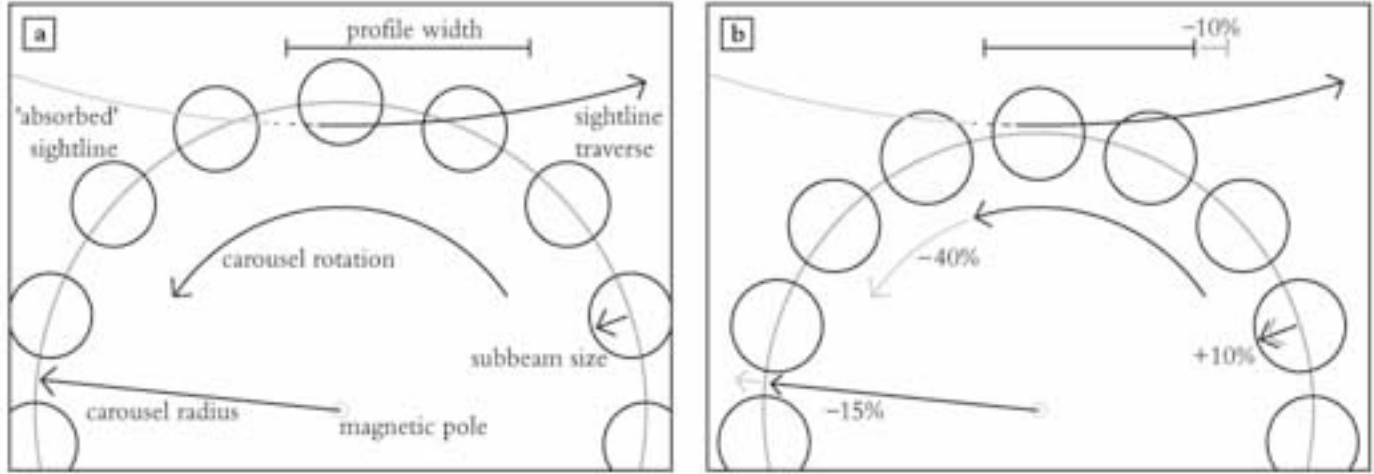

**Fig. 5.** Subbeam carousel geometry in the normal and the post-null configuration. **a)** The normal configuration, with a sight line that is partially obscured by the absorption. **b)** After nulls the carousel is smaller. As their number is unchanged, the subpulses have moved closer together. On the non-absorbed, trailing edge, the pulse window moves inward. The leading edge of the profile is located where the absorption ends and does not move. The subpulses are wider and revolve less fast.

*Post-null polar gap height*

If we follow the radius-to-frequency-mapping argument discussed in the introduction, then the reduction of the carousel size implies that the emission after a null comes from lower in the pulsar magnetosphere than it does normally. The reduction of the subbeam separation supports this idea, but the broadening of the subpulses seems inconsistent.

At this stage a comparison with the pulse profiles and drift patterns observed at higher frequencies seems promising. Emission at higher frequencies is also supposed to originate lower in the pulsar magnetosphere, so similar effects may point at one cause: a decrease in emission height.

Much like the post-null profile, the average profile at higher frequencies moves towards earlier arrival (Rankin et al. 2002). Comparing drift patterns, we see that at higher frequencies the driftrate decreases and that the subpulses broaden and move closer to one another (Davies et al. 1984), strikingly similar to the behaviour we find after nulls.

Only one anomaly remains: when comparing drift patterns at different frequencies, the changes in driftrate and subpulse separation $P_2$ always counterbalance, leaving the carousel rotation time $\hat{P}_3$ unchanged. This means that the subbeams we observe at a certain frequency are cuts through rods of emission, intersections at the emission height associated with that frequency. The rods rotate rigidly but diverge with in-creasing height. At each height $\hat{P}_3$ is the same, while the speed and separation of the subbeams do change.

In contrast to this normal invariance of $\hat{P}_3$, we see it change considerably after nulls, an increase that cannot be explained by a change in viewing depth. We therefore expect that the disturbance that caused the change in this depth, can also account for the 50% increase in $\hat{P}_3$.

Usually, both the subbeam separation $P_2$ and the emission height are assumed to scale with the polar gap height $h$ (Ruderman & Sutherland 1975; Melikidze et al. 2000). The 15% decrease in $P_2$ would thus be due to a identical fractional decrease in the gap height, which should also cause an equal decrease in emission height.

If this change in gap height could also account for the observed increase of the carousel rotation time $\hat{P}_3$, all post-null drift-pattern changes can be attributed to one single cause.

$\hat{P}_3$ is thought to scale as $h^{-2}$. For the inferred 15% gap height decrease this predicts a 40% increase in $\hat{P}_3$, nicely similar to the 50% we find.

With one single cause we can therefore explain both the puzzling well-known phenomena (the driftrate decrease after nulls, the bright first active pulse) and the newly discovered subtle ones (the change in the position of the average profile, the decrease of the subpulse separation and the subpulse-width increase). This post-null decrease in gap height offers a glimpse of the circumstances needed to make the pulsar turn off so dramatically.

## 4. Conclusions

We have shown that the drift of the subpulses directly reflects the actual motion of the subbeams, without any aliasing.

In other pulsars with drifting subpulses this may be different: for those we predict drift-direction reversals or longitude jumps in the post-null drift pattern.

We find that the carousel-rotation time for PSR B0809+74 must be long, probably over 200 seconds. The expected lifetime of the subbeam characteristics is less, which explains why thus far no periodicity from the carousel rotation could be found in the pulse sequence.

The rotation time we find is larger than theoretically predicted, not only in absolute numbers but also after extrapolating the rotation time found in PSR B0943+10. Both the magnitude and the scaling relations that link the carousel-rotation time to the magnetic field and period of the pulsar are therefore incorrect.

When the emission restarts after a null the drift pattern is different, and having determined the alias mode, we identify the underlying changes in the geometry of the subbeam carousel. A combination of a decrease in carousel size and 'absorption' already explains many of the changes seen in the post-null drift pattern.

The resemblance between the drift pattern after nulls and that seen at higher frequencies, thought to originate at a lower height, is striking. Assuming that similar effects have identical causes leads us to conclude that after nulls we look deeper in the pulsar magnetosphere, too.

Both this decrease in viewing depth and the striking increase in the carousel rotation time can be quantitatively explained by a post-null decrease in gap height.

# INTERMITTENT NULLS IN PSR B0818−13, AND THE SUBPULSE−DRIFT ALIAS MODE

**with Gemma Janssen**

We show that all long nulls in PSR B0818−13 are trains of rapidly alternating nulls and pulses (each shorter than one pulse period). Sometimes only the nulls coincide with our pulse window, resulting in one of the apparently long nulls seen occasionally. We show these are seen as often as expected if during such a train the probability for nulls is 1.2 times less than for pulses.

During nulls, the subpulse drift-speed appears to increase. We assume that the carousel of sparks that possibly underlies the subpulses actually slows down, as it does in similar pulsars like PSR B0809+74, and conclude that the subpulse−drift in this pulsar must be aliased. The carousel must then rotate in 30 seconds or less, making it the fastest found to date.

## 1. Introduction

Observations of drifting subpulses are a nice example of undersampling: in some radio pulsars the positions of the pulse components change regularly (as in fig. 1), but as these positions are determined only once per pulse period, the exact underlying motion remains unknown.

Finding this underlying motion, and its relation to other pulsar parameters, could be helpful in determining which, if any, of the proposed mechanisms to generate drifting subpulses (Ruderman & Sutherland 1975; Wright 2003) is correct. We will assume the subpulses are formed by discrete locations of emission ('sparks') that rotate around the magnetic pole of the pulsar, causing the pulse components to move through the pulse windows ('drifting subpulses'). Different directions and speeds of this rotation can produce identical subpulse positions at the observer, so-called aliasing (see also fig. 2 in Chapter 2). To determine the actual speed of the spark carousel, one can check for the periodicity caused as brighter or offset sparks re-appear after one carousel rotation. Deshpande & Rankin (1999) did so for PSR B0943+10, and found a carousel rotation time of around 41 seconds, longer than expected from theory (Ruderman & Sutherland 1975). With the rotation time known, Deshpande & Rankin (1999) could track individual sparks. They noted that sparks retain their characteristic brightness and positions for about 100 seconds. If this number were the same in other pulsars, carousel rotation times longer than 100 seconds could not be found by a periodicity search, as the characteristic traits of the subpulses will have changed before they return into view. For such long carousel rotation periods a different technique is needed.

In some pulsars there is a mechanism that can be used to circumvent the undersampling problem and hence determine the underlying carousel-speed: nulling. In these pulsars, a few percent of the pulses are missing as the pulsars suddenly turns off ('nulls'). After these nulls, the subpulse drift is often affected. In Chapter 1 and 2, van Leeuwen et al.) investigated this drifting-nulling interaction in PSR B0809+74 and find it can only be explained if the underlying carousel rotates in over 200 seconds, again much longer than expected. They predict that in pulsars with shorter carousel rotation times the drift direction should reverse after nulls.

In their 1983 paper, Lyne & Ashworth mention the nulling-drifting interaction of PSR B0818−13 and show the post-null drift behaviour (their fig. 14). For long nulls, there appears to be a reversed drift direction. Combining this with the prediction of Chapter 2 (van Leeuwen et al. 2003) we decided to further investigate the drifting-nulling interaction in PSR B0818−13.

## 2. Observations and data reduction

We have observed PSR B0818−13 from 2000 to 2003 with PuMa, the Pulsar Machine (Voûte et al. 2002) at the Westerbork Synthesis Radio Telescope (WSRT). The 11 observations amount to 20 hours in total or, using the 1.238s period, $5.8 \times 10^4$ individual pulses. All observations were conducted at a time resolution of 0.4096ms and around frequencies in between 328 and 382 MHz, with a bandwidth of several times 10MHz, depending on interference. Each 10MHz band was split in 128 channels and dedispersed offline.

The data were searched for nulls and subpulses in the way described in Chapter 1 (van Leeuwen et al. 2002). For each observation we compose a histogram of the energy of individual pulses, like fig. 2. The discrete population around zero-energy pulses is labeled 'nulls'. We then fit Gaussian profiles to each individual pulse, determine which subpulse-fits are significant and store the subpulse positions, widths and heights along with the raw pulse data. As the subpulse-drift bands form regular structures, we use the predictive capability of the thus-found strong subpulses to detect weaker subpulses over the noise.

## 3. Results: analysis of nulls and subpulse drift

### Statistics of short and long nulls

PSR B0818−13 nulls about 1% of the time. As the histogram of null lengths fig. 3 shows, most of these nulls are very short, with lengths of only one or two periods. What the histogram doesn't show however, is that although roughly 2/3 of these short nulls occur by themselves, many are observed in groups of intermittent pulses and nulls, like the groups seen in fig. 1. As this behaviour is uncommon, Lyne &

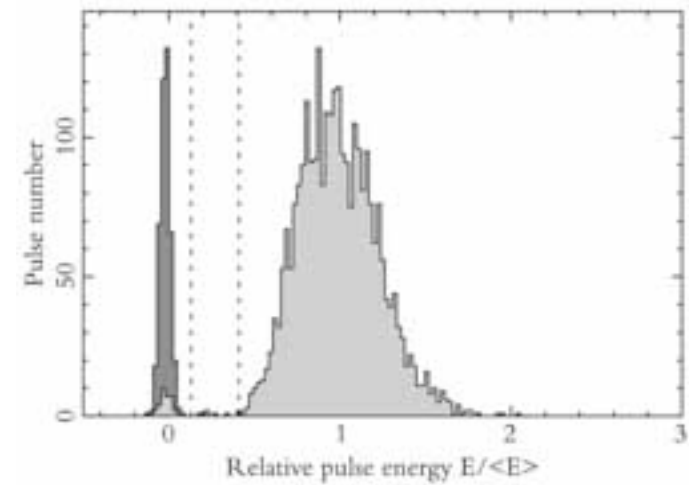

**Fig. 2.** Energy histogram for PSR B0818−13. In light grey we show the energy distribution of the energies found in the on-pulse window. In black we show the energies found in an equally large off-pulse window, caused by system noise. The set of low-energy pulses detached from both the normal and the null distribution are probably half-nulls.

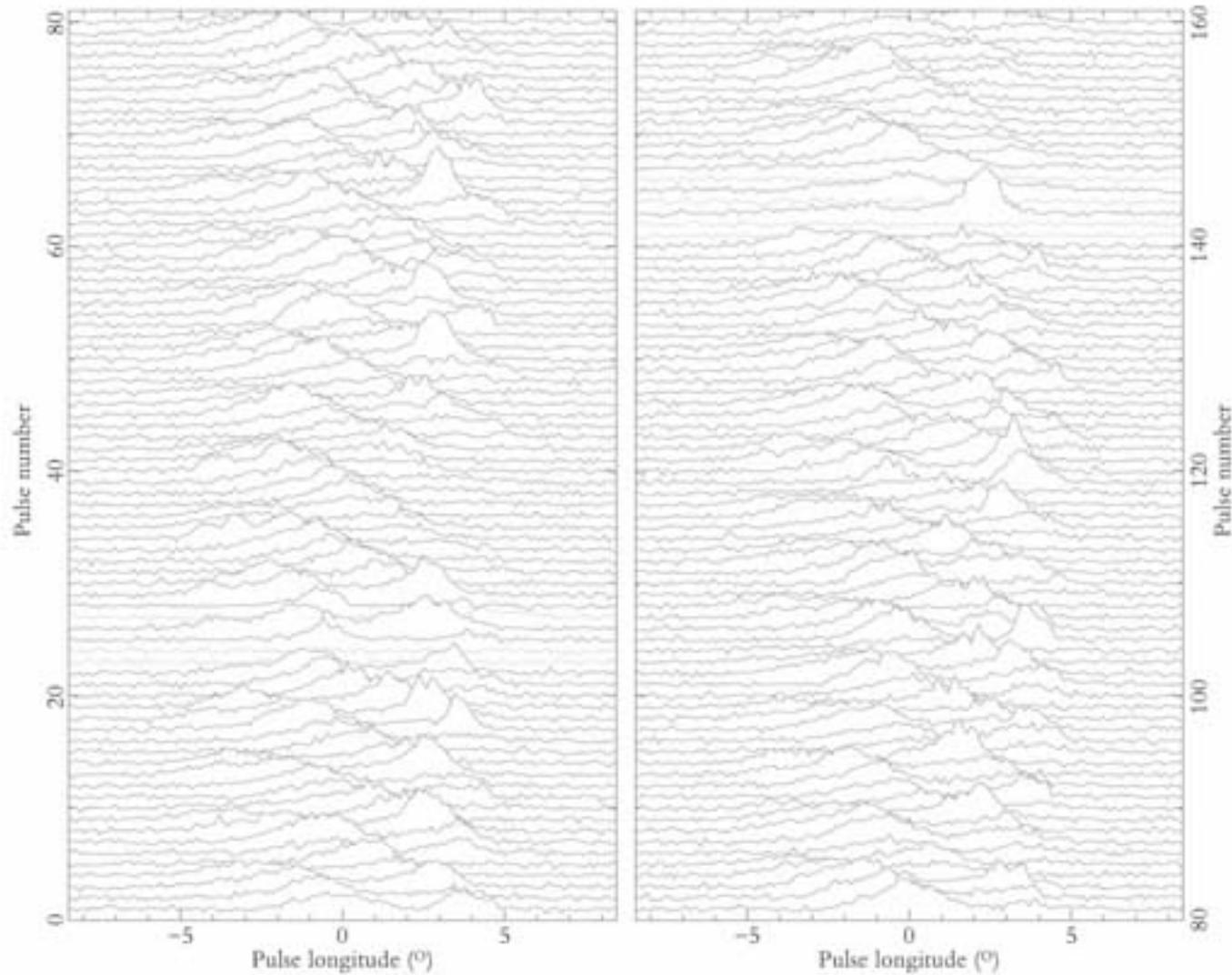

**Fig. 1.** Stacked single pulses with two intermittent nulls in lighter grey.

Ashworth (1983) chose to disregard all intermittent short nulls, and only investigate longer nulls. But has the pulsar really turned off during such a long null? Or are the long nulls in PSR B0818−13 actually intermittent nulls where the series of short nulls happen to line up with our line of sight? This is the question we investigate.

If we assume that, during a train, the pulsar quickly and successively turns on and off, we can use the statistics of the null/pulse occurrence within a train to determine the likeliness ratio of null and pulse occurrence. To classify the different pulse trains, we have defined the first and last nulls of a train to be the start and end of the group.

In fig. 4 we give an overview of all event lengths. From this figure it is clear that the occurrence of null-groups as a whole is not governed by chance: there are too many long events. Within an event the pulsar blinks fairly quickly, but the change back to the normal pulse state usually takes several seconds.

The null group in the left panel of fig. 1, for example, we classify as an event of length 5, while the one in the right panel is an event of length 6. As the first and last nulls act as markers, there are only 6−2=4 positions in the rightmost of fig. 1 group that can be either nulls or pulses. In this case, there are two nulls and two pulses

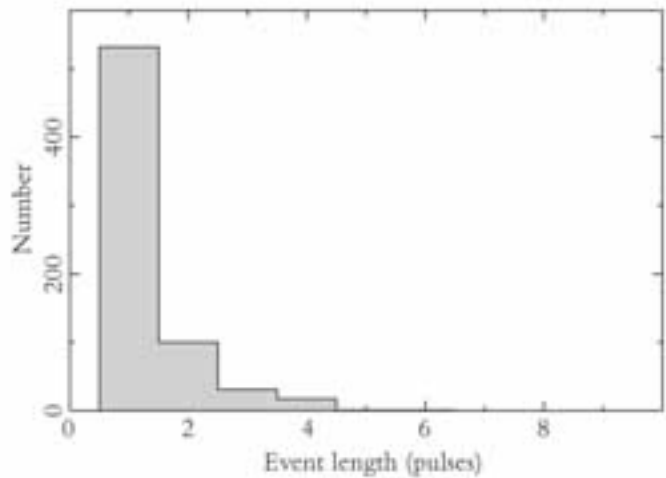

**Fig. 3.** Histogram of the null lengths.

In the histogram, we do indeed see a number of these low-energy pulses (about 0.3%, with energies varying from 10 to 40% of the average pulse energy). For a pulse to have 10–40% of normal energy, the null should start (stop) in the last (first) 20% of the pulse windows, so within ± 2°. If the positions of the starts and stops of null are random, we would expect to see this happen in $\frac{2}{360}$=0.5% of pulses, in good agreement with the values found above. If these low-energy pulses are really half-nulls, we would expect them to occur preferentially near null trains, as normal nulls do. We therefore calculated the mean distance to a null for both normal and low-energy pulses. We find that low-energy pulses are on average about 30% closer to a null than normal pulses, so from this we conclude that at least some of the low-energy pulses are actually half-nulls, where we see the pulsar turning on or off while facing us.

*Profiles around nulls*

The next step in investigating how the pulsar turns on and off is to investigate changes in pulse type or shape before and after nulls. First, we compare the number of sub-pulses in normal pulses with the number observed in pulses that lie in between nulls (cf. pulses 25, 26, 143 and 145 in fig. 1). We find that on average the former contain 1.79±0.01 subpulses, the latter 1.80±0.05. To compare pulse shapes, we have averaged the last and first active pulses around all nulls in our sample, as shown in fig. 5.

The profile of the first pulse after the null is roughly as expected; as the pulsar sometimes turns back on only halfway through the pulse window, we expect the leading edge of the profile to diminish compared to the trailing edge, and this we indeed see. For the last pulse before the null one would expect the opposite (the in this middle part, in a NNPNPN pattern. In table 1 we give an overview of all the different null-pulse-trains present in our sample, in the same pattern notation. In the middle parts of all events, there are 238 possible locations to be divided between pulses and nulls. Of these, 107 are nulls, and 131 are pulses, so we conclude that, on average, pulses are 1.2 times more likely to occur within a group than nulls. Using this ratio, we can predict the chance that a train of a certain length appears to be one long null. If we consider that we probably miss or misidentify events that start or end with pulses in stead of nulls, the occurrence of long apparent nulls agrees quite well with the prediction. For groups of length 5, for example, the first and last positions are nulls by definition, but the middle 3 positions should be nulls, the probability of which is $(\frac{1.0}{1.0+1.2})^3$ =0.09. This is in good agreement with the observed ratio of 1 in 9 that follows from table 1.

This means that at the pulsar the intermittent pulse trains and the long nulls are the same. When the short nulls in such a train all happen to occur when the pulsar faces us, the effect only appears different.

We now know that during the null-trains, the pulsar blinks quickly, producing several nulls that are shorter than a pulse period. The continuous increase towards shorter nulls in null-length histogram fig. 3 already hinted at the existence of a large number of nulls shorter than 1 pulse period. Some of the starts and ends of these nulls must occur within the pulse window. Because of the variability normally already present in the pulse shapes, this dying or growing of the flux is not directly recognisable, but statistically it results in a lower energy for the pulse. In the histogram of pulse energies these 'half-nulls', or 'half-pulses' should be located in between the energy-values of the real nulls and the real pulses (the vertical lines in fig. 2).

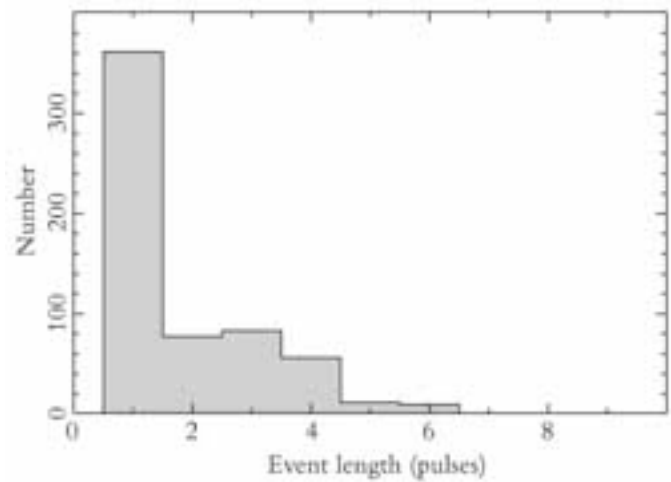

**Fig. 4.** Histogram of the event lengths.

| Event length | pulse–null combinations | # of events | # N | # P | total |
|---|---|---|---|---|---|
| 1 | N | 361 | | | |
| 2 | NN | 76 | | | |
| 3 | NNN | 27 | 27 | 0 | 27 |
| | NPN | 52 | 0 | 52 | 52 |
| 4 | NNNN | 16 | 32 | 0 | 32 |
| | NPPN | 13 | 0 | 26 | 26 |
| | NPNN | 19 | 19 | 19 | 38 |
| 5 | NNNNN | 1 | 3 | 0 | 3 |
| | NPPPN | 2 | 0 | 6 | 6 |
| | NPPNN | 1 | 1 | 2 | 3 |
| | NPNPN | 1 | 1 | 2 | 3 |
| | NPNNN | 4 | 8 | 4 | 12 |
| 6 | NNNNNN | 1 | 4 | 0 | 4 |
| | NPPPPN | 2 | 2 | 6 | 8 |
| | NPPNPN | 3 | 3 | 9 | 12 |
| | NPNPNN | 2 | 4 | 4 | 8 |
| | NPNNNN | 1 | 3 | 1 | 4 |
| | total: | | 107 | 131 | 238 |
| | fraction | | .45 | .55 | 1 |

**Table 1.** Length, pattern and occurrence of all null-groups in our sample. By definition, events always start and end with nulls. N denotes a null, P a pulse. Symmetric patterns are treated as one: NPNN for example is a combination of NPNN and NNPN. Within a group, observing a pulse is 1.2 times more likely than observing a null.

turns off near the end of the null), but this we do not see. The last pulse before the null actually shows a stronger trailing part, contrary to what we would expect.

Finally we note that the first and last active pulse more clearly show two-peaked profiles than the average profile does. As the separation between the peaks is roughly equal to the subpulse separation $P_2$, they could be explained by the pulsar turning on and off at certain preferred subpulse-carousel positions. In this case the subpulses would not smear out in the average, as in the normal profile, but add up. We have not been able to verify this suggestion in an independent way however, so we put it forward with some reservation.

*Profiles around low-energy pulses*

As some of the low-energy pulses are probably partial nulls, we also investigated the profiles of these pulses and of the pulses surrounding them. Similarities between the profiles around nulls and low-energy pulses, would again suggest they are related. We averaged the profiles of 80 low-energy pulses and the pulses that surround them. The result is shown in fig. 6.

First thing to note is the two-peaked profile of the average low-energy pulses themselves, with the peak separation equal to the subpulse separation. This indicates that certain subpulse-positions are overrepresented, a conclusion strengthened by the fact that the last and first active pulses are offset by the same amount as would be expected based on normal subpulse drift. Apparently, normal pulses with certain subpulse positions are less bright than average and contaminate the equally dim half-null sample.

leading edge should become relatively more important as the pulsar sometimes already

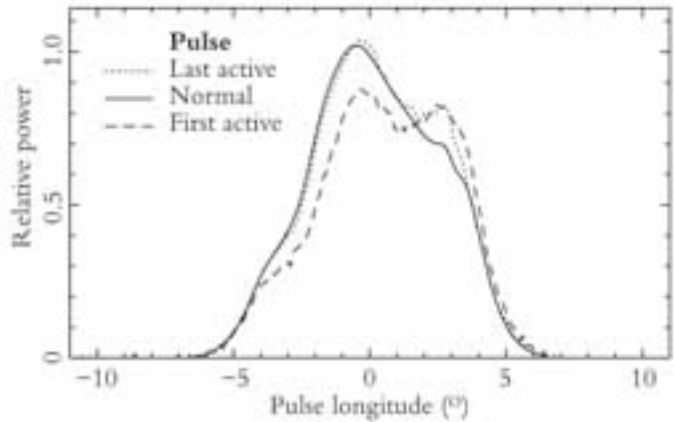

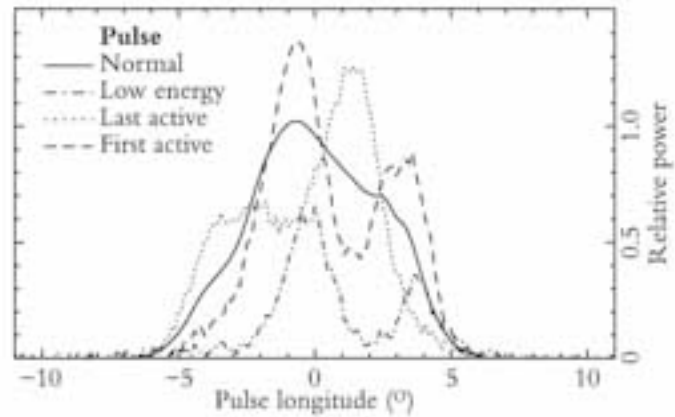

**Fig. 5.** Average profile around nulls. Black line: average profile for normal pulses. Dotted line: average profile for all pulses directly before a null. Dashed line: average profile for all pulses directly after a null. We have used all (500+) nulls, including those in groups.

**Fig. 6.** Average profile around lows, as in fig. 5. Again, the black line represents the average pulse profile. The dotted and dashed lines are the profiles before and after the low in this case. The dashed–dotted line shows the average profile for the low–energy pulses.

**Fig. 7.** Subpulse-pattern longitude jumps over continuous nulls. Each mark is the jump over one null. **a)** Mapped into a region from $-\frac{3}{4}P_2$ to $\frac{1}{4}P_2$ around the expected location in the continuous-drift case. **b)** The same data, duplicated to show subpulse grouping, with simulated pre-null subpulse positions added for clarity. The vertical dashed-dotted line predicts subpulse positions for a total stop of all drift during the null. The dashed lines are the prediction if drifting continuous during the null as it did before, for non-aliased drift. The dotted lines show which subpulses one spark makes if the drift is aliased. **c)** The same data, now with the predicted aliased subpulse path (dotted line), and the actual lower driftrate.

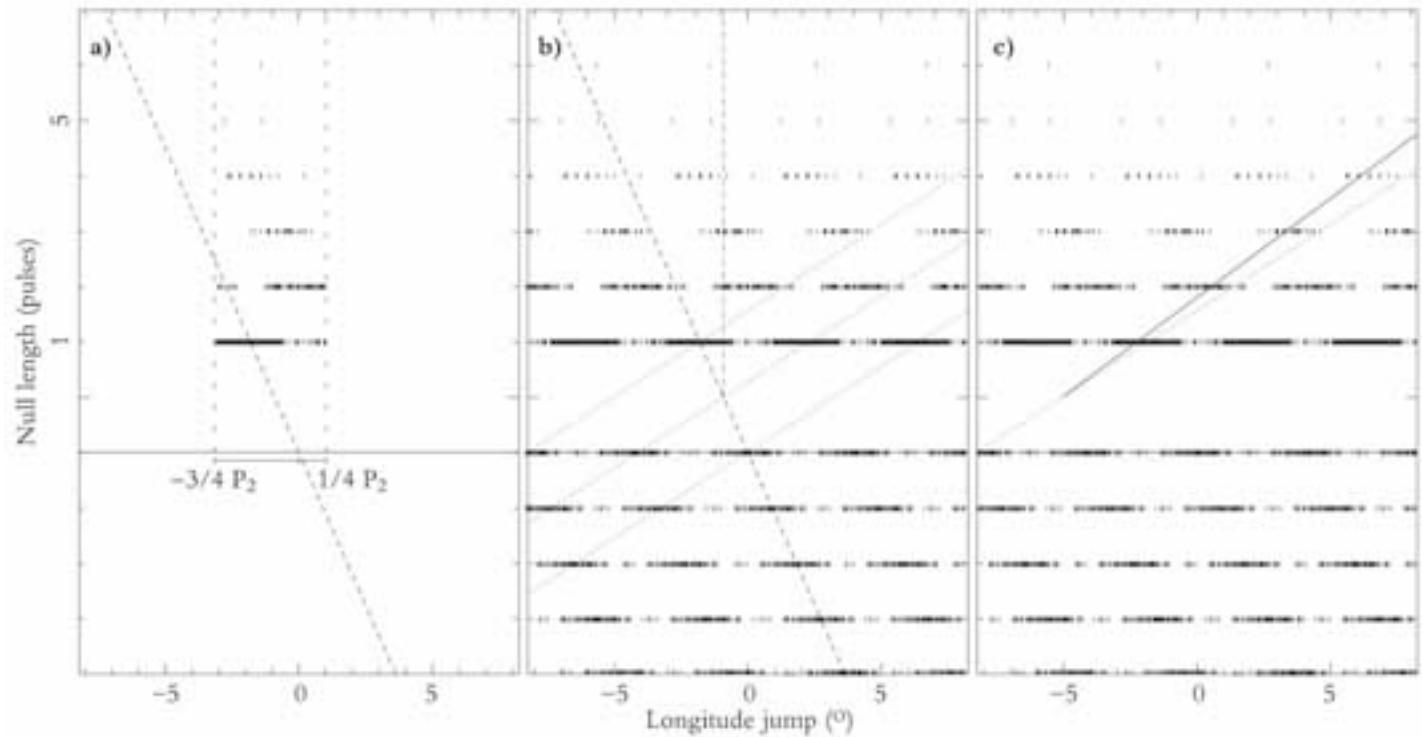

## Position jumps over nulls

Looking at the subpulse drift in PSR B0818−13 in fig. 1, the drift direction seems obvious (slowly towards earlier arrival) but in reality it might be different. The subpulses lack clear, recognisable traits (width, height or spacing) so instead of moving slowly towards earlier arrival, the subpulses might actually move more quickly towards later arrival, and still produce the same image as fig. 1. As we only observe the subpulse position once per pulse period, we undersample the subpulse movement and cannot determine the underlying speeds immediately (also see fig. 2 in Chapter 2).

By focusing on drift behaviour around nulls, we hope to obtain more information about the subpulse-drift alias mode and the rotation speed of the underlying carousel, like we previously did for PSR B0809+74 (Chapter 2). For each null in our sample, we determine the positions of the preceding and following subpulses and calculate the phase jump of the subpulse system over each null.

In fig. 7a we plot the longitude jump of the subpulse system versus null length, for all nulls in our sample. As the subpulses are indistinguishable, it is not clear which pre-null subpulse corresponds to which post-null subpulse. Therefore, in fig. 7a, all the data is mapped into one region that is $P_2$ (the average subpulse separation) wide.

To visualise the data better, we have recreated the subpulse pattern by duplicating the data-points with offsets of N×$P_2$ in fig. 7b. The real grouping of the subpulses is now much more apparent. This diagram is analogous to fig. 1, but is now thought to show the drifting of the subpulses during the null, when they are not actually visible. We have also added some simulated pre-null subpulse positions in the lower half of the diagram, for clarity.

If, during the null, the subpulse would stop drifting, the subpulses should follow the vertical dashed-dotted line in fig. 7b. If the subpulses continue to drift during the null, they should follow the dashed line. Neither of the two is a particularly good fit. After the null, the subpulses appear to outrun the prediction-line, which would indicate the drift-speed has increased during the null.

## The subpulse drift alias mode

For the dashed line in fig. 7b we have assumed that the drifting is not aliased, i.e. that one driftband is made by only one spark. Another option would be that the carousel is rotating the other way (as in fig. 2b in Chapter 2). In this case, a single spark does not move towards earlier arrival and follow the dashed line in fig. 7b. It moves towards later arrival and follows one of the dotted lines.

Compared to these dotted lines of expected positions, the post-null subpulses are lagging, which means the drifting slows down during a null. This would be consistent with the behaviour of the only other well-studied case of drifting–nulling interaction, in PSR B0809+74. There the spark-carousel unambiguously decelerates during a null, even to the point of a total stop. If we assume that our two-pulsar sample behaves similarly instead of oppositely, the spark carousels in both should slow down during nulls, and therefore the real drift direction in PSR B0818−13 must be towards later arrival. It could do so as illustrated by the dotted lines in fig. 7b, but faster too. If we assume the carousel consists of 20 sparks, similar to PSR B0943+10 Deshpande & Rankin (1999) and PSR B0809+74 (Chapter 2), we can place an upper limit on the carousel rotation time. In the slowest aliased case, it takes a subpulse 1.3 pulse periods P to reach the position of its neighbour (i.e. the vertical spacing of the dotted lines in fig. 7b). The whole carousel then rotates in $1.3 \times P \times N_{sparks}$ ~30s. This already makes PSR B0818−13 the pulsar with the fastest carousel rotation time known; if the subpulse drift is more strongly aliased, the rotation time is even smaller.

*Subpulse drift during intermittent nulls*

If the carousel in PSR B0818−13 slows down during a null, then why doesn't it stop like the carousel in PSR B0809+74 does? Again, the answer is in the intermittent nature of the apparent long nulls. What appears to be a long null is actually a series of short nulls: during each of these the carousel may slow down, only to speed up again in between. On average, this comes down to a steady lower carousel speed during nulls. In fig. 7, this would translate to a straight line through the post-null subpulse groups, and this is in good agreement with the data, as the full line in fig. 7c shows. If we assume the alias mode as illustrated by the dotted line in fig. 7c, the average driftrate (full line) during a long apparent null is about $0.87\pm0.05$ of the normal driftrate (dotted line). Using the pulse/null ratio of 1.2 we found previously, this means that during the short nulls that compose this long apparent null, the drift must be less than half the normal driftspeed to produce the apparent driftspeed observed.

If this is correct, and the slow drift-rate seen in the apparent long nulls is due to the subpulse drift starting and stopping every time a short null changes into a short burst, we must find the same lower driftrate over the intermittent null-groups. In fig. 8 we show the subpulse pattern over these intermittent nulls in the same fashion as fig. 8 did for continuous nulls. Again we note that the subpulses arrive earlier than expected. The similarity in figures 7 and 8 is striking; even in their drift-rate, the continuous and intermittent nulls behave identically.

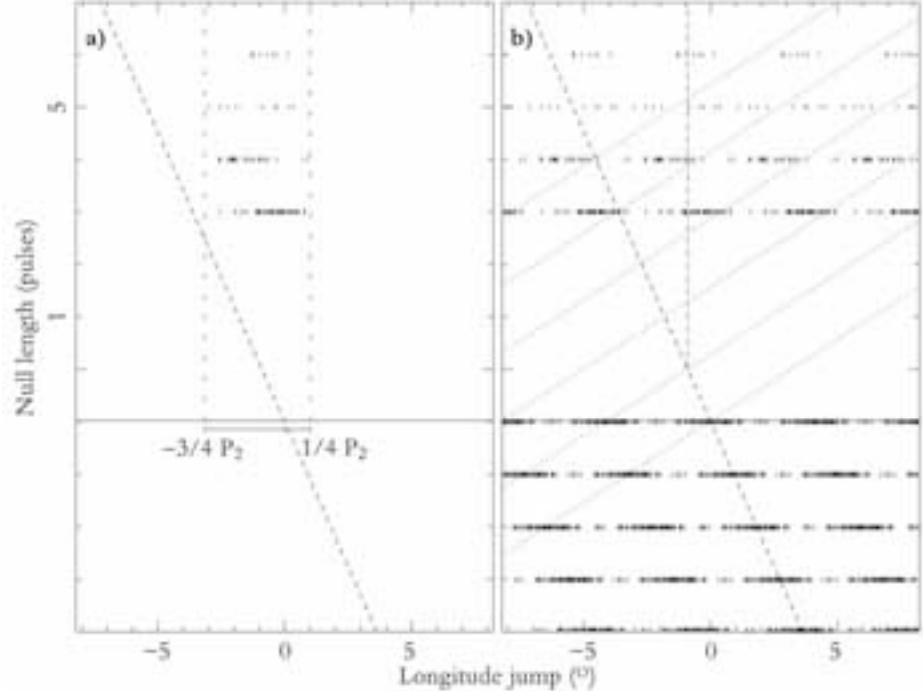

**Fig. 8.** As fig. 7, but now for intermittent nulls.

## 4. Discussion

In a rotating-carousel model, the groups of subsequent short nulls and pulses can be explained in two ways. In the one we have thus far focussed on, the entire carousel quickly stops and starts emitting. In this model all subbeams pause at the same time. In a different model, only some subbeams in a carousel stop emitting while the carousel rotation continues, creating pulse-null patterns similar to those observed. When the off-part of the carousel rotates into view, one observes a null. One pulse period later, an on-part might have moved into the pulse window to produce a normal pulse. Such a configuration would have to meet some very specific criteria, though. If the pattern is caused by on and off-parts of the carousel turning in and out of view, one should also observe half-on half-off pulses, with fewer subbeams. Yet we have found that the pulses in pulse-null groups contain exactly as many subpulses as normal pulses do. Only a carousel with many subbeams (roughly several hundreds) at a very high rotation speed could begin to explain this lack of observed on- to off-part transitions, but with a subpulse separation of 4°, the inferred size of the total carousel appears unrealistic.

When we compare pulsars B0809+74 and B0818−13, the similarities are striking. Both shine normally for 98-99% of the time, to be occasionally interrupted for several seconds and then resume their normal emission. Both show similar changes in driftrate during nulls. It seems clear we see the same mechanism at work in these two pulsars. Yet for PSR B0809+74 we know the long nulls must be real, as the subpulse drift stops during these nulls (Chapter 2), while in PSR B0818−13 we never see long nulls. There we observe periods of intermittent emission and quietude, in groups about as long as the nulls in PSR B0809+74.

Apparently, the mechanism that causes the emission in PSR B0809+74 to fully turn off, only reaches a border-line state in PSR B0818−13. Why the chances for null or pulse occurrence within such a state should be almost equal is unclear. The long timescales involved (even in the intermittent nulls) remains the biggest problem. In a polar-gap model (the basic physics of which are already debatable) for example, the potential gap over the magnetic pole can build up or discharge in microseconds (Ruderman & Sutherland 1975). Why the pulsar would flicker on a timescale of seconds, for 10 seconds in total, as it does in PSR B0818−13 remains unclear; we have not been able to explain this behaviour with any current pulsar theory. What one can do, is study the behaviour of different pulsars to find underlying patterns, which is what we have now done for PSR B0809+74 and PSR B0818−13.

The carousel rotation times found in PSR B0943+10 (41 seconds) and PSR B0809+74 (> 200 seconds) pose a similar time-scale problem; in a simple polar-gap theory, the rotation times are roughly several seconds (Ruderman & Sutherland 1975) and the rotation times found thus-far are much larger. The upper limit of 30 seconds we find for PSR B0818−13 extends the sample, back towards the previously expected values. A theory like the abovementioned also predicts that the rotation times scale with pulse period P and magnetic field strength B as $\frac{B}{P^2}$, which implies that the rotation time of PSR B0818−13 should be 3 times longer than that of PSR B0809+74, instead of the more than 10 times shorter we find. Our determination of the carousel rotation time in PSR B0818−13 is therefore a step to finding the dependencies of carousel rotation times on other pulsar parameters empirically.

## 5. Conclusions

Both the apparently long nulls and the groups of intermittent nulls can be quantitatively explained by trains of short nulls and pulses. The subpulse-drift behaviour over the two is identical: during the nulls, the subpulse drift appears to speed up. Assuming the underlying sparks slow down, as they do in PSR B0809+74, we conclude that the subpulse drift in PSR B0818−13 is aliased. This then leads to a carousel-rotation time of less than 30 seconds, the fastest one found to date.

# UNUSUAL SUBPULSE MODULATION IN PSR B0320+39

**with Russell Edwards and Ben Stappers**

We report on an analysis of the drifting subpulses of PSR B0320+39 that indicates a sudden step of ~180 degrees in subpulse phase near the centre of the pulse profile. The phase step, in combination with the attenuation of the periodic subpulse modulation at pulse longitudes near the step, suggests that the patterns arise from the addition of two superposed components of nearly opposite drift phase and differing longitudinal dependence. We argue that since there cannot be physical overlap of spark patterns on the polar cap, the drift components must be associated with a kind of multiple 'imaging' of a single polar cap 'carousel' spark pattern. One possibility is that the two components correspond to refracted rays originating from opposite sides of the polar cap. A second option associates the components with emission from two altitudes in the magnetosphere.

## 1. Introduction

For a number of pulsars it is known that each pulse is composed of a number of subpulses, and that for each successive pulse the subpulses appear to 'drift' by a given amount across the profile (Drake & Craft 1968). The rate at which subpulses drift across the profile is not constant but rather depends on their pulse longitude. This can be understood in the context of the widely adopted 'carousel' model of drifting subpulses (Ruderman 1972), which postulates the existence of a ring of equally spaced 'sparks' above the polar cap that give rise to 'tubes' of plasma streaming upward from the surface, ducted along magnetic field lines. At a certain altitude these particles emit microwave radiation that is beamed along tangents to the local field lines, giving rise to a beam pattern that consists of a ring of 'subbeams' (bright spots), reflecting the polar cap spark configuration. As the pulsar rotates the distant observer samples emission along a certain path in the beam pattern. Under the right viewing geometry, the sight-line makes a tangential pass along the ring of subbeams, giving rise to the reception of a sequence of one to a few subpulses every time the star rotates. Due to the non-linear mapping between (spin) longitude and magnetic azimuth, the subpulses should appear more closely spaced around the point at which the sight-line makes its closest approach to the magnetic pole. As the carousel slowly rotates about the magnetic pole, the longitude at which the emission from a given spark is seen exhibits a monotonic drift with time, which by extension of the above argument is greater in magnitude further from the magnetic pole.

By stacking the intensity time series modulo the pulse period to produce a two-dimensional array in pulse number and pulse longitude, and taking Fourier transforms along constant-longitude columns (forming a longitude-resolved fluctuation spectrum; LRFS), Backer (1970a,b) was able to measure not only the characteristic periodicity in pulse number ($P_3$, the time taken for one subpulse to drift to the former position of its neighbour), but also its amplitude and phase as a function of pulse longitude. Many subsequent studies employed this method but ignored the phase information, however those that examined the longitude-phase relation (Backer 1970a; Wright 1981; Davies et al. 1984; Biggs et al. 1987; Ashworth 1988) found behaviour qualitatively consistent with that expected under the carousel model.

The longitude-phase relationship expected under the carousel model depends on the number of sparks present, the misalignment angle between the spin and magnetic axes, and their orientation with respect to the line of sight. Hence studies of the relationship have the potential to provide valuable information about the pulsar, particularly in comparison to the observed longitudinal dependence of the polarisation position angle which, under the magnetic pole model of Radhakrishnan & Cooke (1969), depends on the same geometric parameters.

Motivated by these goals, we embarked on a program of analysis of several pulsars with drifting subpulses using the Westerbork Synthesis Radio Telescope (WSRT). One of the target sources was PSR B0320+39. This pulsar was discovered independently by Damashek et al. (1978) and Izvekova et al. (1982) in surveys of the northern sky at 400 and 102.5 MHz respectively. Further observations revealed the emission of very regular drifting subpulses (Izvekova et al. 1982) which occur in two distinct pulse longitude intervals, with a non-drifting component present in the average profile (Izvekoza et al. 1993; Kouwenhoven 2000). In this paper we report a most unexpected result from our study of the longitude-phase relationship of the drifting subpulses of PSR B0320+39.

## 2. Observations and analysis

We observed PSR B0320+39 on 2000 November 3 with the Westerbork Synthesis Radio Telescope (WSRT). Signals from fourteen telescopes were added (after appropriate delays) in each sense of linear polarisation in a 10MHz band centred at 328MHz. These were processed using the Dutch Pulsar Machine (PuMa), operating as a digital filterbank with 64 channels and 819.2$\mu$s sampling in total intensity (Stokes I). For details of the system see Voûte et al. (2002).

In offline analysis we de-dispersed the data and binned the resultant time series into an array in pulse longitude and pulse number using the ephemeris of Arzoumanian et al. (1994). A total of 3400 pulses in 3701 longitude bins was recorded, with a region of 512 bins including the on-pulse used in further analysis. The samples were normalised to produce an average profile with a zero baseline and a peak value of unity. The first 100 pulses are shown in a greyscale form in Fig. 1.

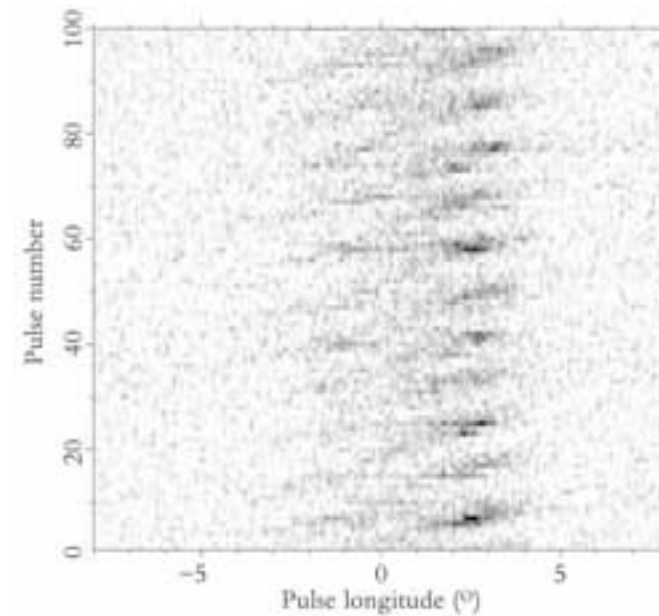

**Fig. 1.** First 100 pulses of 328 MHz timeseries for PSR B0320+39, binned in pulse longitude and pulse number.

The techniques employed in the further analysis are based on those presented in detail by Edwards & Stappers (2002). We computed the two-dimensional Discrete Fourier Transform (DFT) of the first 2048 pulses of the pulse longitude − pulse number array to yield what we call the two-dimensional fluctuation spectrum (2DFS). This is mathematically equivalent to the harmonic resolved fluctuation spectrum (HRFS; Deshpande & Rankin 1999; Edwards & Stappers 2002)[1], and the form of the power spectrum found here for PSR B0320+39 matches that found by Kouwenhoven (2000) for the HRFS. For the purposes of clear display we also computed a spectrum using data from which we subtracted the average profile from each pulse, and padded with zeros in the longitude axis, in order to improve the resolution. The resolution in the time-associated axis was unnecessarily high (2048 elements), so to improve the signal-to-noise ratio we smoothed the power spectrum by convolution with a 7×1-bin ($v_t \times v_l$) boxcar before plotting the relevant portion of the spectrum[2] in Fig. 2. The spectrum is dominated by a pair of components associated with the drift modulation; two are present because the spectrum derives from the Fourier transform of real-valued data (see Edwards & Stappers 2002). Also present with high significance are a pair of responses corresponding to the second harmonic of the subpulse response. The features near $v_t = 0.02$ and $0.04$ cycles/period are significant, however they also appear in an analysis of off-pulse data, indicating that they are caused by periodic interference.

After performing the 2D DFT, the complex spectrum was shifted in both axes by an amount which placed the primary response to the drifting subpulses at zero frequency. We then masked other components present in the spectrum (namely the complex conjugate 'mirror' of the primary response, the pair of second harmonics and the DC component) by multiplication of the coefficients with a transfer function:

$$f(v_t) = \begin{cases} 0 & |v_t - v_{tc}|/w < 0.5 \\ 2|v_t - v_{tc}|/w - 1 & 0.5 \le |v_t - v_{tc}|/w \le 1 \\ 1 & |v_t - v_{tc}|/w > 1 \end{cases} \quad (1)$$

where $v_t$ is the time-associated axis, $v_{tc}$ is the centre frequency of the component to mask, and $w$ is the frequency extent of the region of zero transmission. The results that follow derive from filters with $w = 0.05$ cycles/period, although as expected they were found to be insensitive to $w$ unless it becomes too small to mask adequately or so large that it masks some of the desired signal. We then computed the inverse DFT of the result.

[1] Without the use of 'twiddle factors' (Cooley & Tukey 1965), the 2DFS will differ from the HRFS. However, it is easily shown that this merely amounts to the provision of slightly different sampling lattices over the same continuous function, due to the assignment of integer pulse numbers (2DFS) versus integer harmonic numbers (HRFS).

[2] The sampling interval used allows longitude-associated frequencies up to 1850 cycles/period, however all significant power is present in the portion of the spectrum shown.

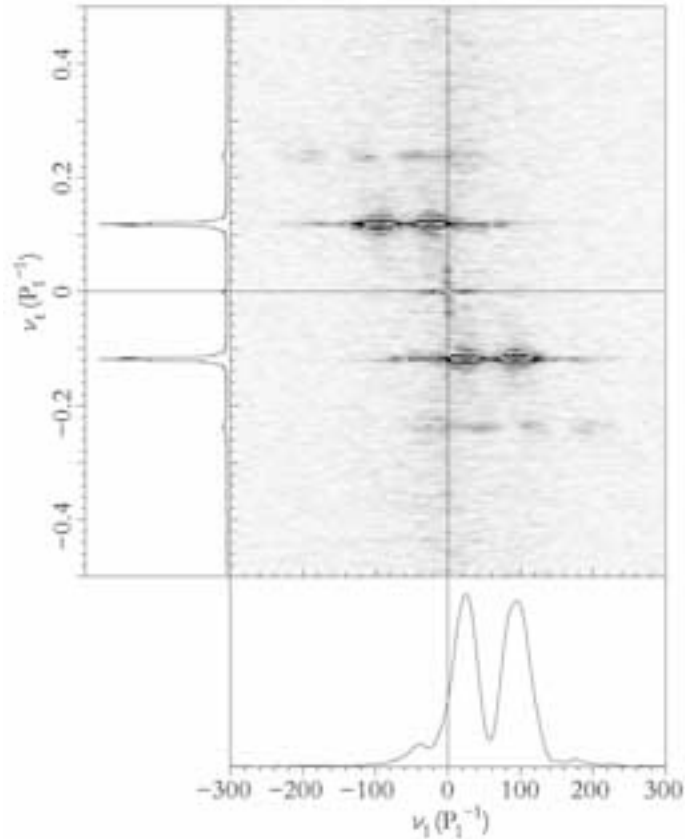

**Fig. 2.** Main panel: Two-Dimensional Fluctuation Power Spectrum of PSR B0320+39. The greyscale saturates at 0.05 the peak power, while the contours show the same spectrum at levels 0.05 and 0.5 times the peak. Left panel: 'collapsed' spectrum formed by averaging rows of the 2D spectrum. Bottom panel: power as a function of $v_l$ for $v_t = -0.118$.

We represent the subpulse signal as the real part of the product of a pure two-dimensional sinusoid and a two-dimensional complex modulation envelope. By the convolution theorem, this envelope is the result of the inverse transform performed above. It describes the deviations from a pure periodicity of constant amplitude that arise due to amplitude modulation at the pulse period, scintillation or nulling, non-uniform spacing of subpulses in pulse longitude (e.g. due to sight-line curvature) and slow variations in the drift rate. Using the iterative scheme of Edwards & Stappers (2002) we decomposed the two-dimensional envelope into the product of two modulation envelopes that vary in pulse longitude and pulse number respectively. Such a decomposition is able to model the behaviour of a rotating carousel, where the geometry defines the longitude-phase dependence and the carousel rotation determines the time-phase dependence, and its basic assumptions − that the longitudinal spacing of the subpulses is time invariant, and that the time spacing of the subpulses is longitude independent − are in agreement with previous studies; see Edwards & Stappers (2002). The validity of this approach in the case of PSR B0320+39 was confirmed by computation of the 2DFS of the difference between the observed signal and that predicted using the real part of product of the two one-dimensional envelopes. No significant power remained in the portion of the spectrum previously occupied by the primary drifting subpulse response.

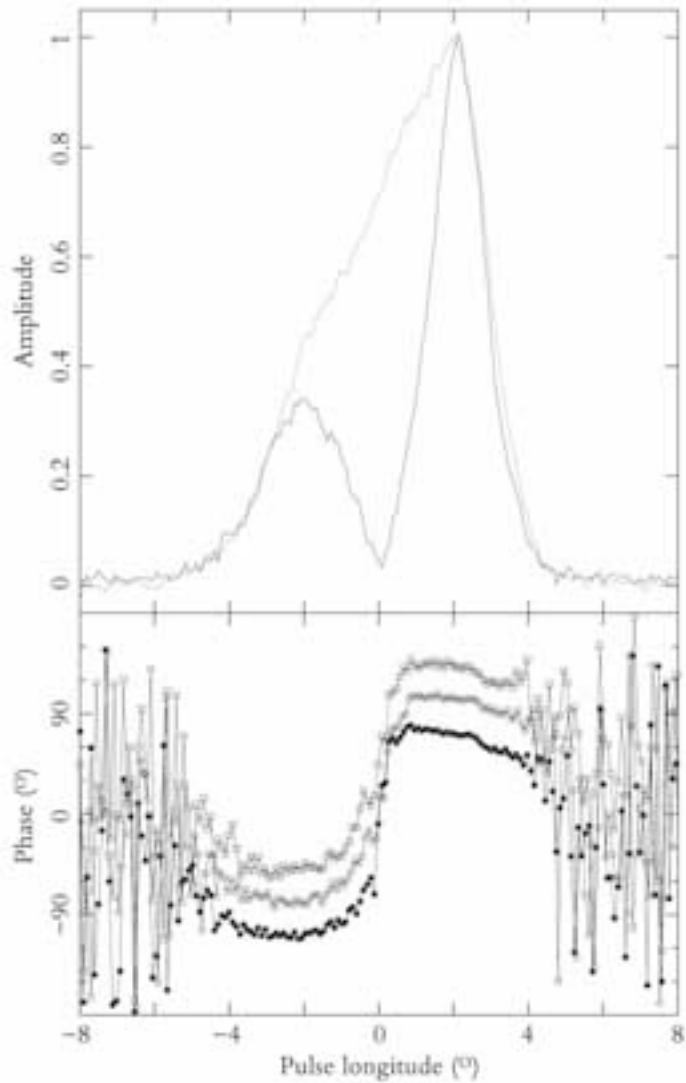

The spectral shifting performed earlier has the effect of removing a constant phase slope in each of the envelopes. Since the slope varies over the envelope, the choice of value to use in the shifting is somewhat arbitrary and was refined after viewing the resultant envelopes. The periodicity with pulse number was well defined and a value of $\hat{P}_3 = -P_1/0.118$ was used. The response to the drifting component in the other axis is very broad, and in fact extends well into negative frequencies (Fig. 2). This complicates the choice of value for $1/\hat{P}_2$ to be used in shifting the spectrum in the longitude axis. We chose a value of $\hat{P}_2 = P_1/60$ to give the resultant envelope a relatively small phase slope over most of the pulse[3].

The inferred longitude-dependent envelope (Fig. 3) shows that nearly linear subpulse modulation occurs in two distinct longitude intervals, with a striking phase offset of ~180° between them. In terms of the longitude-time diagram, the effect of the offset is to cause the drift band peaks in one component to extrapolate to troughs in the other and vice versa. To check this result we computed the (complex) LRFS and examined the amplitudes and phases in frequency bins near $\hat{P}_3 = P_1/0.118$. Since the modulation is not a pure periodicity, the power is spread over several bins in the spectrum. Although of a much lower signal-to-noise ratio, the shape of the phase envelope in each bin was consistent with that inferred with our preferred technique, which makes optimal use of all modulation power. Similarly, division of the observation into segments in time (which results in wider frequency bins and potentially also in reduced frequency drift in each segment) and examination of the phase relation in the peak bin of each segment also gave consistent results (regardless of the segment size). In order to improve the signal-to-noise ratio of these other methods, we also used an iterative algorithm based on multiplication with the complex conjugate of a template envelope to determine the correct phases with which to add envelopes from different frequency bins or time segments (producing in effect coarse-grained versions of the algorithm of Edwards & Stappers 2002). The resultant summed envelopes were again consistent with the result of our preferred algorithm (Fig. 3). As a final check, we folded the longitude-time data about the pulse number axis (modulo $1/0.118$ pulses) to produce an average profile as a function of drift phase and pulse longitude (Fig. 4). As with the LRFS methods, much of the modulation was lost due to the unstable drift motion, however some residual modulation remained and is sufficient to confirm the unusual phase behaviour.

**Fig. 3.** Inferred modulation envelope (solid line and filled circles) and average pulse profile (dotted line). Also shown is the phase envelope computed by coherent addition of a set of 69 rows ($0.108 < \nu_r < 0.128$) of the 3400-pt LRFS over which significant power was seen (unfilled circles), and the phase envelope computed by coherent addition of the peak row from each of twenty-six 128-pt LRF spectra resulting from time-segmentation of the sequence (with this transform length the majority of modulation power falls in a single row). The nominal phase slope of $1/\hat{P}_2 = 60°/°$ has been subtracted from the phase envelopes before plotting.

---

[3] Since this value does not lie on the natural grid of the DFT, the shift was actually performed in the time-longitude domain by multiplication with a complex exponential.

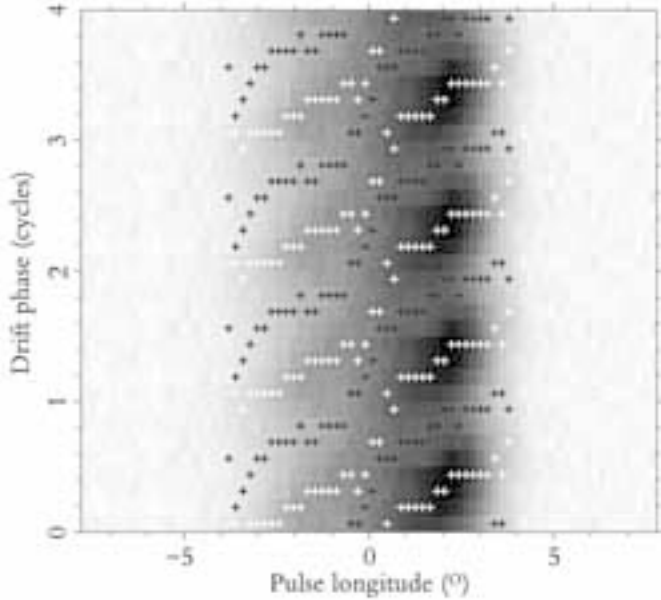

**Fig. 4.** Data folded at $P_3$, as a function of pulse longitude and drift phase. Maxima and minima in each longitude bin are indicated with black and white crosses respectively. The data are plotted four times to guide the eye.

Knowing that there is a $\sim 180°$ offset between the two halves of the subpulse phase envelope, we may explain two otherwise curious facts about PSR B0320+39. The first concerns the values of $P_2$ previously reported, which are around $P_1/120$ (Izvekoza et al. 1993 report $3.1°$ of longitude, i.e. $P_2 \simeq P_1/116$ at 406 MHz; Kouwenhoven 2000 reports 24.5 ms, i.e. $P_2 \simeq P_1/123$, at 328 MHz). These strongly differ from our nominal $\hat{P}_2 = P_1/60$ because they are based on measurements of actual subpulse separations, which will nearly always be taken between subpulses on opposite sides of the phase shift. Since the subpulse in the trailing component leads (or lags) the position it would take in a continuous linear phase envelope by $\sim 180°$ of subpulse phase, the derived spacing differs from that expected from the typical phase slope by a factor of $1/2$ (and probably sometimes $3/2$). The second curiosity explained by the phase offset is the local minimum in the subpulse HRFS response at around 60 cycles/period (Fig. 2). The complex longitude-dependent modulation envelope (once the constant phase slope is subtracted) is quite close to an odd real-valued function, which causes its Fourier transform to have local minimum in its power at DC. This is the cause of the local minimum around 60 cycles/period, the typical longitudinal drift frequency.

Without the phase offset the envelope would be close to a real-valued even function and its Fourier transform would not have a strong minimum at its centre.

Furthermore, we note that the equivalence of the 2DFS and HRFS provides an explanation for the apparent 'secondary' component at 0.118 cycles/period, suggested by Kouwenhoven (2000) to be due to the ambiguity between which drift band joins to which between the two components. In fact, the subpulse response is very broad in the longitude-associated frequency axis, and extends significantly into negative frequencies. The symmetry of the two dimensional Fourier transform of real functions is such that the power at any frequency pair $\nu_l, \nu_t$ is equal to that at $-\nu_l, -\nu_t$ (where $\nu_l$ and $\nu_t$ are the frequencies in the longitude-associated and time-associated axes respectively). If only the positive frequencies in $\nu_l$ are plotted, as is the case with the HRFS, no information is lost but individual components that span the $\nu_l = 0$ axis will appear as two spatially separated components at $\nu_t$ values of opposite sign. When the full spectrum is plotted as in Fig. 2 the proper interpretation becomes apparent.

## 3. Discussion and interpretation

### Polar cap models

The phenomenon of drifting subpulses has been used repeatedly in investigations of pulsar emission as a potentially strong clue about the geometry of the system. The fact that those pulsars that show nearly linear drifting subpulses tend to have a double, 'unresolved double' or 'single' average pulse profile morphologies, with many showing a transition from resolved to unresolved double as the observing frequency is increased, is usually seen as strong support for some form of 'hollow-cone' polar cap model (Radhakrishnan & Cooke 1969; Ruderman & Sutherland 1975; Backer 1976; Rankin 1983, 1986; Lyne & Manchester 1988). PSR B0320+39 shares these characteristics of regular drifting subpulses and a transition from a resolved double to an unresolved double average profile morphology between 102 MHz (Izvekova et al. 1982)[4] and 328 MHz (this work; also Damashek et al. 1978; Izvekova et al. 1993).

The local steepening of the phase envelope of PSR B0320+39 in the transition region is inconsistent with the carousel model in its most basic form. For a single ring of 'sparks', as long as the sparks are elongated in either magnetic azimuth or opening angle (or neither, but not both), the subpulse modulation seen in a ring of constant opening angle has the same phase as that in any other ring, if measured from the same fiducial azimuth. Therefore, the observed longitude-dependent subpulse phase envelope is directly tied to to the magnetic azimuth sampling of the sight-line via multiplication by the number of sparks N (Edwards & Stappers 2002). The only way to produce a sudden swing in magnetic azimuth (and hence subpulse phase) is to pass close to the magnetic pole. Such a configuration is inconsistent with the usual interpretation of unresolved double profile morphologies, and with the gradual linear polarisation position angle swing observed in this pulsar (Suleimanova & Pugachev 2002). To prove this assertion we attempted to fit the observed subpulse phase swing with the geometric formula of Edwards & Stappers (2002), using a $\chi^2$ minimisation algorithm. From the figure provided by Suleimanova & Pugachev (2002) we conservatively estimate the range of polarization position angle gradients consistent with the data as $d\chi/d\phi = 12 \pm 3°/°$ (where $\chi$ and $\phi$ are the position angle and pulse longitude respectively). The parameters of the fit were required to be consistent with this gradient via the geometric model of Radhakrishnan & Cooke (1969). The result (Fig. 5) clearly fails to describe the observed behaviour.

---

[4] Due to an improved system response the profile of Kuz'min & Losovskii (1999) shows much stronger component separation than that of Izvekova et al. (1982); see fig. 6.

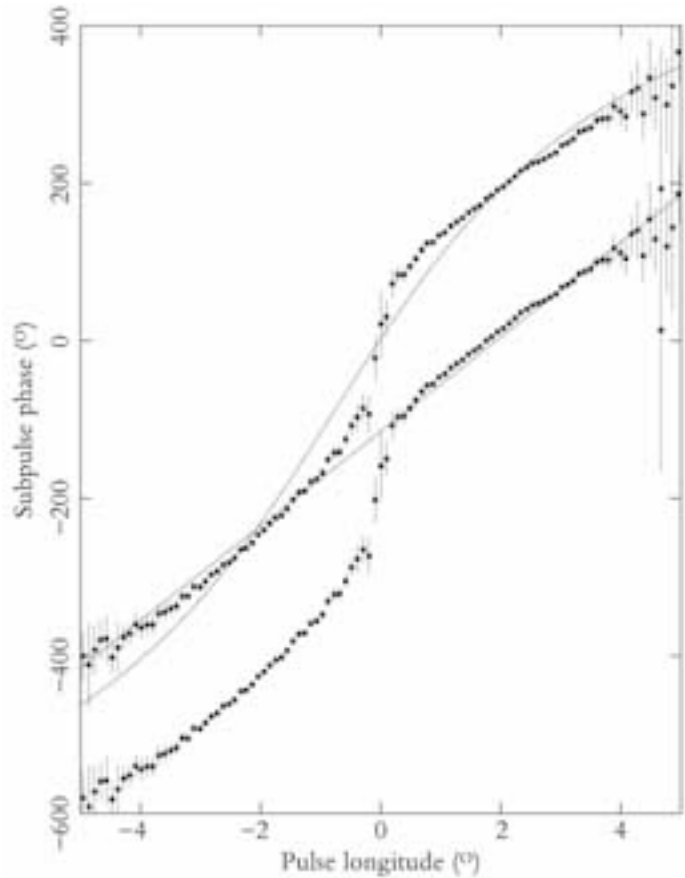

**Fig. 5.** Inferred subpulse phase envelope, including the 60°/° phase slope previously removed (indicated by the straight line). The data are plotted twice with an offset of 180° to faciate evaluation of the central feature as the transition between two offset regions of linear drift. Two-sigma error bars are included. The curved line represents the best fit to the data under the standard carousel model, given the polarimetric constraint (see text).

One way to produce a jump in subpulse phase is to adopt a nested ring polar cap spark configuration (Gil & Sendyk 2000) and to assign the two halves of the profile to rings of different opening angles and numbers of sparks. However, this would result in a measurably different $P_3$ unless the circulation rates of the two carousels were different by a fortuitous ratio (i.e. the inverse of the ratio of the numbers of sparks in the rings). In addition, the transition from resolved to unresolved double profile with increasing frequency could not be explained by standard radius-to-frequency mapping (RFM; Cordes 1978). Some other mechanism such as the finite beam–width of the elementary emission mechanism ($2\gamma^{-1}$), the finite band–width of emission at a given radius (with attendant RFM in the effective range of radii visible), or refractive broadening would be required to produce the profile evolution.

*Double imaging of magnetospheric origin*

Given the failure of the standard model to account for the observed subpulse drift behaviour of PSR B0320+39, we develop in this section a pair of explanations based on a simple and plausible extension of the model. Namely, we argue that the most important features of the longitude-dependent subpulse modulation envelope − a local reduction of amplitude accompanied by a rapid swing in the phase angle − are the result of destructive interference between two superposed drifting subpulse signals of nearly opposite phase. The observed behaviour and this explanation of it can be understood by analogy with the phenomenon of orthogonal polarisation mode transitions, where the pulsar signal is seen to consist of the incoherent sum of two signals with nearly orthogonal polarisation states (e.g. Manchester et al. 1975; Stinebring et al. 1984). In both cases the total intensity of the two signals simply add in the average profile, while the near anti-parallel nature of the vectors representing the complex subpulse modulation signal (in the Argand plane) or the polarised part of the signal (in the Poincaré sphere) explains the reduction in magnitude (subpulse amplitude, polarised intensity) and the swing in orientation (subpulse phase, polarisation position angle) during a transition of dominance from one mode to the other. Other than the obvious features of the shift in phase and the attenuation in amplitude, the continuity of both the average profile and the absolute value of the gradient of the subpulse amplitude envelope serve as supporting evidence for this hypothesis.

Since the polar cap cannot be physically populated with overlapping spark systems, we argue that the superposed drifting components of PSR B0320+39 are associated with a form of 'double imaging' of the polar cap excitation pattern. Whereas the polar cap may consist of a single, drifting ring of excitation points (e.g. sparks), the physics that map this to a radiated beam pattern must do so under two different transformations that are received in incoherent addition by the observer. This gives rise to a subpulse pattern that is the sum of two overlapping components with differing peak pulse longitudes, and phases that differ in their contribution at a given pulse longitude by a constant offset of ∼180°. This also implies that the average pulse profile is the simple sum of two overlapping components approximately equal to the corresponding subpulse amplitude envelopes.

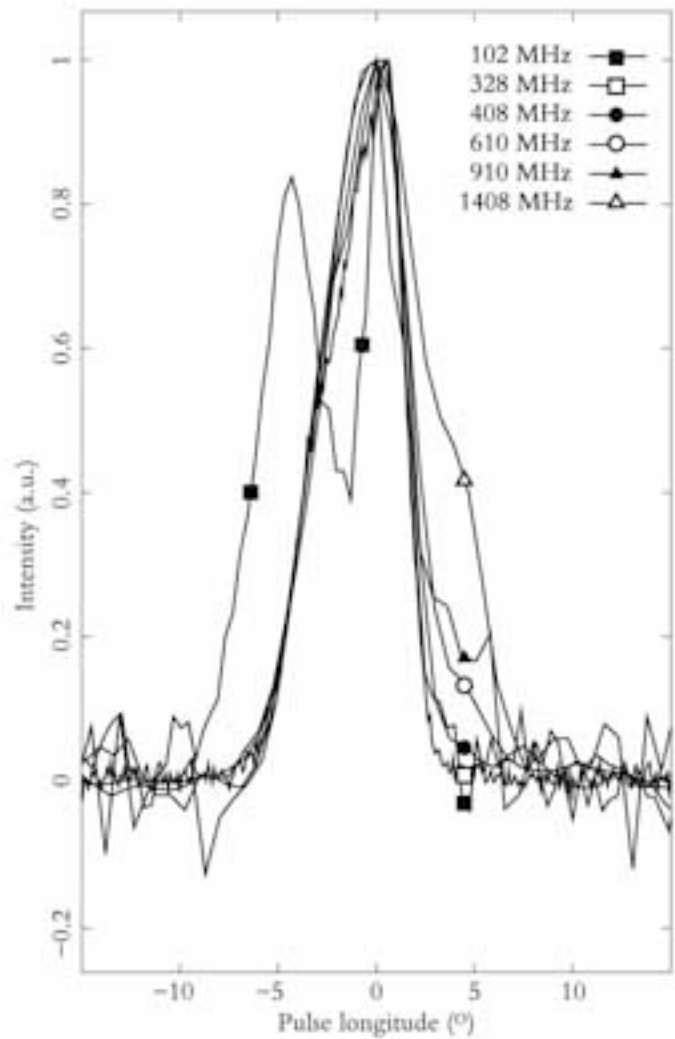

**Fig. 6.** Average total intensity pulse profiles of PSR B0320+39 over a frequency range of 102–1408 MHz. The profiles are normalised to the same peak intensity and (in the absence of timing information) were aligned by eye to match the model proposed here. The 328 MHz profile is from this work. The 102 MHz profile is that of Kuz'min & Losovskii (1999), and the remaining profiles are from Gould & Lyne (1998). All profiles other than the 328 MHz one were obtained via the EPN profile database.

In this section we discuss two potential mechanisms for such an effect. The first associates the images with different ray paths in a refractive magnetosphere (Petrova & Lyubarskii 2000), while the second attributes them to emission originating at two discrete altitudes in the magnetosphere (Rankin 1993). Before going into the details of each scenario, given that they both produce a beam consisting of a pair of nested cones, we are are in a position to examine the consistency between these models and the observed pulse profile morphology. This is important because, in assigning the two components of PSR B0320+39 to cones of different opening angles, we are abandoning the standard 'single cone' explanation favoured for most pulsars with regular drifting subpulses. We suggest instead that a cut through an outer cone (from the outside to the inside) produces a leading component, while a trailing component is produced by a pass along the edge of the inner cone. At most radio frequencies the difference in the opening angles of the cones is small and the components overlap, but radius-to-frequency mapping causes the outer cone to expand at lower frequencies, giving the components sufficient separation to be resolved at 100 MHz. Clearly one should also expect a third component to be present on the trailing edge of the profile as the sight-line passes across the outer cone for the second time. This is actually observed at frequencies at and above 600 MHz (Gould & Lyne 1998), giving further support to our assertion that more than one cone of emission is visible from this pulsar. As shown in Fig.6, the pulse profiles observed for PSR B0320+39 match this picture well. We now turn to the specifics of the two mechanisms.

*Refraction*

Refractive effects have long been considered potentially important in pulsar magnetospheres (e.g. Melrose 1979; Barnard & Arons 1986), particularly in the production of circular polarisation (Allen & Melrose 1982) and superposed orthogonally polarised radiation (McKinnon 1992). Only recently, however, has the potential for refraction to explain pulse profile morphologies seen detailed study (Lyubarskii & Petrova 1998; Petrova & Lyubarskii 2000; Petrova 2000). Based on the simple argument that the magnetospheric plasma density has minima both at the magnetic pole and at the edges of the cone defined by the last open magnetic field lines, Petrova & Lyubarskii (2000) showed that rays emitted poleward of the peak of the plasma distribution could be refracted *across* the pole to exit with an angle to the pole of the opposite sign to that with which they were emitted. Rays emitted outward of the plasma peak are simply refracted to somewhat greater opening angles. This is illustrated schematically in Fig. 7. Clearly, at any pulse longitude the observer may potentially receive two rays, one from either side of the magnetic pole.

In an axisymmetric plasma distribution, refraction occurs purely in the plane of the magnetic field lines. Petrova (2000) shows that the magnetosphere tends to concentrate rays from within a given set of co-planar field lines into two components of different peak opening angle, corresponding to the different ray paths described above.

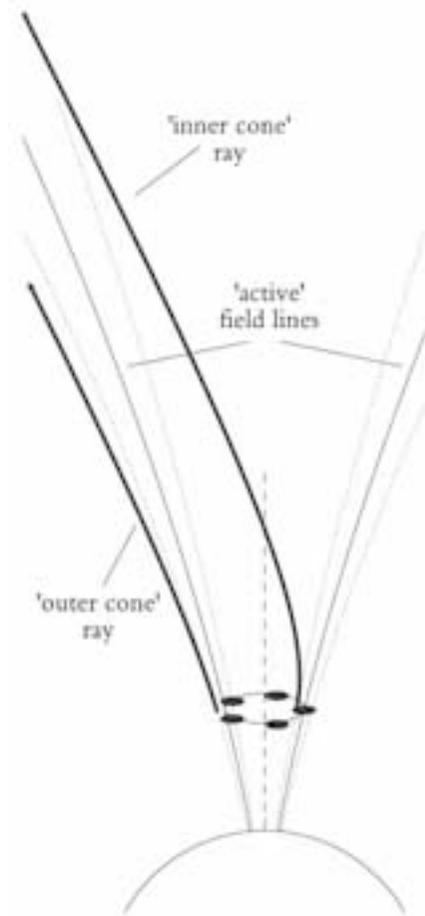

**Fig. 7.** Schematic illustration of the simultaneous observability of rays from opposite sides of the polar cap, via magnetospheric refraction. The 'active' field lines mark the peak of the plasma distribution at a given altitude. The dotted lines indicate the approximate edge of the emitting flux tubes. The left hand ray originates outside the peak of the plasma distribution, and is refracted further outward. The right hand ray originates inward of the peak plasma density but is refracted across the pole to exit at the same angle as the other ray. A carousel-like intensity pattern with 5 subbeams is shown to illustrate the origin of the subpulse phase difference between the two components associated with these ray paths.

gitude interval over which the phase transition occurs might be explained as a retardation effect or a refractive delay. Alternatively, inclusion of magnetospheric rotation effects or refraction under a moderately non-axissymmetric plasma distribution (Petrova & Lyubarskii 2000 model both but only in the context of polarisation wave mode coupling; see also Blaskiewicz et al. 1991) could potentially provide the kind of symmetry-breaking needed to produce subpulse phase offsets other than 0 or 180°. The form of the beam pattern and the manner in which a rotating carousel polar cap system maps to drifting subbeams under such conditions is yet to be investigated, to our knowledge.

We note that this model can also explain a related result concerning the drifting subpulses of PSR B0809+74. In chapter 5 we find that over the course of each subpulse, the polarisation state jumps in a regular manner from one polarisation state to another, almost orthogonal state. This is similar to the behaviour seen in PSR B0320+39, in that the observed total intensity drifting subpulse pattern can be understood as the sum of two components with a phase offset between them. However, in PSR B0320+39 the drift components cannot be associated with orthogonal polarisation modes since the linearly polarised average profile at 408 MHz peaks in intensity at the centre of the profile (Gould & Lyne 1998), where maximal depolarisation ought to take place, were the components orthogonally polarised. The single-pulse polarimetric work of Suleimanova & Pugachev (2002) confirms that the situation is similar at 103 MHz: significant power is present in two modes only in the profile wings. The model of Petrova (2001) for the origin of orthogonal polarisation modes offers a simple explanation for this disparity. Under that model the secondary polarisation mode is produced via the conversion of power from the ordinary mode in regions of propagation that are quasi-longitudinal to the magnetic field. If radiation in one of the conal components experiences significant conversion while the other does not, as could well occur since the ray trajectories involved are very different between the two, the situation is as seen in PSR B0809+74. If the configuration is not conducive to conversion then both components have similar polarisation, as in PSR B0320+39. Since the conversion process also alters the orientation (i.e. position angle, ellipticity) of the modes, the longitude-dependent non-orthogonality of the polarisation modes reported in Chapter 5 might also receive an explanation via this effect.

*Dual emission altitudes*

A second potential explanation for how we receive the sum of two related subpulse signals is that the two signals arise at different altitudes in the magnetosphere. This is the essence of the 'double conal' model of Rankin (1993), so-called because the production of radiation cones at two distinct altitudes in the magnetosphere gives rise to a beam pattern that consists of two nested cones. The radial separation of the emitting regions and the differential tangent angle between the active field lines and the magnetic axis that arises as a result causes the polar cap plasma flux cone to produce

These are associated with inner and outer components of the conal beam, which in terms of explaining the morphology of pulsar average profiles serves the same purpose as the double cone of Rankin (1993). This grouping of rays occurs within a given plane of field lines; all possible origins for visible rays must be co-planar, along with the line of sight and the magnetic axis. Therefore, to the extent that the plasma distribution is axisymmetric, the geometric conveniences of the basic magnetic pole model (Radhakrishnan & Cooke 1969) are preserved, including the polarisation behaviour, and the mapping of azimuthal intensity modulations (e.g. from a plasma flow initiated by a polar cap spark carousel) to observable drifting subpulses.

Moreover, if the polar cap pattern consists of a ring of an *odd* number of points of excitation, the subpulse pattern seen in the component associated with pole-crossing rays should be close to 180° out of phase with that in the other component. The slight deviation from purely out of phase signals required to explain the finite lon-a wider radiation beam at the upper emission region, compared to the lower one. If the height difference is large, up to four well-resolved conal profile components are produced, while reduced vertical separation causes the components to overlap and produce a 'boxy' profile shape. Fig. 8 illustrates this model as it might apply for PSR B0320+39 at 102 MHz, with the conal beams overlapping to a lesser degree than at higher frequencies.

Clearly this model is able to explain how we see superposed drifting patterns, but we have not yet offered an explanation for why there should be a subpulse phase offset between them. They arise from the same pattern of sparks (or plasma flux production) on the polar cap, yet we observe a phase difference of $\sim 180°$ near the profile centre. For the subpulse pattern to be delayed by $P_2/2 \simeq 30$ ms due to simple retardation (Cordes 1978) requires that the emission regions are separated by $\sim 9000$ km, 1–2 orders of magnitude greater than the values typically inferred from profile morphologies (e.g. Gangadhara & Gupta 2001; Mitra & Rankin 2002; Gupta & Gangadhara 2003). Including differential aberration reduces this by no more than a factor of 2 (Cordes 1978). Hence we consider this explanation unlikely. An alternative explanation worth pursuing in the future, but outside the scope of this work, is that a significant electric field exists throughout the magnetosphere, causing particles to accelerate upward mostly along magnetic field lines, but with some azimuthal E × B drift (e.g. Wright 2003). The drift would cause the beam pattern radiated at a given height to be rotated in magnetic azimuth by a certain amount, which may account for the subpulse phase offset seen here.

In any case, under this model the phase difference must ultimately be due to the different altitudes at which the components originate. Since RFM must apply to the outer (i.e. upper) component to explain the 102 MHz pulse profile, the difference in subpulse phase between the components ought to be a strong function of frequency. Although this is the first study to examine the longitude dependence of subpulse phase in this pulsar, the fact that the longitude-dependent subpulse amplitude envelopes measured at 102 MHz and 408 MHz by Izvekoza et al. (1993) drop to close to zero in the profile centre, just as at 328 MHz, suggests that at all three frequencies the subpulses add destructively in this interval, and therefore that the phase difference between them is close to 180°, regardless of frequency. The origin of the phase offset therefore remains a mystery under this scenario.

## 4. Conclusions

The drifting subpulses of PSR B0320+39 occur in two distinct intervals of pulse longitude. Using a sensitive new technique we have shown that the extrapolation of subpulse drift from a given longitude interval is $\sim 180°$ out of phase with the subpulses of the other. This was confirmed by analysis of the longitude-resolved fluctuation spectrum and by folding the data into profiles corresponding to different phases of the subpulse drift.

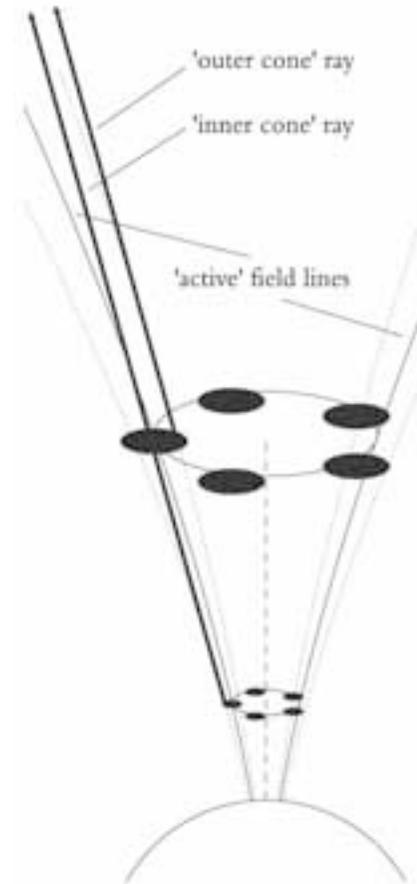

**Fig. 8.** Schematic illustration of the simultaneous observability of rays from two different emission heights (see also Fig. 7). Due to the curvature of the magnetic field lines along which the radiating particles stream, the ray from the upper region must come from a more inward field line that that from the lower region, if the two are to make the same angle to the magnetic axis.

This suggests the presence of two components of nearly opposite drift phase, and based on the forms of the average pulse profile and the longitudinal dependence of the amplitude of subpulses, we argued that the modes are superposed at all longitudes. The adding of intensity contributions from the two modes results in attenuation of the modulations in the longitude interval where the modes are of similar intensity, and produces a steep transition of $\sim 180°$ in the longitudinal subpulse phase envelope at the point where the component intensities are equal.

Two specific models are suggested to explain this behaviour. Both produce a nested pair of conal beams, one via magnetospheric refraction, the other via emission at two distinct altitudes. The leading and trailing components are associated with outer and inner cones respectively. The second cut of the outer cone does not give rise to a component at 102 MHz or 328 MHz, but does produce a third, trailing component at frequencies above 600 MHz. The component separation increases rapidly at frequencies below 328 MHz, in accordance with the expectations of radius to frequency mapping. The refractive scenario can explain the observed phase offset as a simple consequence of having an odd number of subbeams in the carousel, whilst under the dual-altitude emission model the phase offset is more difficult to understand. Planned observations of the time and frequency dependence of the phase offset, and the behaviour of subpulses in the third component present at and above 600 MHz, should prove most helpful in the further evaluation of these models.

# PULSAR 'DRIFTING'–SUBPULSE POLARIZATION: NO EVIDENCE FOR SYSTEMATIC POLARIZATION–ANGLE ROTATIONS

**with Ramachandran, Joanna Rankin, Ben Stappers and Marco Kouwenhoven**

Polarization-angle density displays are given for pulsars B0809+74 and B2303+30, which exhibit no evidence of the systematic polarization-angle rotation within individual subpulses previously reported for these two stars. The 'drifting' subpulses of both pulsars exhibit strikingly linear and circular polarization which appears to reflect the characteristics of two nearly orthogonally polarized emission 'modes' - along with the severe average-profile depolarization that is characteristic of their admixture at comparable overall intensities.

## 1. Introduction

In this chapter we provide straightforward, definitive evidence to the effect that pulsar 'drifting' subpulses exhibit polarization reflecting virtually only the projected magnetic field direction. We find that the linear position angles (hereafter PAs) of such subpulses are oriented—like almost all other pulsar radiation—either parallel to or perpendicular to this direction. That is, drifting-subpulse polarization closely follows the rotating-vector model (hereafter RVM) first articulated by Radhakrishnan & Cooke (1969) and by Komesaroff (1970).

Several influential papers, however, have suggested or reported precisely the opposite situation—that a characteristic rotation of the PA can be observed within some subpulses. In perhaps the first report of individual pulse polarization—the oscilloscope images reproduced by Clark & Smith (1969)—such an effect is clearly suggested, and other investigators writing within the first several years after the pulsar discovery (e.g., Lyne et al. 1971) emphasize the great variability of individual-pulse polarization in contrast to the usual great stability of the average or profile polarization, including the PA traverse. This latter compendium, along with Manchester's (1971) work, nonetheless demonstrated that most pulsars could at least partially be reconciled with the RVM, so that questions could be entertained regarding the manner in which the apparently disorderly subpulse polarization diverged from the more orderly average characteristics.

In this context, pulsar B0809+74 with its bright, regular, beautiful sequences of drifting subpulses has developed as the canonical example of extraordinary polarization behavior. Both Lyne et al. and Manchester determined that the pulsar's average linear polarization at meter wavelengths was low, with a PA characterized by two shallow, negatively rotating segments offset by about 90°. It was then interesting to assess whether its individual subpulses exhibited rotations over and above what might be expected from changes in the projected magnetic field direction, or namely, whether their behavior seemed compatible with the RVM model above.

It was difficult, anywhere, in 1971 to measure the polarization of individual pulses reliably, and in their now classic paper Taylor et al. used the NRAO 92-m telescope with a four-channel, 235-MHz polarimeter to record Stokes parameters I, Q and U of sequences from pulsar B0809+74. They noted that the pulsar's polarization properties "are closely linked with [its] bands of drifting subpulses" and that "successive bands of subpulses displayed nearly identical polarization behavior". Obvious also was that the individual subpulses were much more highly polarized than the pulsar's average profile. Such 'drift' bands in this pulsar are, of course, regularly spaced by very nearly 11 rotation periods, so Taylor et al. were able to average some 10 complete drifting cycles to improve their signal-to-noise (hereafter S/N) ratio. This original and inventive technique then permitted the authors to remark that "the position angle is moving with the subpulses and is not fixed on the rotating star." The paper's fig. 1 indicates that most subpulses exhibit a characteristic PA rotation irrespective of where they fall in longitude—that is, most subpulses have a similar leading-edge PA which then ro-tates in a consistent manner throughout its duration. Taylor et al.'s interpretation then appears virtually inescapable.

Unfortunately, this conclusion is incorrect as we will show below, but the authors of the above paper had no reasonable means of deducing this circumstance at the time. With 30-years hindsight we can now identify major clues to this three-decade-old deception: a) the total rotation within each subpulse (in their fig. 1) is usually close to 90°; and b) their resolution (between 3 and 4°) represents a signicant fraction of the subpulse width. Indeed, only slowly did evidence emerge to the effect that pulsar PAs tend to assume two mutually orthogonal orientations. Manchester et al. (1975) were the first to give histograms suggesting a bimodal PA distribution in certain stars, but it was not for a further five years that the generality of this phenomenon began to be appreciated fully (Backer & Rankin 1980). By this time, however, most interest in PSR B0809+74 and its drifting-subpulse polarization had subsided.[1] Only now are we beginning to apprehend fully the crucial role played by the orthogonal polarization modes within drifting-subpulse sequences.

## 2. Polarization-modal structure of conal single profiles

We can now see clearly that conal single ($S_d$) pulsar profiles are strongly affected by the interaction of the two orthogonal polarization modes. Of those stars known to have discernible *drifting* subpulses (e.g., see Rankin 1993) every single one exhibits low fractional linear polarization in the metre-wavelength region, and many show '90° flips' (accompanied by complete depolarization) at longitudes where the dominant mode switches—all of which results in segmented or disorderly PA behavior bearing almost no resemblance to any underlying geometrical (RVM) origin.

We can see this general behavior in all of these pulsars which have received adequate investigation: B0031−07 exhibits low linear polarization and PA 'flips' below 1 GHz (Gould & Lyne 1998; hereafter GL); in PSR B0809+74 we see a similar behavior (Manchester 1971; Lyne et al. 1971); PSR B0820+02 seems to show two 'flips' at 408 MHz (GL); PSR B1923+04 has a single flip (Hankins et al. 2001); and in PSR B2016+28, perhaps the best studied of all the $S_d$ pulsars, we see consistent modal dominance effects near the center of its profile (e.g., see GL). For 2016+28 the polarization histograms in Backer & Rankin (1980) and Stinebring et al. (1984)—see their figs. 18 and 31, respectively—leave no doubt about the characteristics of this overall modal behavior. Studies of pulsar B0943+10 give a detailed, completely consistent example of the situation wherein one mode is consistently stronger than the other (Suleymanova et al. 1998; Deshpande & Rankin 2001). We also see this situation clearly documented in PSR B2303+30 (Backer & Rankin 1980), and GL

---

[1] Sadly, the one other report of systematic PA rotation during subpulses in pulsar PSR B2303+30 (Gil 1992) is also incorrect; we now know that the Arecibo Observatory polarimetry of the early 1970s was flawed by the use of Gaussian-shaped IF filters. This had the effect of correlating polarized power at adjacent longitudes in an insidious manner.

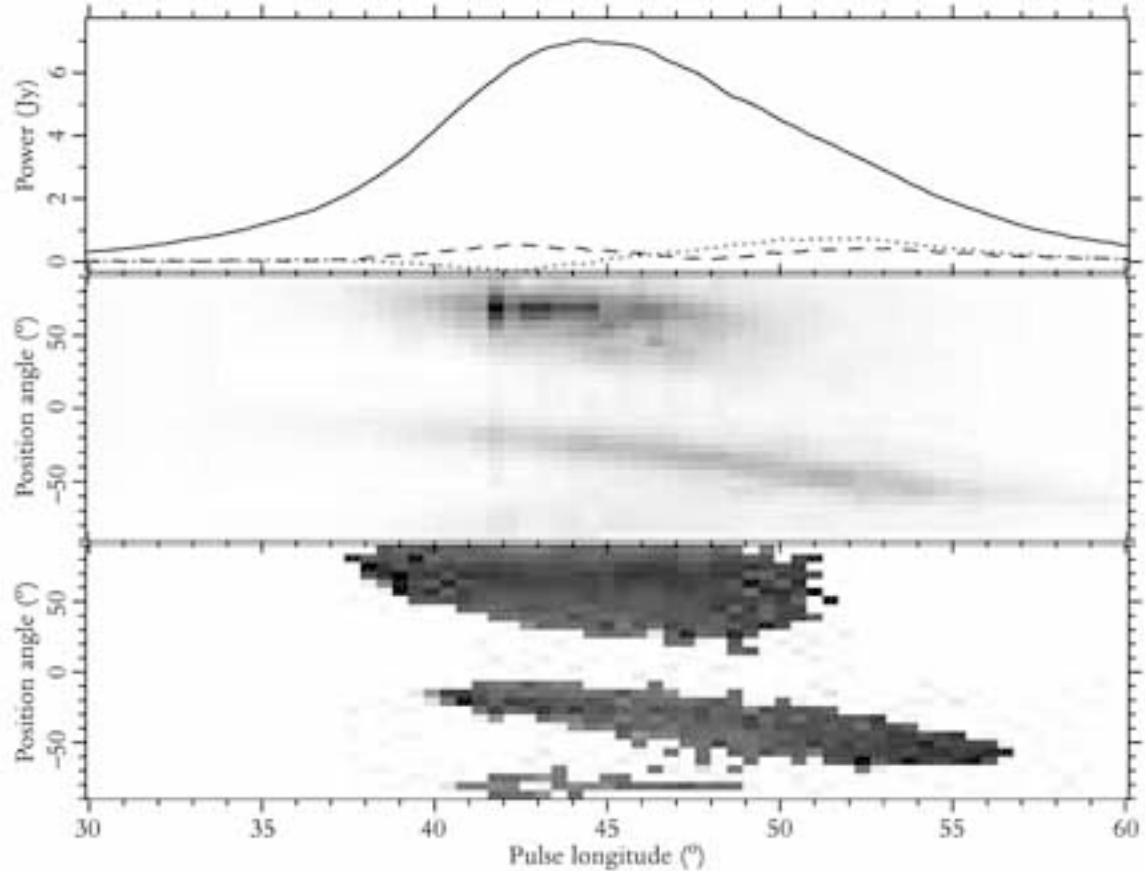

**Fig. 1.** Polarisation-angle-density diagram for PSR B0809+74 at 328 MHz. The top panel gives the usual average Stokes parameters: total power $I$ (solid line); average linear power $L\ (=\sqrt{Q^2 + U^2})$, corrected for the statistical bias (dashed curve); and average circularly polarized power $V$ (dotted curve). The usual box showing the resolution and off-pulse noise rms level has not been plotted because it would be almost invisible. The central panel displays (in grey scale) the polarization position-angle distribution, where the values in each pixel are weighted properly by the square of their S/N level (see text). Note the two distinct 'tracks' corresponding to the two orthogonal polarization modes (separated by about 90°). A sequence of length 3600 pulses was used. The bottom panel shows (again in gray scale) the fractional linear polarization distribution for those samples falling above a five standard-deviation noise threshold and weighted as in the central panel. The peak of the distribution is some 60%.

and Hankins et al.'s work on PSRs B1540+06, B1612+07, B1923+04, B1940−12, B2110+27, and B2303+30 reveals compatible behavior. Indeed, that PSR B1923+04 exhibits both behaviors (modal flips in some observations and not in others) suggests that modal power variations are probably common in $S_d$ stars, where they are often comparable in intensity—and slight changes in their relative amplitude would produce different average polarization effects.

## 3. Observations and analysis

Our 328-MHz observation of PSR B0809+74 was conducted with the Westerbork Synthesis Radio Telescope (WSRT), using its pulsar backend, PuMa. WSRT is an east-west array with fourteen equatorially mounted 25-m dishes. For this observation (which was made on the 26th November 2000), the delays between the dishes were compensated, and the signals were added in phase to construct an equivalent 94-m single dish having a sensitivity of about 1.2 K Jy⁻¹. We also calibrated the telescope array for polarisation measurements following the procedure given by Weiler (1973) and Weiler & Raymond (1976; 1977). With a bandwidth of 10 MHz, centred around 328 MHz, the signals were Nyquist-sampled, and Fourier transformed to synthesize a filterbank of 64 complex frequency channels. Stokes parameters were computed on-line in this frequency domain. Finally, after some averaging, 2-bit (4-level) data samples were recorded for all Stokes parameters and frequency channels, with a time resolution of 819.2 $\mu$sec. During the offline analysis, we removed the interstellar-dispersion delay between the signals of various frequency channels, producing multi-bit, floating-point sample values. For this purpose, we used a dispersion measure (DM) of 5.7513±0.0002 pc cm⁻³ as determined by Popov et al. (1987).

The 1380-MHz observations were also conducted with the WSRT with its pulsar backend, PuMa. All the Stokes parameters were recorded as mentioned above, with a bandwidth of 80 MHz and 512 frequency channels.

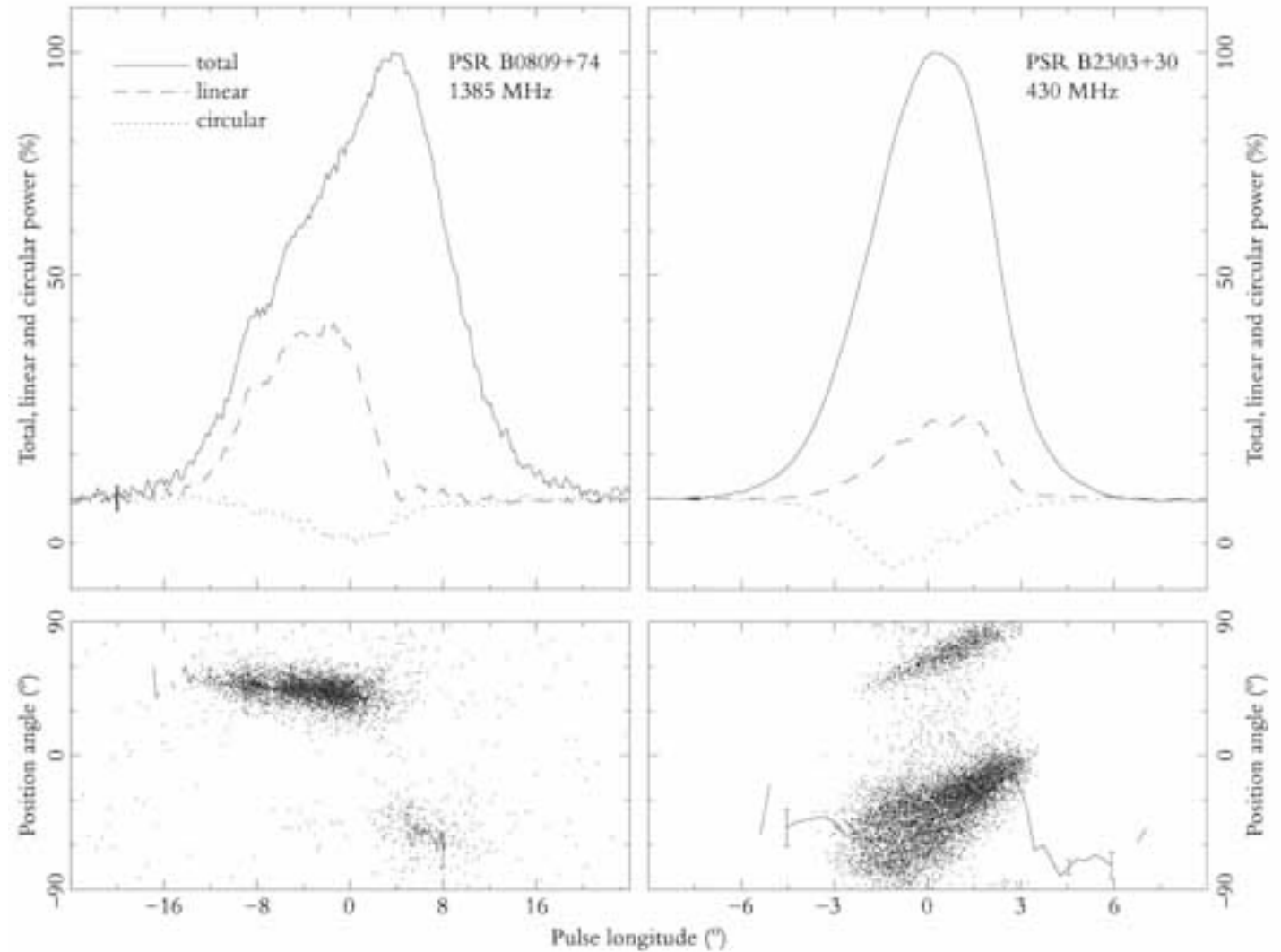

**Fig. 2.** Polarisation-angle histograms. **left)** PSR B0809+74 at 1365 MHz. The top panel gives the usual average Stokes parameters: total power $I$ (solid line); linear power $L$ $(= \sqrt{Q^2 + U^2})$ (dashed curve); and circularly polarized power $V$ (dotted curve). A box showing the resolution and off-pulse noise rms level is just visible at the left of the diagram. The lower panel represents the histogram of sample position angles with estimated errors less than $30°$ (see text), together with the superposed average PA traverse. Note the two distict tracks corresponding to the two orthogonal polarization modes (separated by about $90°$). A 2708-pulse sequence was used. **right)** PSR B2303+30 at 430 MHz. Again, note the two well separated modal PA tracks. The display used 2370 pulses and a polarization-angle threshold of $8°$.

The 430-MHz observation of PSR B2303+30 was recorded at Arecibo on 15 October 1992, using the 40-MHz correlator, which dumped the ACF/CCF's of the right- and left-handed channel voltages at 1206 $\mu$sec intervals. Using 10-MHz bandwidth and retaining 32 lags, dispersion delays were reduced to negligible values. The time resolution was then about $0.°276$ longitude (one rotation period equals $360°$ longitude). After 3-level correction, the ACF/CCFs were calibrated and Fourier transformed to produce Stokes parameters, which were then corrected for dispersion delays, Faraday rotation, and feed effects (see Suleymanova et al. 1998).

Figures 1 & 2 (at 328 and 1385 MHz, respectively) represent the first available polarization PA histograms for pulsar B0809+74. The top and top-left panels in these two figures show the usual average polarization properties: total power (Stokes parameter $I$; solid curve), linear power, corrected for the statistical bias (dashed curve), and circular power ($V$; dotted curve). The middle panel of Fig. 1 and the bottom-left

panel of Fig. 2 then give the PA distribution information: in the former as a grey scale with the values contributing to each pixel carefully weighted in order to maximize their significance; and in the latter as a display of sample values comprising a density plot. In the first figure, the values contributing to a pixel were weighted according the square of their S/N values—after taking care to estimate the PA error according to the procedure given in Rankin & Rathnasree (1997; see footnote 11); whereas in the latter the PA values are given as dots corresponding to all data samples for which the error in the PA, $\sigma_\chi$ ($= \sigma_{on}/L$) was less than $30°$. The on-pulse noise level $\sigma_{on}$ was estimated as $\sigma_{off}(I_{sys} + L)/I_{sys}$, where $I_{sys}$ is the total power corresponding to the system temperature and $\sigma_{off}$ is the standard deviation of the noise in Stokes parameter $I$ well away from the pulse window. (Clearly, $\sigma_{off}/I_{sys} = (2\tau\Delta\nu)^{-\frac{1}{2}}$, where $\Delta\nu$ is the total bandwidth and $\tau$ the effective integration time.). Finally, the bottom panel of Fig. 1 gives the density of the fractional linear polarization, also as a grey scale—where a fully black pixel represents a value of about 60%. Here, in addition to S/N weighting as above, a threshold of $5\sigma_{on}$ was imposed to improve the quality of the display. This detailed analysis and display was possible for the 328-MHz sequence, because the average S/N was relatively high, but was not practical for the other two sequences.

For both pulsars in Fig. 2 the two parallel tracks—corresponding to the two orthogonal modes, separated by about $90°$—can be very clearly seen. For PSR B0809+74 at the lower frequency, the two modes have comparable power throughout most of the pulse, resulting in severe depolarization; whereas at the higher frequency we see little secondary-mode (herafter SPM) emission during the first half of the pulse, so that most of the depolarization occurs after the peak of the average profile. We can also see here that the PA-traverse rate associated with this SPM emission is $-2°/°$ or steeper, a fact that cannot readily be determined from the usual average polarimetry (e.g., the otherwise very well measured profiles recently of GL) as well as one having considerable importance for future efforts to understand the configuration of this pulsar's rotating subbeam system. These PA tracks can now be assessed and interpreted according to the RVM of Radhakrishnan & Cooke (1969)— and the initial evidence here is that they are nearly, but not precisely, orthogonal.

We further note that while the PA histograms clearly exhibit the modal nature of the polarized power in pulsars B0809+74 and B2303+30, these distributions remain remarkably complex. The widths of the primary polarization-mode (hereafter PPM) and SPM distributions is very different. When the two modal contributions have about equal power, some of the radiation will be randomly polarised (as can be clearly seen in the figures); whereas, when one dominates the other, its PA determines the ensemble PA. Statistical theory discussions of this phenomenon are rather complex (see Davenport & Root 1958; Moran 1976; and especially McKinnon & Stinebring 1998).

Finally, the color polarization display in Figure 3 provides a completely different way of looking at the polarization characteristics of the subpulses. The four columns give the total power, fractional linear polarization (corrected for statistical bias), polarization angle, and fractional circular polarization of the first 200 pulses of the 328-MHz sequence of Fig. 1—all color-coded according to the respective color scales at the far left of the display. Immediately, we see the drift bands—with the individual subpulses drifting from left to right (the pulses are numbered from the bottom to top), the three nulls, and the fairly consistent, about 40% level of fractional linear polarization. More arresting, however, are the associated polarization angle and circular-polarization behavior, where we find two strongly preferred values for both the PAs—nearly orthogonal values coded chartreuse and magenta, respectively, and respective 40% negative and positive fractional circular values, coded purple and orange.

Careful inspection of the pulse-sequence PA behavior in the figure shows that there are very frequent $90°$ flips —usually from about $-20°$ (magenta) to $+70°$ (chartreuse)—but no good example of systematic rotation. Of course, we see variations which may be partly modal and partly the effect of the noise, but were the pulsar subpulses are strongest (e.g., between number 18 and 27) the modal flips are clearly seen in every pulse. We do see the longitude of these flips occuring at progressively earlier phases, so as to remain approximately parallel to the drift-band and thus within the subpulses as they drift—and it was a smeared out signature of this phenomenon which Taylor et al. recorded 30 years ago.

The full story is, however, much more complex: there is a progressive mixing of the two polarization modes as the subpulses drift across the profile window—as subpulses at the extreme edges tend to exhibit only one of the modes. In this context, the constancy of the fractional linear polarization is striking; indeed, one gets the impression that most of individual samples are not modally depolarized—only the contrasting (dark blue) ones which usually lie close to the PA flips. Also intriguing is the circular, which is highly correlated with the modal linear power, while exhibiting a slightly displaced distribution along the drift bands. We will return to these questions in much more detail in subsequent papers on PSR B0809+74.

## 4. Discussion

Polarisation position-angle displays for PSR B0809+74 and PSR B2303+30, computed from well measured sequences and given above, show clearly that the PAs trace out two well defined, nearly parallel trajectories (separated by approximately $90°$) as a function of longitude. As such, they individually trace curves which are completely compatible (within their errors) with the Radhakrishnan & Cooke (1969) RVM, and thus entirely exclude the large ($\sim 90°$), extraordinary, subpulse-related PA rotations previously reported for these two stars. If such systematic rotations occurred, the character of the distributions would have to be very different—and the Taylor et al. result—about $90°$ PA rotation across subpulses occurring over a subpulse-width-order interval of longitude *is* completely explained by the the prominent, stongly longitude-segregated regions of modal power, together with their subpulse-width-order instru-mental resolution. Of course, we can say nothing about any small extraordinary PA rotation within the respective PPM and SPM tracks—but this question is well beyond the limited scope of our paper.

**Fig. 3.** Color display of a 200-pulse portion of the 328-MHz observation in Fig. 1. The first column of the displays gives the total intensity (Stokes parameter $I$), with the vertical axis representing the pulse number and the horizontal axis pulse longitude, colour-coded according to the left-hand scale of the top bar to the left of the displays. The second and third columns give the corresponding fractional linear polarisation ($L/I = \sqrt{Q^2 + U^2}/I$) and its angle ($\chi = \frac{1}{2}\tan^{-1} U/Q$), according to the scales at the top-right and bottom-left of the left-most panel. The last column gives the fractional circular polarisation ($V/I$), according to the scale at the bottom-right of the left-hand panel. Plotted values have met a threshold corresponding to 2.5 standard deviations of the off-pulse noise level. Note the strikingly modal character of the polarized power, with the linear assuming nearly orthogonal angles—color-coded as either chartreuse or magenta—and with the corresponding circular falling at about 40% negative (purple) and positive (orange), respectively. It is also notable that the PAs exhibit (apart from noise variations) virtually only these angle values, so that with a single pulse or subpulse, we see not primarily rotation, but modal 90° flips. Can it be any wonder that polarization behavior of such remarkable complexity caused so much early confusion?

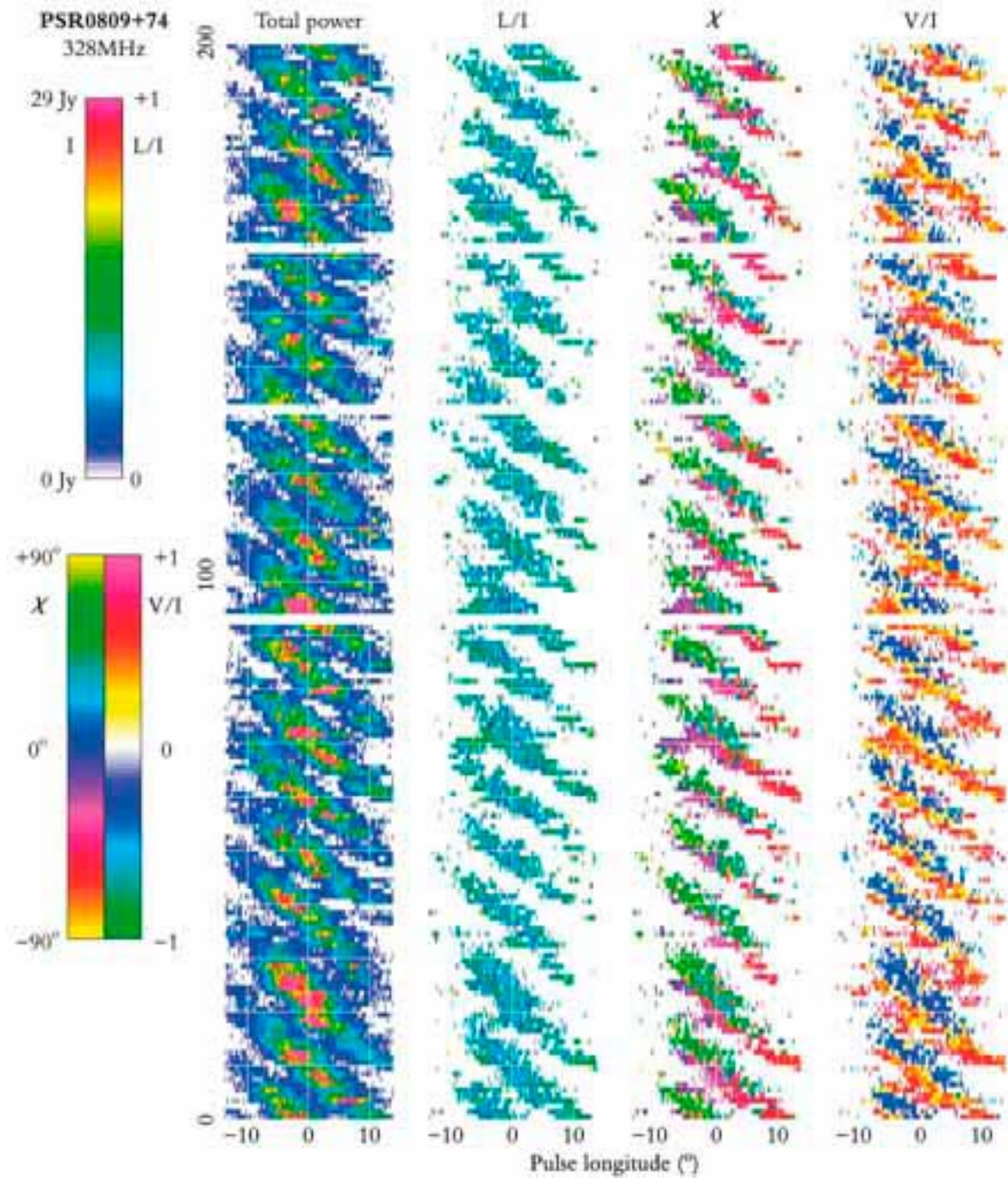

Nonetheless, we also see here that the character of the modal emission is exceedingly complex, entailing at least four factors: a) its overall angular and/or temporal distribution—probably following from its generation within a rotating subbeam system, b) our viewing geometry, which determines the manner in which the profile window weights the various contributions, and c) the level of instrumental noise. The PA distributions for PSR B0809+74's two modes at 328 MHz, for instance, are quite different; the PPM's PA width is much broader than the SPM and their separation may not be just $90°$. For all these reasons the depolarization in this pulsar is very complex, such that the individual pulse polarization—which is typically 40% linear *and* circular—is reduced in their aggregation to hardly 5 and 10%, respectively.

Consequently, it should perhaps neither be surprising that PSR B0809+74's subpulses appeared to exhibit the extraordinary PA rotation on the basis of early observational and analytical methods, nor that it has taken fully 30 years to discern that this interpretation was incorrect. Thus, we now hope and trust that this truer, more universal characterization of the pulsar's polarized emission will facilitate broader and more comprehensive physical explanations of its overall nature.

# |V| : NEW INSIGHT INTO THE CIRCULAR POLARIZATION OF RADIO PULSARS

**with Aris Karastergiou, Simon Johnston, Dipanjan Mitra and Russell Edwards**

We present a study of single pulses from nine bright northern pulsars to investigate the behaviour of circular polarization, V. The observations were conducted with the Effelsberg 100-m radio telescope at 1.41 and 4.85 GHz and the Westerbork radio telescope at 352 MHz. For the first time, we present the average profile of the absolute circular polarization |V| in the single pulses. We demonstrate that the average profile of |V| is the distinguishing feature between pulse components that exhibit low V in the single pulses and components that exhibit high V of either handedness, despite both cases resulting in a low mean. We also show that the |V| average profile remains virtually constant with frequency, which is not generally the case for V, leading us to the conclusion that |V| is a key quantity in the pulsar emission problem.

## 1. Circular polarisation in pulsars

Radio emission from pulsars is highly polarised. Occasionally the linear polarisation $L$ is extraordinarily high, sometimes close to 100% of the total radio emission, and the circular polarisation $V$ is amongst the highest observed in natural sources of electromagnetic radiation. Typically, however, it is only a small fraction of the total pulsar emission, although in some cases it can be much larger (i.e. ∼60% in PSR B1702−03 in Radhakrishnan & Rankin 1990). The polarisation of pulsar emission has been attributed to the emission mechanism itself and to propagation effects in the pulsar magnetosphere (e.g. Melrose 2000). Simultaneous, multi-frequency observations in full polarisation have shown irregularities in the polarisation of single pulses at different frequencies, which have been considered direct observational evidence of propagation effects (Karastergiou et al. 2001, 2002; Karastergiou, Johnston & Kramer 2003).

Average pulse profiles of radio pulsars are known to stabilise after the integration of a sufficient number of pulses. This also holds for the polarisation profile, although the stabilising time-scales for polarisation are longer than for total power (Rathnasree & Rankin 1995). Inspecting such average profiles, Rankin (1983a, b, 1986) claimed that there is a difference in the circular polarisation in central or *core* components from that in outer, *cone* components, thought to originate in cones of emission around the magnetic axis. More specifically, core components exhibit a higher degree of $V$ in the integrated profile than cone components, often characterised by an "S"-shaped swing from one to the other handedness (Clark & Smith 1969). On the other hand, cone components are weakly circularly polarised in comparison to their core counterparts. In fact, it was such differences, combined with a difference in the spectral behaviour of core and cone components, that led Rankin to suggest that there is a fundamental difference in the emission mechanism between core and cone components.

Despite the efforts, it remains unclear whether $V$ is intrinsic to the emission mechanism, propagation generated, or even both. Cordes, Rankin & Backer (1978) discovered that in PSR B2020+28, the handedness of circular polarisation was associated with the polarisation position angle of the linear polarisation. In individual pulses from this pulsar, where the position angle occasionally jumps abruptly by 90° at certain pulse longitudes, these jumps where accompanied by a sense reversal in $V$. The idea that $V$ is tied to the orthogonal polarisation mode (OPM) phenomenon in pulsars was taken further by Stinebring et al. (1984a, b) and more recently McKinnon & Stinebring (2000) and McKinnon (2002). An investigation of the same issue in PSR B1133+16 using simultaneously observed single pulses at 1.41 GHz and 4.85 GHz, demonstrated that the $V$ – OPM association can be seen at both these frequencies, it is however stronger at the lowest of the two (Karastergiou et al. 2003).

In many cases, the longitude-resolved distributions of $V$ in the single pulses are much broader than the instrumental noise, despite having a mean very close to zero. It can be easily inferred that such broad $V$ distributions with an almost zero mean, denote a number of highly circularly polarised single pulses. Integrated profiles of pulsars with a very low mean $V$, however, obscure the information that single pulses may be individually highly circularly polarised.

In this letter we present high quality single-pulse data from 9 pulsars, observed at 3 widely spaced frequencies. We investigate the behaviour of $V$, by defining $|V|$ in the single pulses and looking at differences between $V$ and $|V|$ in components that have been classified in the literature as of core and cone origin. We also take advantage of the three observed frequencies and discuss the frequency evolution of $|V|$ as compared to $V$.

## 2. The data

### The observations

The data presented here were obtained with the 100-m radio telescope in Effelsberg and the Westerbork Synthesis Radio Telescope. They consist of single pulse sequences from nine bright northern sky pulsars. The Effelsberg observations were made at 4.85 GHz and 1.41 GHz between July and September 2002. Details of the receivers and the calibration procedures of the multiplying polarimeters used can be found in von Hoensbroech & Xilouris (1997). In short, both systems have an equivalent system flux density of ≈20 Jy. For the 4.85 GHz observations a bandwidth of 500 MHz was used, imposing restrictions on the resolution due to dispersion smearing. At 1.41 GHz, where dispersion smearing is significantly greater, bandwidths of 10, 20 and 40 MHz were used depending on the dispersion measure of the pulsar.

The 352 MHz observations, 30 minutes per pulsar, were conducted with the WSRT on March 8th and 9th, 2003. The WSRT consists of fourteen 25-m dishes in an east-west array, which can be combined to form a 94-m single dish equivalent. The equivalent flux density of the system was ≈120 Jy. We applied the schemes described by Weiler (1973) and Edwards et al. (2003) to calibrate the data. Bandwidths of 10 MHz were processed by the backend system PuMa (Voûte et al. 2002), acting as a digital filter-bank. The data were 4-bit sampled at between 0.2048 and 0.8192 ms, for between 32 and 1024 channels, depending on the pulsar period and dispersion measure, and de-dispersed offline.

### Absolute circular polarisation

We define the quantity $|V|$ in the single pulses as the absolute value of $V$ in every phase bin of every individual pulse (Karastergiou et al. 2001). By taking the absolute value of the $V$ data stream, we are also altering the statistics of the instrumental noise, which is no longer normally distributed and has a non-zero mean. We subtract this positive mean from every single pulse and integrate $|V|$ in all the pulses to produce an average $|V|$ profile. Subtracting the aforementioned offset from the single pulses has the result of an $|V|$ profile which can be closer to zero than the profile of $V$. This happens in cases of low $|V|$ in the single pulses, where subtracting the offset has the effect of removing part of the signal. Full details of the data analysis and the statistics of $|V|$ are outlined in Karastergiou et al. (2004).

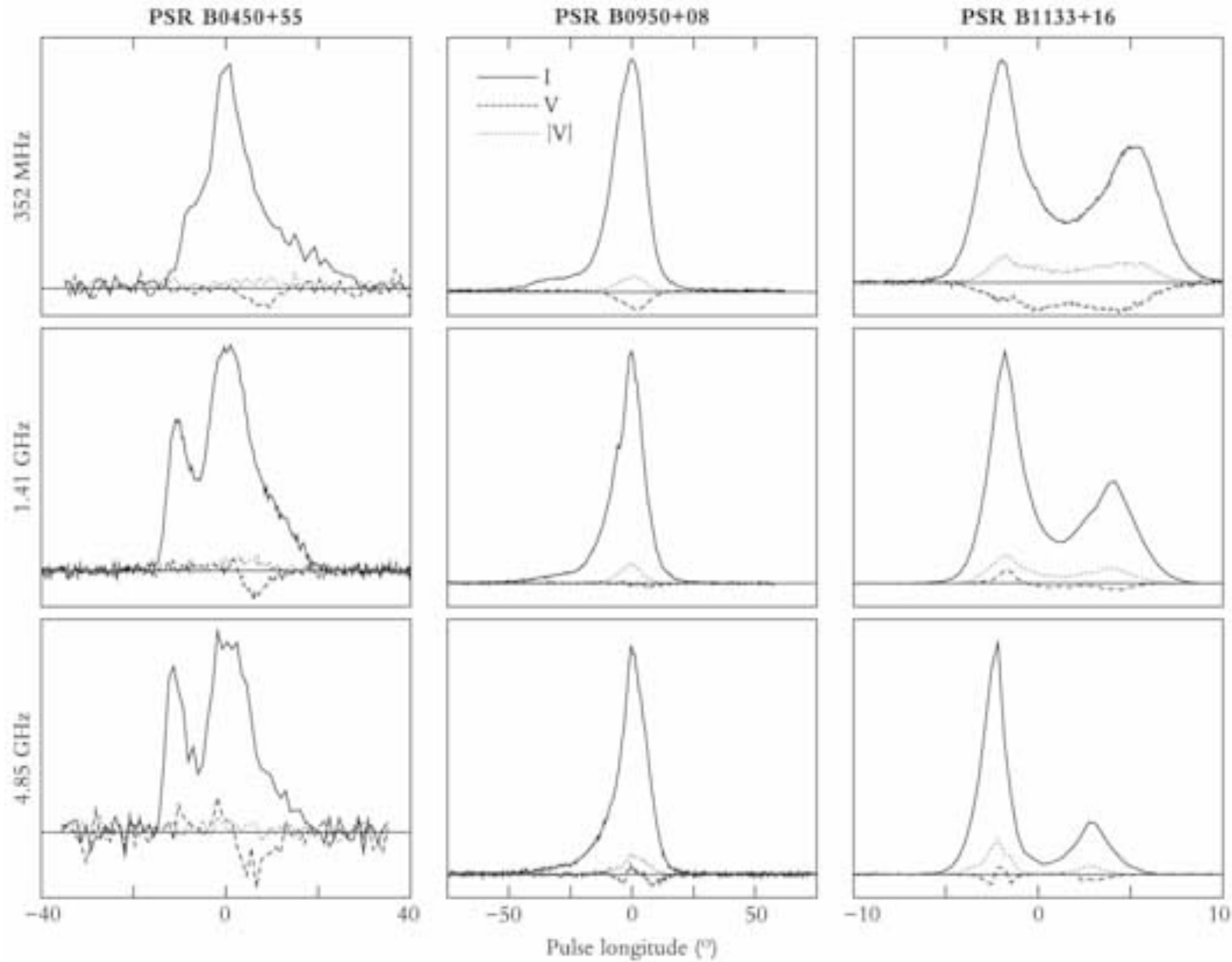

**Fig. 1.** Profiles of PSRs B0450+55, B0950+08 and B1133+16 observed at three frequencies. The solid line represents the total power, the dashed line represents $V$ and the dotted line $|V|$. For clarity, the linear polarisation is not plotted.

## 3. Results

The nine pulsars that we observed display a variety of features in both their total power profiles and their polarimetric properties. Our single-pulse observations at three, widely-spaced frequencies give us the opportunity to trace the properties of $V$ in the single pulses as a function of frequency. Figs. 1, 2 and 3 show the integrated pulse profiles of the observed pulsars, together with the profiles of $V$ and $|V|$. Positive and negative values of $V$ denote left and right handed circular polarisation respectively. The alignment of the profiles at the three frequencies observed is only approximate.

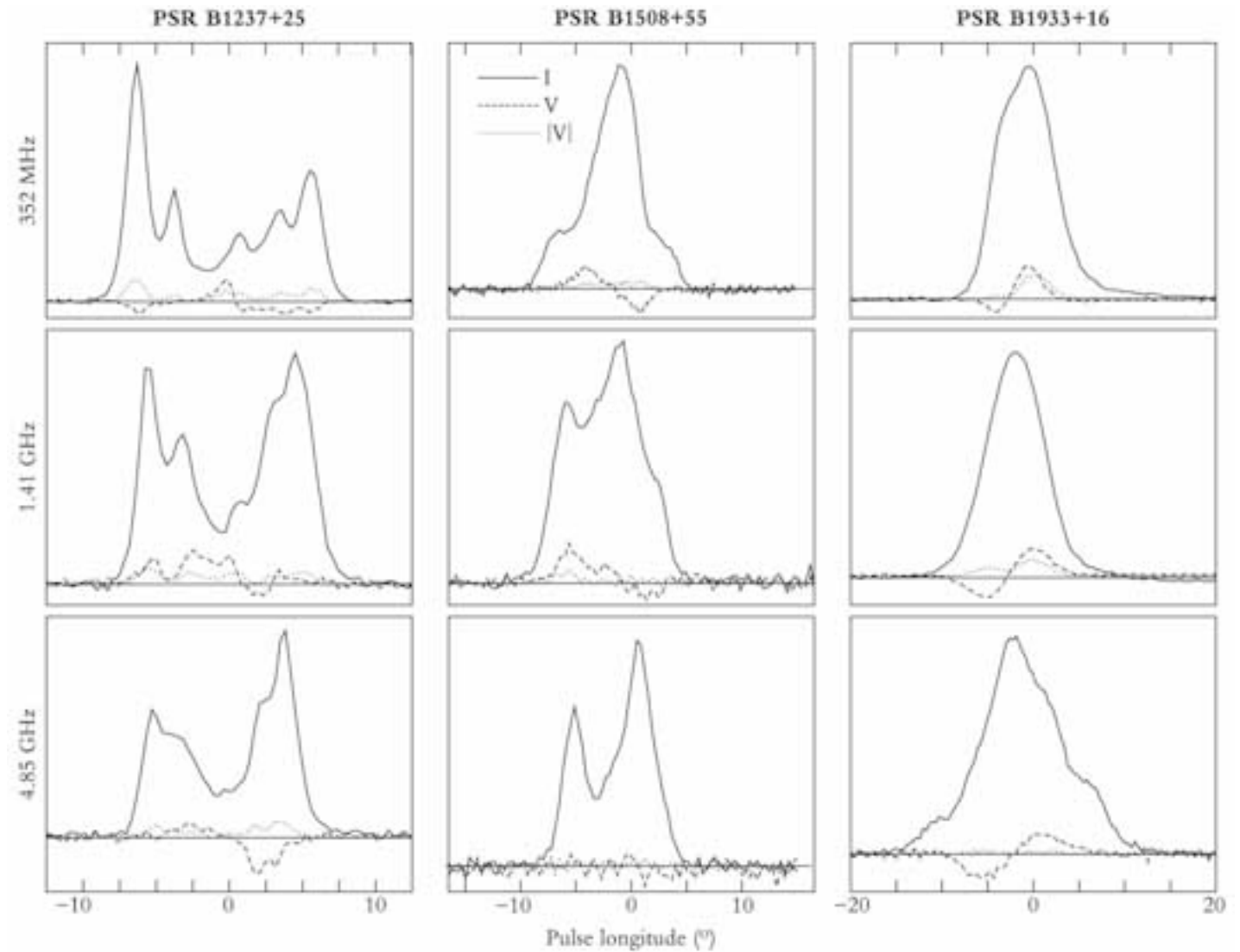

**Fig. 2.** Same as Fig. 1 for PSRs B1237+25, B1508+55 and B1933+16.

*Circular polarisation in core components*

PSRs B0450+55, B1237+25, B1508+55, B1933+16 and B2111+46 (in Figs. 1, 2 and 3) all have identifiable core components, demonstrating steep position angle swings (Lyne & Manchester 1988) and, in most cases, a swing in the *V* profile from one to the other handedness located at the same pulse phase. This can be seen most clearly in PSRs B1933+16 and B2111+46. In both these pulsars, the swing in the *V* profile is evident at all three frequencies. The profile of $|V|$ shows a minimum which coincides exactly with *V* changing handedness. This is evidence that *V* in the single pulses not only has a mean of zero, but is also tightly distributed around its mean, because a broad distribution would include pulses of high S/N of either handedness which would inevitably increase $|V|$. This indicates that the $V = 0$ phase remains fixed in the single pulses. We interpret this as direct evidence that the *V* swing is a constant feature of the single pulses.

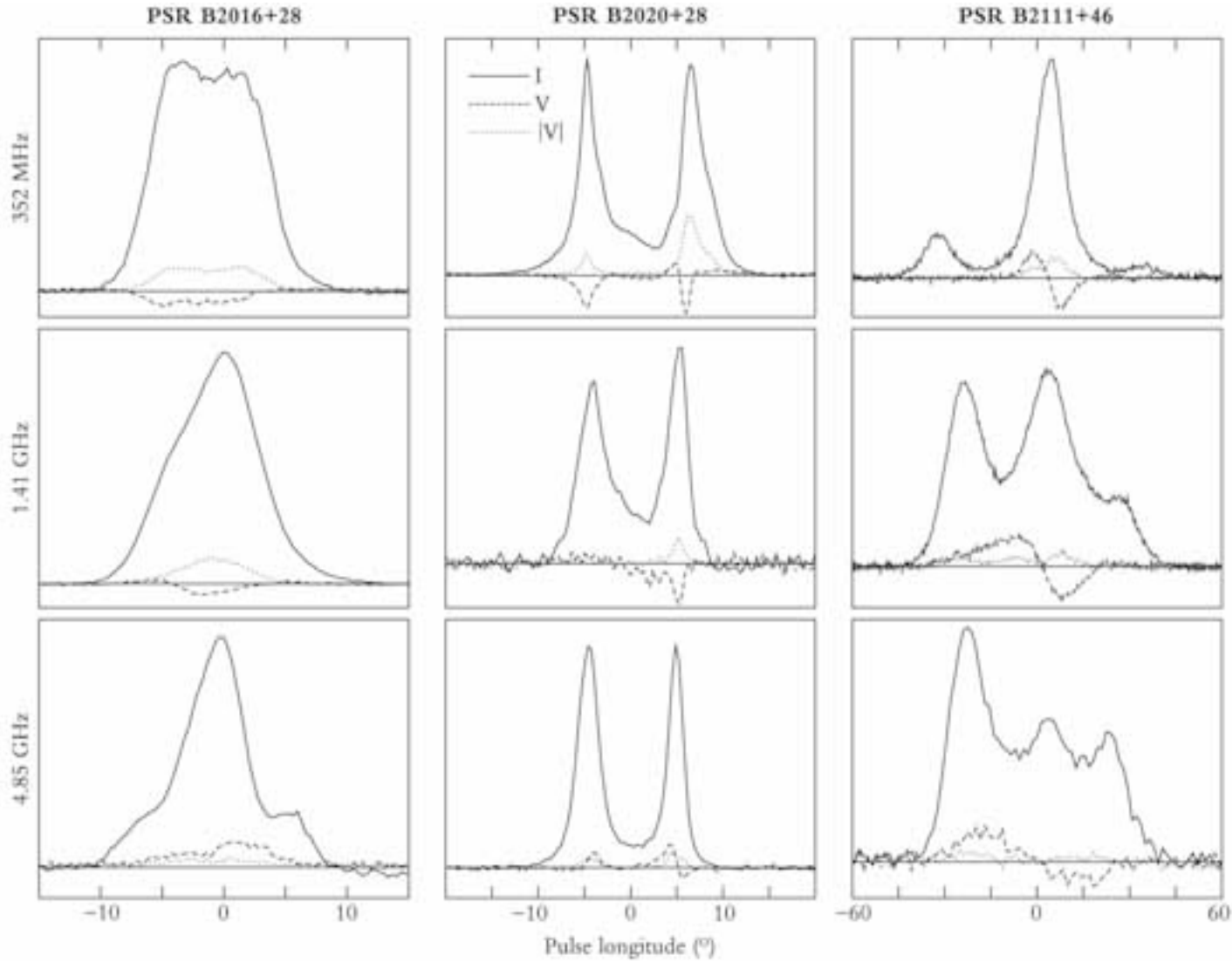

**Fig. 3.** Same as Fig. 1 for PSRs B2016+28, B2020+28 and B2111+46.

We also notice that the swing in $V$ in the core component of both these pulsars sweeps across a wider pulse phase range with frequency. Apart from the two afore-mentioned pulsars, the $V$ and $|V|$ signature described here can also be seen in all but the highest frequency in PSR B1508+55.

*Circular polarisation in cone components*

PSRs B0950+08 and B1133+16 in Fig. 1 both consist of cone components according to Lyne and Manchester (1988). The fact that the $V$ profiles at 1.41 and 4.85 GHz are very weak is in accordance with the statement that $V$ is generally low in cone components (e.g. Radhakrishnan & Rankin 1990). However, as can be seen in Fig. 1, $|V|$ is quite high in these components at these frequencies, making it comparable to $V$ in the core components of other pulsars in our sample. The mechanism, therefore, that is responsible for circular polarisation in the single pulses appears to be almost equally efficient between core and cone components.

*Frequency dependence of $V$ and $|V|$*

Apart from the observation that the cone components in PSRs B0950+08 and B1133+16 show high $|V|$, our data permit an investigation of $V$ and $|V|$ as a function of frequency. At the lowest frequency, $V$ is negative in both pulsars and the profile of $|V|$ is approximately a mirror image of that of $V$. At 1.41 GHz, $V$ is significantly less than at 352 MHz. The profile of $|V|$ however is similar to that at 352 MHz. This can be caused by $V$ in the single pulses being symmetrically distributed around zero, albeit with cases of large deviation from the mean. The fraction $|V|/I$ in the single pulses is approximately the same between these two frequencies and it is the symmetrical distribution of $V$ around zero in the single pulses, that causes the $V$ profile to be almost zero. Moving further up in frequency, the 4.85 GHz profile of $V$ has a more complex structure, with a number of phases of transition between left and right handedness. Despite this, the profile of $|V|$ remains the same even at this frequency for both pulsars. The constance of the $|V|$ profile with frequency regardless of the changes in $V$ constitutes evidence that $|V|$, the flux density of the circularly polarised component of the radiation, is a fundamental quantity of the emission process.

PSR B2016+28 in Fig. 3 shows the $|V|$ profile to be wider than the $V$ profile at the two lowest frequencies. Especially at 1.41 GHz, at either side of the $V$ component the $|V|$ component drops to zero slower. In the leading edge of the component, it appears as thought $V$ is slightly left handed, but despite the change in handedness, the $|V|$ profile is approximately the same fraction of $I$ as at 352 MHz. This change of handedness appears to be part of a smooth change with frequency, which is responsible for the 4.85 GHz profile being on average slightly left handed across the pulse. It seems, therefore, that this pulsar resembles PSRs B0950+08 and B1133+16. In fact, Figs. 1, 2 and 3 show that $|V|$ is much less dependent on frequency than $V$ itself. The 4.85 GHz data may be slightly harder to interpret, because of a lack of S/N in the single pulses, but there is no pulsar that behaves contradictory to the above statement at 4.85 GHz.

*The orthogonal polarisation mode phenomenon and $|V|$*

Both PSR B0950+08 and B1133+16 are pulsars that show OPM jumps in the integrated polarisation profiles. The OPM phenomenon has been shown to be more prominent at higher frequencies (Stinebring et al. 1984a, b, Karastergiou et al. 2002) and it is our opinion that the $V$ profile changes seen in these two and other pulsars from our sample, are OPM related. A full discussion on this is beyond the scope of this letter, it is however part of the more detailed approach to the behaviour of $|V|$ in

Karastergiou et al. (2004). The orthogonal modes have been shown to be associated to a particular handedness of circular polarisation (Cordes et al. 1978 Stinebring et al. 1984 a, b), however it has been shown recently that this association becomes weaker at higher frequencies for PSR B1133+16 (Karastergiou et al. 2003). This results in a randomisation of the handedness of the circular polarisation which effectively reduces the mean $V$, but leads to a relatively high $|V|$. The relevance of the OPM phenomenon can be also seen in PSR B2020+28 in Fig. 3, especially at 352 MHz. In the trailing component of the pulse, $V$ sharply turns from positive to negative and back to being slightly positive. These sudden swings, which do not seem to affect the wider component of the $|V|$ profile occur together with OPM jumps in the PA swing. It therefore seems that also in this case the OPM phenomenon has the effect of reducing the mean $V$ while $|V|$ remains largely unaffected. This is also a good example of a $V$ swing without a $|V|$ minimum at the cross-over point, as is the case in the core components. Obviously, in PSR B2020+28, at the pulse phase of the change in the sign of the mean $V$, the single pulses will be somewhat circularly polarised, with cases of either handedness and a mean of zero.

## 4. Conclusions

We have demonstrated that accompanying the integrated profile of $V$ with the profile of $|V|$ provides important information on the behaviour of circular polarisation in pulsar radio emission. Making use of this simple quantity, which requires single pulse observations, has immediately revealed new properties in a group of otherwise well studied bright stars. We have shown that:

1. In the pulsars that exhibit a sign change signature in the core components, a minimum in the $|V|$ profile is seen to coincide with the point $V$ changes handedness. We consider this direct evidence that the phase where the mean $V$ is 0, must have $V \approx 0$ also in the single pulses and, subsequently, that the swing signature is also constant in the single pulses. Multiple frequency observations demonstrate that this fact is true at all observed frequencies.
2. In a number of cases, the profile of $|V|$ resembles the profile of $I$, regardless of the complicated structure seen in $V$. This is also seen to be frequency independent. Complicated $V$ profiles are seen in pulsars where the OPM phenomenon is observed, and, as such, appear predominantly towards the higher frequencies.
3. In pulsars with cone components, we have shown that the very weak $V$ profiles that may be due to the OPM phenomenon, show comparatively high $|V|$. This demonstrates that the single pulses may occasionally be highly circularly polarised, but average out to zero.

Points (i) and (iii) indicate that the difference in terms of $V$ between core and cone components is related to how constant the $V$ features are in the single pulses. This is also in accordance with the observation that core components are more stationary in phase with respect to their cone counterparts that often tend to jitter. Such jittering of a component which is partially left- and partially right-hand circularly polarised will have the effect of reducing $V$, while $|V|$ should survive. All three points mentioned above warrant further investigation using $|V|$ to determine more accurately the nature of $V$ in pulsars. They convincingly suggest once again, that in the effort for progress in solving the pulsar emission mechanism problem, single pulse observations carry vital information that should not be underestimated and lost due to integration.

# MAGNETIC FIELD DECAY VERSUS PERIOD–DEPENDENT BEAMING

with Frank Verbunt

Several recent papers conclude that radio-pulsar magnetic fields decay on a time-scale of 10 Myr, apparently contradicting earlier results. We have implemented the methods of these papers in our code and show that this preference for rapid field decay is caused by the assumption that the beaming fraction does not depend on the period. When we do include this dependence, we find that the observed pulsar properties are reproduced best when the modelled field does not decay.

When we model the magnetic fields of new-born neutron stars by a single distribution sufficiently wide to explain magnetars, the magnetic field and period distributions we predict for radio pulsars are wider than observed. A bimodal distribution does explain the observed period and magnetic field distributions, independent of the use of a critical field line.

## 1. Introduction

Already before the first pulsars were discovered, the theoretical prediction was made that a neutron star with a strong magnetic field would lose rotation energy by the emission of electro-magnetic radiation, and thus that its rotation period would increase (Pacini 1967). The subsequently observed increase in the periods of all four first detected pulsars supported the idea that pulsars are neutron stars with strong ($\gtrsim 10^{12}$ G) dipolar magnetic fields. Neutron stars in high-mass X-ray binaries also showed evidence of such high fields (e.g. Trümper et al. 1978).

Soon after, several lines of evidence appeared to indicate that the strength of the magnetic field disappears on a time scale of several million years: the lack of old, long-period pulsars and the weaker magnetic fields seen in pulsars with higher characteristic ages $\tau_c \equiv P/(2\dot{P})$ (Gunn & Ostriker 1970); pulsars apparently stopped emitting before they reached a large distance from the galactic plane (Taylor & Manchester 1977); the accreting neutron stars in low-mass X-ray binaries, which are generally old, show no signs of strong magnetic fields. Subsequently, however, all of these lines of evidence were shown to be misleading. The absence of long-period pulsars can be explained by the decrease of luminosity and beaming fraction (Lyne & Manchester 1988; Rankin 1993) with period. The large range of period derivatives $\dot{P}$ compared to a relatively narrow range of periods P fully explains the anti-correlation of derived field strength B $\propto \sqrt{P\dot{P}}$ with characteristic age $\tau_c$ (Lyne et al. 1975). The apparent maximum distance of pulsars to the galactic plane is an artifact of the distance determination from the dispersion measure caused by an electron layer with a finite scale height (Bhattacharya & Verbunt 1991). Several neutron stars with strong magnetic fields were discovered to reside in old binaries, in particular Her X−1 (Lamb 1981; Verbunt et al. 1990).

The main remaining argument was the careful population study of pulsars by Narayan & Ostriker (1990), which found that a single population of pulsars whose field does not decay cannot explain the observed distributions of periods and period-derivatives. A re-investigation of this population synthesis by Bhattacharya et al. (1992), however, contradicted this result. Various other arguments led to a growing perception that the magnetic field of neutron stars does not decay spontaneously during the observed life time of radio pulsars: the pulsars with the highest characteristic ages move towards the plane, indicating their field is retained during the $\gtrsim 70$ Myr it took the galactic gravitational field to reverse their initial velocity away from the galactic plane, where they were born (Lorimer et al. 1997; Hartman et al. 1997). The birth rate of neutron stars derived from the observed number of pulsars is uncomfortably high if they can be observed for a few million years only (Blaauw 1985); a longer life time solves this (Hartman et al. 1997). The birth rate problem is much exacerbated if the few pulsars with very low radio luminosity (such as PSR J0108−1431, Tauris et al. 1994) represent a numerous, hitherto undetected population. The angular momentum of pulsars with respect to the galactic center shows that pulsars with high characteristic age are also old dynamically (Hansen & Phinney 1997). The field strengths of several accreting neutron stars have been estimated from their cy-clotron resonance scattering features, and the results lie in a narrow range between $(1-4)\times 10^{12}$ G for both young systems and systems as old as $10^8$ years (Makishima et al. 1999).

Thus, observations both of radio pulsars and of neutron stars in X-ray binaries appeared compatible with the idea that the magnetic field of a neutron star does not decay spontaneously on a time scale shorter than $\sim 30-100$ Myr, or alternatively that its decay may be induced by mass transfer in a binary. It is therefore interesting that in the last few years several studies challenge this idea, and suggest a return to the original idea of rapid spontaneous field decay. A population synthesis by Gonthier et al. (2002) finds decay times of around 10 Myr, as does a pulsar velocity study by Cordes & Chernoff (1998) that is used for input in the population synthesis of Arzoumanian, Chernoff, & Cordes (2002). In this paper, we investigate whether we can reproduce these new results, and whether we can explain the differences with earlier results. In Sect. 2, we describe the adaptations made to the computer code used in Hartman et al. (1997) to address the new investigations. In Sect. 3 we apply the renewed code to radio pulsars only, in Sect. 4 to both normal radio pulsars and high magnetic field 'magnetars'. A discussion of our results is given in Sect. 5.

## 2. The pulsar population synthesis code

The method of our population synthesis code for pulsars is as follows. We assume distributions for the parameters of newly born pulsars, in particular their location in the galaxy, their velocity, their pulse period and their magnetic field strength (or equivalently, period derivative). We then choose an age of the pulsar between zero and the Hubble time and evolve the pulsar properties. The pulsar position and velocity change under the influence of the gravitational field of the galaxy. The pulsar period and period derivative change according to $P\dot{P} = 10^{39}B(t)^2$, where B is the magnetic field, which we assume to decay exponentially on a time scale $\tau$: $B(t) = B(0)\exp{-t/\tau}$. From the new period we derive a beaming fraction, and compare it with a random number to decide whether the pulsar is retained. If so, we compute the radio luminosity from the period and period derivative. If the pulsar is above the death line in the period vs. period derivative diagram, the flux is computed from luminosity and position in the galaxy, and it is then checked whether any of four large surveys would have detected the pulsar. This procedure is repeated until we have simulated 2000 detections, and the properties of this simulated population are compared with those of the real pulsar population. An important feature is that we also test the real pulsars against out surveys simulations. Only the detected set is used for comparison. This reduces the impact of the simplification one necessarily makes in describing the surveys.

Our assumptions for the initial distributions and for the laws of their evolution can be varied to search for an optimum solution. To determine the optimum we compare specifically the observed and simulated distributions of the pulse period, the magnetic field strength, the projected dispersion measure $DM\sin b$ and the radio luminosity

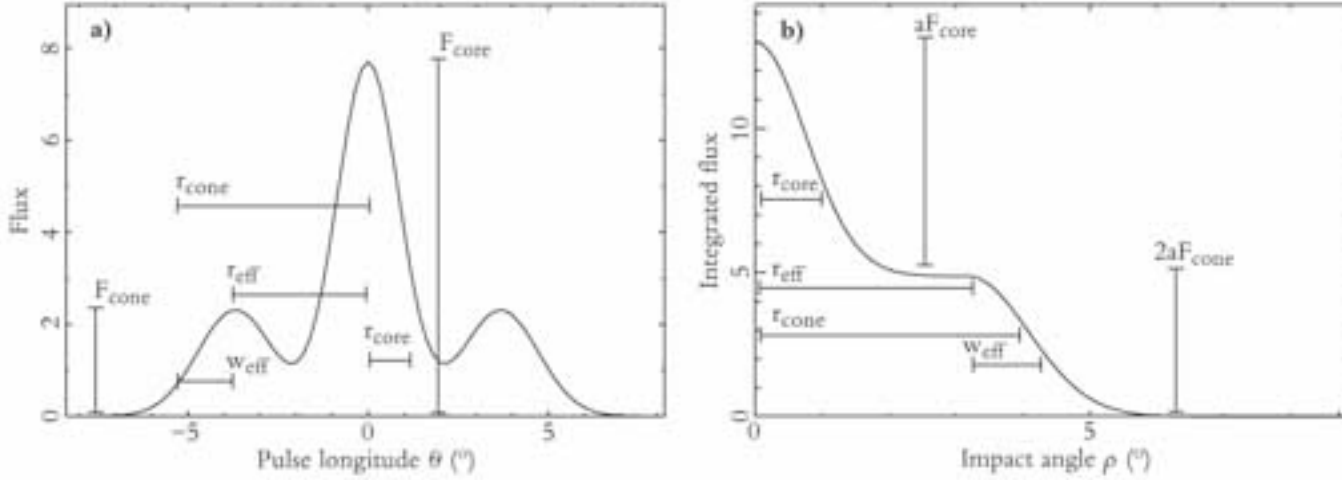

**Fig. 1. a)** parameters that determine the pulse shape and **b)** their influence on the integrated flux, depending on impact angle.

at 400 MHz $L_{400}$. For details we refer to Bhattacharya et al. (1992); Hartman et al. (1997).

### The gravitational field of the galaxy

In the course of making a parallellized version of our computer code we discovered an error in the gravitational potential of the galactic bulge as used in Hartman et al. (1997; their eq. 8), which we have now corrected to:

$$\Psi_{b,n}(R,z) = \frac{-G\,M_{b,n}}{\sqrt{b_{b,n}^2 + R^2 + z^2}} \qquad (1)$$

In our tests, this change does not appear to affect the outcome of the simulations.

### The luminosity law and beaming fraction

We have tried to implement the features which Arzoumanian et al. (2002) suggest are the largest improvement over older population syntheses: their beaming and luminosity models. The exact relation between the peak beam fluxes and the integrated pulsar luminosity is not given. Our implementation is as follows. The beaming model assumes that all pulse profiles consist of a core and a conal component. The core and conal peak flux ratio depends on pulse period (in seconds) as

$$F_{core}/F_{cone} = \frac{20}{3}P^{-1} \qquad (2)$$

So with increasing period, the profile evolves from core-dominated to cone-dominated. The opening angles of the core and cone evolve with period as

$$\rho_{core} = 1.5° \, P^{-0.5} \qquad (3)$$

$$\rho_{cone} = 6.0° \, P^{-0.5}. \qquad (4)$$

The effective radius and width of the cone (cf. fig. 1a) are

$$r_{eff} = 0.75\,\rho_{cone} \qquad (5)$$

$$w_{eff} = 0.30\,\rho_{cone} \qquad (6)$$

For a given angle between the line of sight and the magnetic pole (the impact angle $\rho$) we can use these equations to calculate the flux $f(\theta,\rho,P)$ at all pulse longitudes $\theta$, as in fig. 1a. For the simulated detection, we are interested in the integrated flux along our scan across the beam $F(\rho,P) = \int_0^{2\pi} f(\theta,\rho,P)\,d\theta$. In the example in fig. 1a, the integrated flux is composed of the core flux ($\propto F_{core}$) and twice the conal flux ($\propto 2F_{cone}$). For different cuts of the line of sight through the pulsar beam, we approximate the integrated flux $F$ (cf. fig. 1b) as:

If $\rho < r_{eff}$
$$F(\rho,a) = 2aF_{cone} + aF_{core}\exp\left\{-\rho^2/\rho_{core}^2\right\}$$
else
$$F(\rho,a) = 2aF_{cone}\exp\left\{-(\rho - r_{eff})^2/w_{eff}^2\right\} + aF_{core}\exp\left\{-\rho^2/\rho_{core}^2\right\}$$

Our normalisation constant $a(P)$ is such that the total flux over all impact angles is unity, $\int_0^{\frac{\pi}{2}} F(\rho,a)\,d\rho = 1$. With this normalisation, $F(\rho,a)$ denotes which fraction of the total flux is in our line of sight.

We have assumed that the flux depends on $P$ and $\dot{P}$ as:

$$F_{tot} = P^\alpha \, \dot{P}_{15}^\beta \, F_0 \text{ mJy kpc}^2 \qquad (7)$$

We take $\alpha = -1.3$ and $\beta = 0.4$, in accordance with Arzoumanian et al. (2002). We tried a range of values for $F_0$ from $10^{-3}$ to $10^3$ and determined that a value of $F_0 = 2$ produces the best results for our simulated period, magnetic field, dispersion measure and luminosity distributions.

Arzoumanian et al. (2002) then state they apply the effects of interstellar dispersion and scattering, without giving further information. From their fig. 2 we deduce that the 'standard candle' luminosity L is smeared by a wide Gaussian distribution $\sigma_{\log L}$ = 0.7. We apply the same smearing to the fluxes from eq. 7.

For each simulated pulsar, we draw random angles between the rotational axis and the magnetic pole, and between the rotational axis and observer's direction to the determine the impact angle $\rho$. We calculate which fraction of the flux is emitted in our line of sight. From the pulsar P and Ṗ, we determine the total flux. We combine the two with the distance to get the flux at the telescope.

### Quantum critical field strength

Regimbau & de Freitas Pacheco (2001) evaluated whether wide neutron-star initial magnetic fields distributions could explain both the properties of radio pulsars and 'magnetars'. To test whether we can reproduce these results, we have incorporated the quantum critical field strength of $9.5 \times 10^{13}$ G that could stop neutron stars with higher fields from shining in radio (Baring & Harding 2001).

## 3. Effects of beaming models in pulsar studies.

### Core-conal beam models

We have used our code, with the implementation of the beaming model of Arzoumanian et al. (2002), to search for the fits to the observed pulsar population for a range of decay times. We will refer to this core-cone model as model CC. Magnetic field decay times of 10 and 100 Myr produce fits of similar quality, both acceptable (see fig. 2). As in Hartman et al. (1997), we use a Kolmogorov-Smirnov test to determine the probability that the modelled and real distributions are the same, and use the product $Q_{ks}$ of the probabilities for the periods, magnetic field strengths, projected dispersion measure (DM sin b) and luminosity as a figure of merit for the overall solution.

We note that the figure of merit for the best models CC for decay times 10 and 100 Myr are about a factor 10 times less likely then the best model, model A for 100 Myr, from Hartman et al. (1997). Similarly, the distribution of characteristic ages $\tau_c$ for models CC is acceptable, but is by itself about a factor of 10 less likely that in the best model A for 100Myr. All models A that had decay times of 10Myr were found by Hartman et al. (1997) to be rather less likely.

As their period evolution is slowest at long periods, one expects many pulsars to have long periods, more than observed (Lyne et al. 1985). This shortage can be ex-plained by any mechanism that makes it harder to detect pulsars at long periods. Lyne & Manchester (1988), Rankin (1993) and Kramer et al. (1998) all find that pulsars with longer periods have significantly narrower beams. The volume from which they are visible (the beaming fraction) is therefore smaller, making it less likely for them to be detected. In models A with long decay times, this period-dependent beaming explains the relative dearth of long period pulsars; models A which combine magnetic field decay and a period-dependent beaming fraction find too few long period pulsars.

The reason that models CC do not favour long decay times, is that their beaming fraction dependence on period is less than in models A. Models A use isotropic beams that scale as $P^{-0.5}$. Although the opening angles of the narrow core and wide conal beam also diminish with period as $P^{-0.5}$, their combined behaviour is different in this model. At short periods, the beam will be core dominated and narrow. Yet towards longer periods, the wider conal components become more prominent: at periods longer then 0.6 seconds, all profiles will have conal components stronger than 10% of the peak flux, substantially increasing the volume from which they can be detected. In fig. 3b we plot the impact of this transition from core-dominated to cone-dominated on the pulse width (and hence the beaming fraction). In fig. 3a we plot the observed distribution of pulse width versus period, as determined by Gould & Lyne (1998) at 600 and 1400MHz. These authors note the widths of the observed profiles do not change in this range, so we assume them be applicable to the pulse widths at 400MHz as simulated by Arzoumanian et al. (2002) in fig. 3b.

Apparently, the strong preference for a short torque decay time found by Cordes & Chernoff (1998) is no longer present in their recent work (Arzoumanian, Chernoff, & Cordes 2002). This could be because Cordes & Chernoff (1998) did not take selection

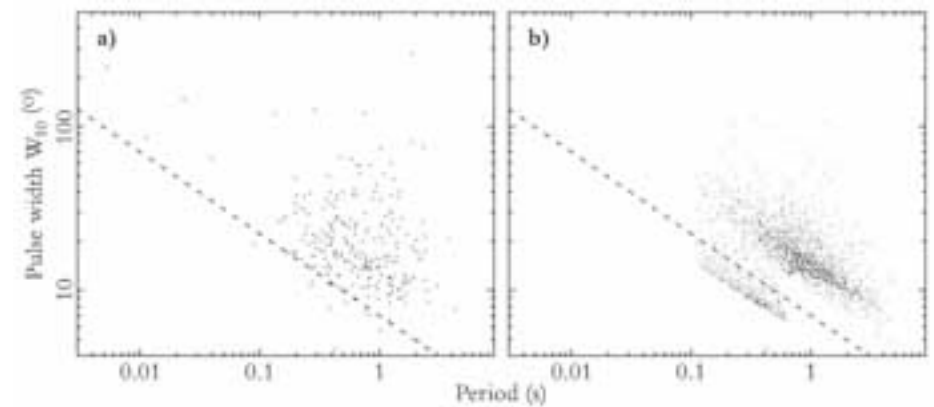

**Fig. 3.** Pulse width versus period for **a)** observed sample (Gould & Lyne 1998) **b)** the simulated detected pulsars using the beaming model of Arzoumanian et al. (2002).

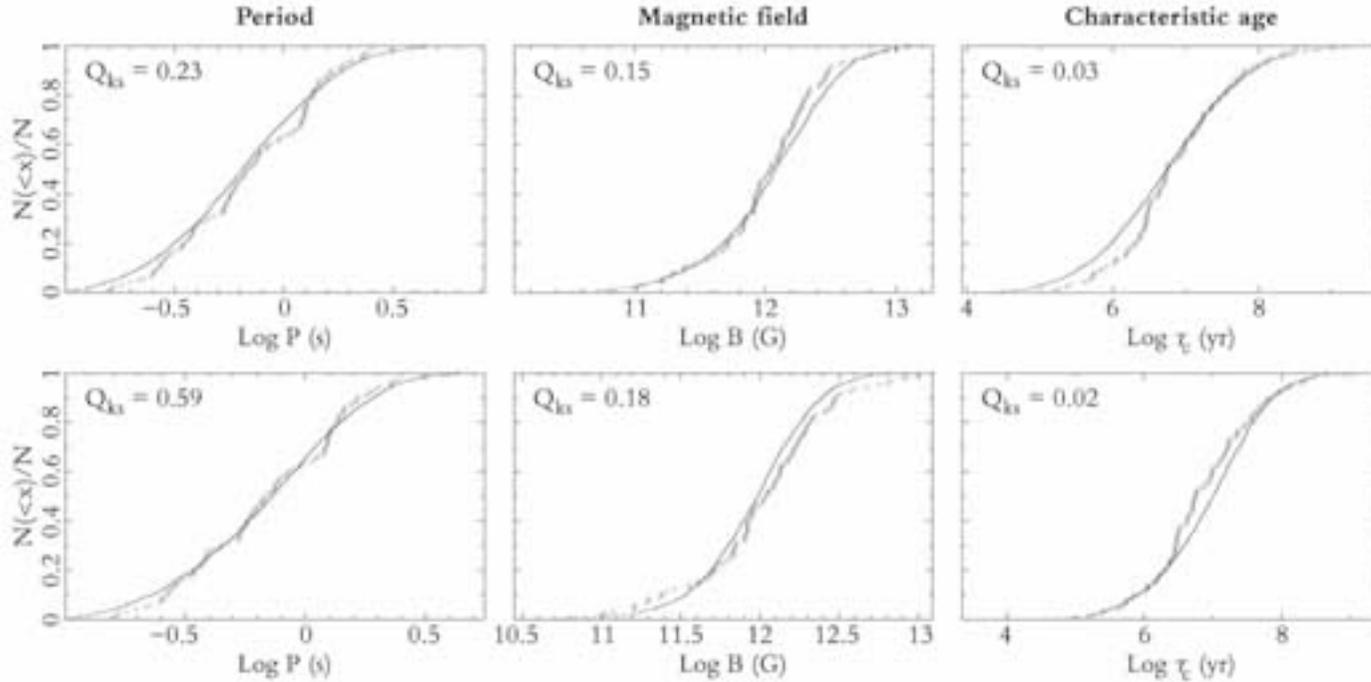

**Fig. 2.** Comparison of period and magnetic field distributions of real and simulated pulsars for model CC with decay times of 10 Myr (top row) and 100 Myr (bottom row)

effects into account, or because of the uncertainties in the proper motion, velocity and scale-height distributions they used to determine the decay time.

Apart from the different beaming models, models CC and A differ also in their description of pulsar luminosities. Model A follows the Narayan & Ostriker (1990) luminosity function in computing the luminosity as the product of a model luminosity, which is a function of P and Ṗ, and a random number chosen from a distribution describing the spread. Arzoumanian et al. (2002) argue that all pulsars with the same P and Ṗ have the same intrinsic (angle-averaged) luminosity and that the observed luminosity depends on the observation angle and interstellar scintillation. This we have used in model CC (see eq. 7). As the width of the resulting observed luminosity distribution is of the same magnitude as the intrinsic spread from Narayan & Ostriker (1990), the two descriptions are computationally very similar. We therefore think that the different results of models CC and A are not due to the different descriptions of the luminosity, but to the different beaming models.

*Period-independent beaming*

Gonthier et al. (2002) simulate the detection of galactic populations of radio and gamma-ray pulsars and conclude that magnetic fields decay in 5 Myr. They assume that the beaming fraction of all radio pulsars is identical. If we follow them and elim-inate the period dependence of the beaming fraction from model A (Hartman et al. 1997) we find that this variant of model A also favours short decay times (see fig. 4). In this case, magnetic field decay is needed to make pulsars turn off towards longer periods. Yet if we do allow for a beaming fraction that decreases with period as $P^{-0.5}$, as is commonly accepted (Vivekanand & Narayan 1981; Lyne & Manchester 1988; Rankin 1993), models without magnetic field decay are favoured (see fig. 5).

## 4. Effect of initial magnetic field distributions in studies of pulsars and magnetars

In their model for the formation of neutron stars, Duncan & Thompson (1992) propose that in some neutron stars, those born spinning faster than P∼3 ms, dynamo processes are efficient enough to amplify the magnetic field. These high-field pulsars (B>$10^{14}$G) they call 'magnetars'. They are observed mostly as soft gamma-ray repeaters and anomalous x-ray pulsars. In the neutron-star initial magnetic field function this dichotomy leads to two distinct components, one for radio pulsars and one for magnetars.

To investigate what the influence of such a bimodal distribution would be on the radio pulsar populations, we have tested scenarios in which half of the pulsars are born

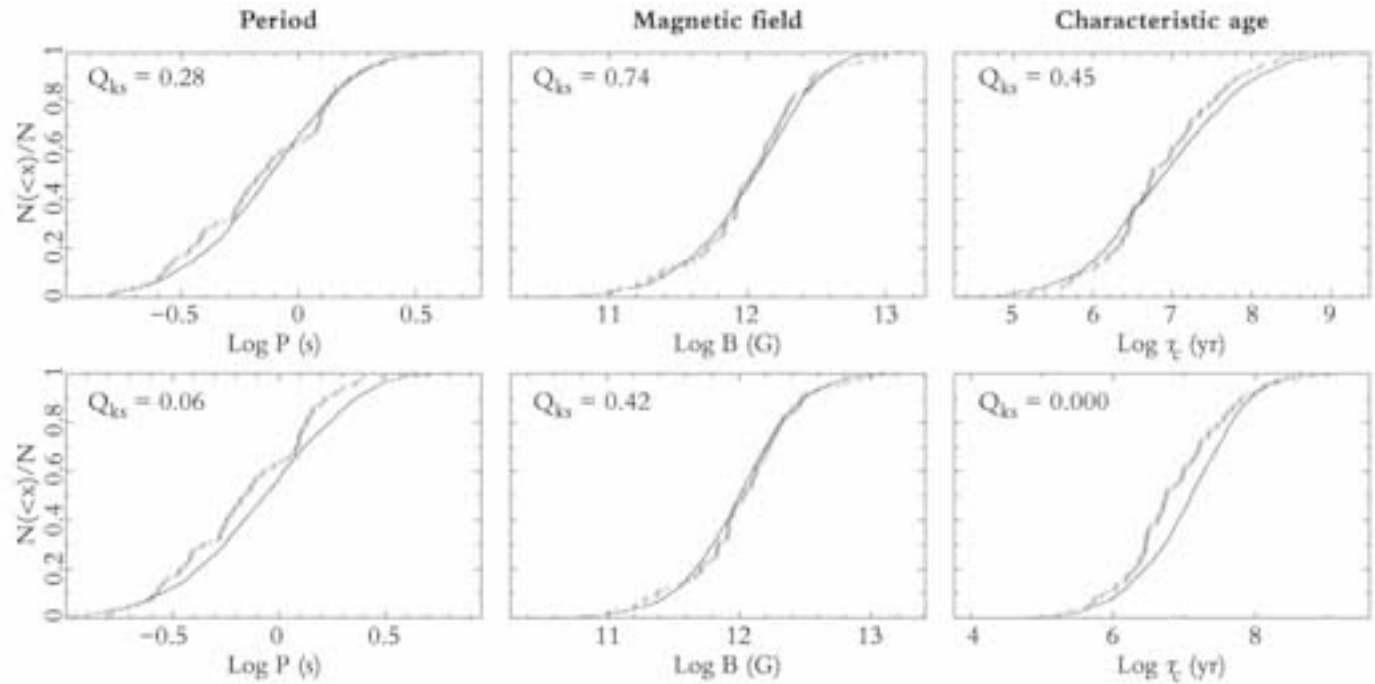

**Fig. 4.** Comparison of period and magnetic field distributions of real and simulated pulsars for a model with period-independent beaming, with decay times of 10 Myr (top row) and 100 Myr (bottom row)

with a magnetic field of $10^{12.3\pm0.3}$ G (the best fit from model A), while the magnetic field for the other half is $10^{14.0\pm0.8}$ G.

The absence of radio pulsars with fields over roughly $10^{13}$–$10^{14}$ G seems to support models in which neutron stars with field strengths higher than the quantum critical field strength of $9.5\times10^{13}$ G do not emit in radio (Baring & Harding 2001). Yet fig. 6a shows how a model without such a quantum critical field limit still produces almost only normal-field pulsars and actually predicts very acceptable distributions of our test quantities. The explanation lies in the rapid evolution of these high-field pulsars to long periods, where their luminosity and beaming fraction are low. Due to this rapid evolution, only very few high-field pulsars are visible at any given time, but some are. Indeed, only recently PSR J1847-0130, a pulsar with a very high inferred magnetic field of $9.4\times10^{13}$ G was discovered, challenging all but the highest quantum critical field predictions.

Regimbau & de Freitas Pacheco (2001) evaluate a similar model, using a continuous field input distribution of $10^{13.5\pm0.9}$ G, no quantum critical field strength cutoff and no deathline. They find that, using such a wide input distribution, the simulations fit the observed period and magnetic field distributions quite well. Yet when we use the input distribution they propose, the outcome does not fit the observed sample well (see fig. 6b). In particular, this model predicts a much wider distribution of magnetic fields than is observed. The reason for this is that the shorter lifetime of pulsars with $B_i=10^{13}$ G (compared to those with $B_i=10^{12}$ G) is compensated by their larger initial numbers. As a result, the simulated distribution contains too many long-period and high-B pulsars.

## 5. Discussion

As described in the Introduction, a variety of studies of radio pulsars and neutron stars in X-ray binaries indicates that the magnetic field of a neutron star does not decay spontaneously on a time scale of less than ∼100 Myr. In this paper we have investigated why several recent population studies of radio pulsars suggest a shorter time scale for spontaneous field decay.

Our main result is that the models allow a trade-off between the period dependence of beaming of the radio emission on one hand, and the decay of the magnetic field on the other hand. The reason behind this is as follows. In a diagram where the same magnetic field $B^2 \propto P\dot{P}$ is plotted as a function of pulse period P, the number of pulsars does not increase with period as follows from the equation for B which indicates a lower $\dot{P}$, hence longer life time at large P. The observed number of pulsars at long periods can be suppressed either by field decay, which moves pulsars to lower B before they reach long periods, or by a reduction of the detection probability at longer periods, e.g. by more narrow beams at longer periods. Thus we can explain

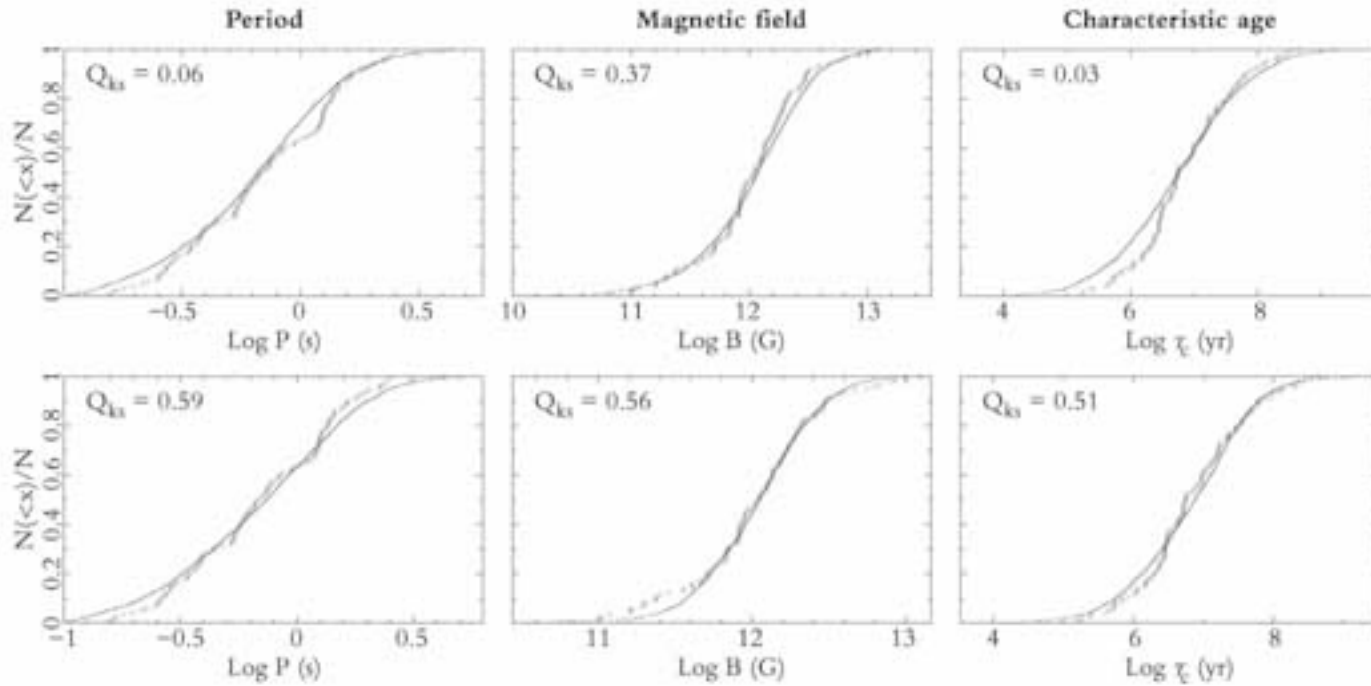

**Fig. 5.** Comparison of period and magnetic field distributions of real and simulated pulsars for model A with decay times of 10Myr (top row) and 100Myr (bottom row)

the difference between the strong preference for rapid field decay in the studies by Cordes & Chernoff (1998) and Gonthier et al. (2002), who do not include a period dependence of beaming, and the absence of such a preference in the study by Hartman et al. (1997) and Arzoumanian et al. (2002), who do.

It appears to us that beaming as a function of the period is the preferred solution, both because observations of radio pulsars indicate such a dependence (e.g. Lyne & Manchester 1988; Kramer et al. 1998), and because a long time scale for field decay is in better agreement with studies of X-ray sources. However, we do think that a more detailed study into the period dependence of beaming of radio emission, through the study of pulse profiles of individual pulsars, is required before definite conclusions can be drawn. Such a study can also address the question whether the axis of the magnetic field aligns closer with the rotation axis as a pulsar evolves towards longer periods, and provide an important handle for the discrimination between field decay and torque decay, in the equation $B^2 \propto P\dot{P}\sin\alpha$, with $\alpha$ the angle between rotation and dipole axis. It is worth noting that the distribution of characteristic ages $\tau_c \equiv P/(2\dot{P})$ is reproduced acceptably only in the models with period dependent beaming and long decay time, even though this distribution – which essentially provides the correlation between P and B – does not enter our figure of merit (see fig. 5). We consider this another argument in favour of a long time scale for field decay.

A reduced detection probability for radio pulsars at longer periods may also be the consequence of a reduced luminosity. Such an effect is partially included in our population synthesis studies, through the dependence of the luminosity on period. However, we find that the best figure of merit of the models is determined mainly by the distributions of the periods P and the magnetic field strengths B, and much less so by the distributions of the projected dispersion measures DMsin b and luminosities L. The reason for this is that the distribution of DMsin b also depends on the velocity distribution of the pulsars at birth, whereas the distribution of L shows a very wide range at the same P,Ṗ combination, due to either an intrinsic spread (as in our simulations) and/or to a wide range of interstellar scintillation (as in the simulations by Arzoumanian et al. 2002). With suitable changes in the assumed velocity distribution and in the luminosity law, one can therefore adapt the population synthesis to produce acceptable fits to the distributions of DMsin b and of L. A more accurately observed distribution of current pulsar velocities, as is coming into existence thanks to improved measurements (e.g. Brisken et al. 2002), will provide a better constraint of the population synthesis.

Contrary to what one might expect, we do not find a strong dependence of the birth rate on the time scale of field decay. A more important factor in the birth rate could be the number of pulsars at very low luminosity. That such pulsars exist is indicated both by single nearby pulsars as PSR J0108−1431 (Tauris et al. 1994) and by the the measured values less than 3 for the braking index $n \equiv 2\text{-}P\ddot{P}/(\dot{P})^2$ – which persists at higher ages, this would invalidate any population synthesis.

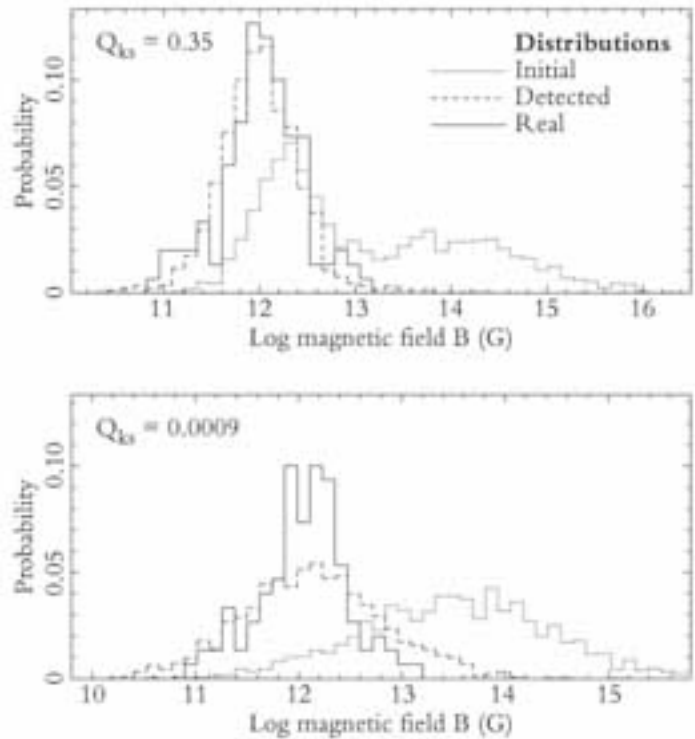

**Fig. 6.** The magnetic field distribution of real and simulated radio pulsars, using a bimodal (top) or a wider single (bottom) initial magnetic field distribution.

Finally, the observed distribution of the magnetic fields of radio pulsars peaks near B$\simeq 10^{12}$ G (see fig. 6), i.e. well below the critical field strength. We have shown that this implies that the magnetars, neutron stars with magnetic field in excess of the critical field, can not be formed from a single, broad Gaussian distribution of the magnetic field strengths of neutron stars at birth. This contradicts the result by Regimbau & de Freitas Pacheco (2001); we have not been able to find an explanation for this. In this respect, it is worth noting that our population synthesis is the only one in which the detection routines are tested directly with the observed pulsars, i.e. it is verified that our simulation would detect the actually observed pulsars. If we assume a bimodal distribution of the initial magnetic field strengths as suggested by theory (Duncan & Thompson 1992), we find that the models can explain the observed period and magnetic field distributions. This is independent of the use of a critical-field line or of a photon-splitting line as a limit to the detectability, i.e. the population synthesis of radio pulsars at the moment does not constrain the presence or absence of such limiting lines in the B−P diagram.

detections in very long observations of central pulsars in supernova remnants (Camilo 2003). That they do not represent very large numbers is indicated by the failure of deeper surveys to detect many more nearby, faint pulsars like PSR J0108−1431. This is important because the birth rates that we find in the models are compatible with the observed rates (see e.g. the discussion in Hartman et al. 1997); much higher predicted birth rates would be unacceptable.

The population synthesis computations assume, through the relation $B^2 \propto P\dot{P}$, that the period derivative $\dot{P}$ can be derived from the dipole model. Recently this assumption has been questioned on the basis of the observation of young ($<10^5$ yr) pulsars with appreciable discrepancies between real age and characteristic age (e.g. Kaspi & Helfand 2002). We note that the discrepancy for young pulsars is easily explained if the period $P_o$ of the pulsar at birth is too close to the current period P to warrant the implicit assumption $P_o{}^2 \ll P^2$. Hence it is not obvious that the discrepancy is important for the much longer time scales ($>$ Myr) in population synthesis. However, if the discrepancy indicates a strong deviation of a dipole equation for $\dot{P}$ – as indicated by

# FINDING PULSARS WITH LOFAR

**with Ben Stappers**

We investigate the number and type of pulsars that will be discovered with the low–frequency radio telescope Lofar. We consider different search strategies for the Galaxy, globular clusters and other galaxies. An all-sky Galactic survey can be optimally carried out by incoherently combining a large number of Lofar stations (large groups of dipoles coherently combined to form a number of beams) at an observing frequency of 140MHz. In a 60–day all-sky Galactic survey, Lofar will find approximately 1500 new pulsars, probing the local population of pulsars to a very deep luminosity limit. For targets of smaller angular size, like globular clusters and galaxies, the Lofar stations can be combined coherently, increasing the sensitivity even further. Searches of nearby globular clusters can find large numbers of low luminosity millisecond pulsars (eg. over 10 new millisecond pulsars in a 10–hour observation of M15). If the pulsar population in nearby galaxies is similar to that of the Milky Way, a 10–hour observation can find the 10 brightest pulsars in M33, or pulsars in other galaxies out to a distance of 1.2Mpc. Giant pulses from Crab–like extragalactic pulsars can be detected out to several Mpcs.

# 1. Introduction

Since the discovery of the first four pulsars with the Cambridge radio telescope (Hewish et al. 1968), an ongoing evolution of telescope systems has doubled the number of known radio pulsars roughly every 4 years: from a large flat receiver with a fixed beam on the sky (the original Cambridge radio telescope) to focusing dishes (Arecibo − Hulse & Taylor 1975), often steerable (Green Bank Telescope), on both hemispheres (the Parkes telescope − Manchester et al. 2001), with large bandwidths and multiple receivers to form more beams on the sky (Parkes, Arecibo). The type of pulsars discovered has changed accordingly. From slow, bright, single and nearby (the original four pulsars) to fast (young and millisecond pulsars), far-away (globular clusters), dim (pulsars in supernova remnants) and in binaries.

The next step in radio telescope evolution will be the use of large numbers of low-cost receivers that are combined to form an interferometer or a single dish. These telescopes, the Allen Telescope Array (Tarter et al. 2002), Lofar and SKA (Kramer 2004), create new possibilities for pulsar research.

In this paper, we investigate the prospects of finding radio pulsars with Lofar. We outline and compare strategies for targeting normal and millisecond pulsars, both in the disk and globular clusters of our Galaxy, and in other galaxies.

# 2. Lofar – The Low Frequency Array

With antennae being tested, funding secured and a small-scale test-telescope coming up, the outlines of the final Lofar design have become clearer. We have evaluated and simulated this 'semi-compact' configuration, and will describe it here in some detail.

We assume that the telescope is situated in the Netherlands and is split in two parts: in the inner 2km 'compact core', the signals from all dipoles are sent to a central processor that forms 8 beams on the sky, each with a bandwidth of 32MHz (depending on the processing capabilities of the final Lofar backend more beams may be possible). This compact core consists of 64 100-meter stations, each containing 50 antennae. Each antenna is formed by 16 dipoles. Outside the compact core, the 45 outer stations (25 of which are within a 12km diameter) each contain 100 antennae. At these outer stations the dipoles are combined locally into 1–8 beams, with bandwidths of 32–4MHz; only these beams are sent back to the central processor.

The high-frequency dipoles operate in a 110 - 240MHz frequency range, but are optimised for 140MHz. The maximum collecting area is 16m$^2$ per antenna, falling off as $\nu^{-2}$ towards higher frequencies $\nu$ and as $\nu^2$ towards lower frequencies. The compact-core gain G at 140MHz is $\frac{A}{2k} \times 10^{-26} \frac{Ws}{Jym^2} = 21.4$ K/Jy, with A the effective area, and k Boltzmann's constant in W s m$^2$. The dipoles are pointed towards zenith; moving away from zenith, the gain drops until it reaches zero at a zenith distance z = 60°. We model this drop-off as cos($\frac{3}{2}$z). The 120K antenna noise (Corey & Kratzenberg 2003) and 100K ground noise combine to a system temperature of about 220K.

Data is baseband recorded and dedispersed either coherently or in a digital filter-bank, depending on the requirements and computational power available; dispersion should therefore not pose a problem. For objects with known dispersion measures (some globular clusters for example), acceleration searches (Ransom 2000) will be used to find pulsars in binaries.

# 3. Pulsar searches

## Galactic disk surveys

To detect the local pulsar population efficiently, a survey should have a large field of view, bandwidth, collecting area and time per pointing, while minimising the system noise. For telescopes in which the receivers are spread out over a large area, like Lofar, one should also evaluate whether to add the beams of these receivers coherently or incoherently. In the coherent case, the beam on the sky will be smaller, but in the incoherent case the sensitivity of the system decreases. If all stations in the 2km compact core are combined coherently, the beam area is $\pi(\frac{0.61\lambda}{D} \times \frac{360}{2\pi})^2 = 0.0044\text{deg}^2$ at 140MHz. Yet each of the individual roughly 100-meter stations forms a beam on the sky of 1.8 deg$^2$. If added incoherently, the resulting beam is as large as that of the 64 individual stations, but at the cost of a decrease in sensitivity of $\sqrt{64}$ compared to the coherent case. As the 400 times larger beam area now allows for a 400-fold increase in the time per pointing, the incoherent approach is $\sqrt{400/64}$=2.5 times more sensitive than the coherent approach, for the same total amount of time (also see Backer 1999).

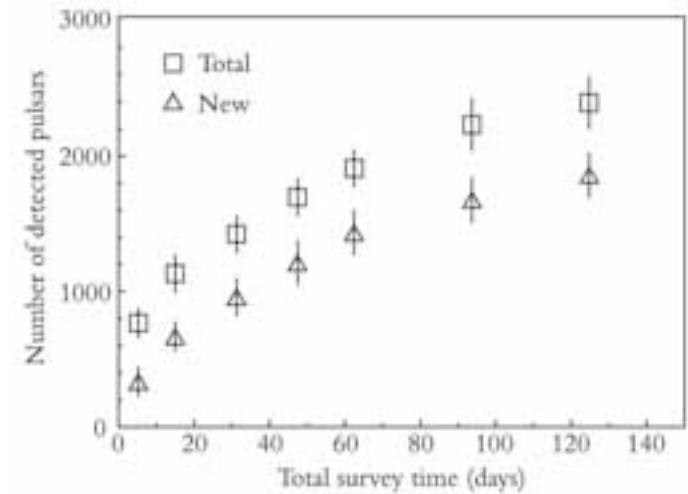

**Fig. 1.** Number of total and new pulsars detected with Lofar, versus total survey time (time per pointing x 1500 pointings / 8 beams). The 'new' detections were not detected by any of the (future) surveys from table 1.

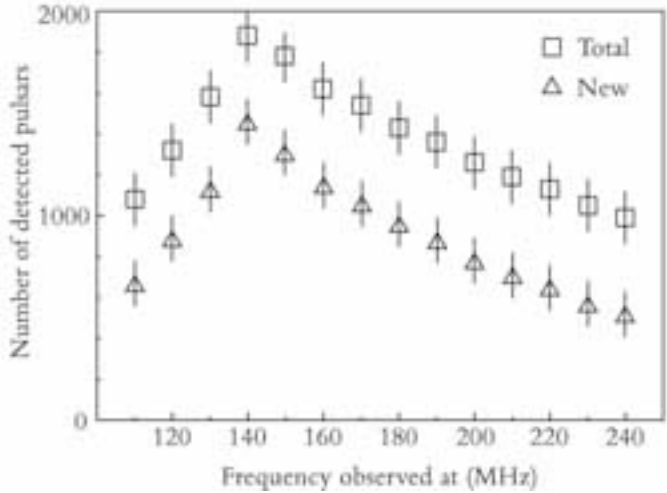

**Fig. 2.** Number of total and new pulsars detected with Lofar in a 60-day survey, versus frequency observed at. The antenna optimisation for frequencies of 140MHz is the dominant factor. To higher (lower) frequencies, the effective collecting area decreases as $\nu^{-2}$ ($\nu^2$).

In the incoherent case, one can also add the signal of the outer stations to increase the sensitivity, without loss of field of view. While the stations in the compact core produce 8 beams each 32MHz wide, the 45 outer stations are limited to 32MHz bandwidth in total. To ensure uniform final beam-patterns, we split the 45 stations in 8 group of 5 stations. We divide these eight groups over the eight compact core beams, each group producing a different 32MHz beam. This again increases the sensitivity by a factor of $\left(\frac{64 \times 50 + 5 \times 100}{64 \times 50}\right)\left(\frac{64}{69}\right)^{\frac{1}{2}} = 11\%$. It is unclear if this increase justifies the added complexity; the antennae could also be used for radio interference mitigation.

Choosing the optimum observing frequency presents a new trade-off. To higher frequencies, the beam size decreases as $\nu^{-2}$. The Lofar effective collecting area falls off as $\nu^{-2}$ above 140MHz and as $\nu^2$ below. At these low frequencies the spectral index of the pulsar luminosity function is $-1.5$ (Malofeev et al. 2000), so towards higher frequencies the pulsar brightness decreases as $\nu^{-1.5}$, but the sky background decreases even faster as $\nu^{-2.6}$ (Lawson et al. 1987). At the low frequencies, the dispersion smearing of the pulses increases. This dispersion smearing has forced recent surveys to observe at higher frequencies, around 1400MHz. As we coherently dedisperse the signal, we can remove the dispersion smearing even at frequencies as low as 140MHz. A different, but in this case more important, effect is scatter broadening: the signal passes through the interstellar medium using different paths, resulting in a significant broadening of the pulse profile. Currently this scatter broadening cannot be removed. The degree of scattering increases sharply towards lower frequencies, as $\nu^{-4.5}$ (Ramachandran et al. 1997; Shitov 1994). At 140MHz, this scatter broadening $t_{scatt}$ depends on the dispersion measure DM as (Shitov 1994):

$$t_{scatt} = 5.2 \times 10^{-7} \, DM^{3.8} \, ms \tag{1}$$

When pulses are smeared out, the number of harmonics in the Fourier transform of the pulse signal decreases, causing the detection probability to drop. Although the sky background and scatter broadening increase towards lower frequency, the total sensitivity is mainly determined by the effective collecting area: 140MHz is the most efficient frequency to observe at (cf. fig. 2).

The minimum detectable flux $S_{min}$ depends on the signal-to-noise ratio at which one accepts a signal as real (SNR), the system temperature ($T_{sys} = T_{antenna} + T_{ground} + T_{sky}$), the zenith-distance dependent gain ($G_z$), the number of polarisations ($N_{pol}$), the bandwidth ($d\nu$), the integration time (t), the number of incoherently added beams (N) and the pulse width (W) and period (P). As it depends on zenith distance and sky noise, $S_{min}$ varies per pointing, but a typical value for Lofar using a 60 minute integration time is

$$
\begin{aligned}
S_{min} &= SNR \left(\frac{T_{sys}}{G_\alpha}\right) \left(\frac{N}{N_{pol} \, d\nu \, t}\right)^{\frac{1}{2}} \left(\frac{W}{P-W}\right)^{\frac{1}{2}} \\
&= 10 \left(\frac{1000K}{24K/Jy}\right) \left(\frac{69}{2 \, 32 \times 10^6 Hz \, 3600s}\right)^{\frac{1}{2}} \left(\frac{0.05s}{0.45s}\right)^{\frac{1}{2}} \\
&= 2.4 \, mJy
\end{aligned} \tag{2}
$$

To determine the number and type of pulsars Lofar can find, we have used the above characteristics to model the detection of a large number of simulated pulsars. For this, we have used a population synthesis code that simulates the birth, evolution, death and possible detection of radio pulsars, as described more extensively in

| Survey | Year | N | $\nu$ (MHz) |
|---|---|---|---|
| Jodrell | 1972 | 51 | 408 |
| UMass-Arecibo | 1974 | 50 | 430 |
| MolongloII | 1978 | 224 | 408 |
| UMass-NRAO | 1978 | 50 | 400 |
| Parkes Multibeam | 1997-2000 | 1050 | 1374 |
| GBT | | | 340 |
| LOFAR | | | 140 |

**Table 1.** Name, year, number of detected pulsars (N) and frequency observed at ($\nu$) of the seven surveys simulated. The time per pointing for the GBT is chosen such that the total survey time is equal to that of a Lofar survey with 1-hour pointings. Data from Davies et al. (1977); Hulse & Taylor (1975); Manchester et al. (1978); Damashek et al. (1978); Manchester et al. (2001); Hobbs & Manchester (2004).

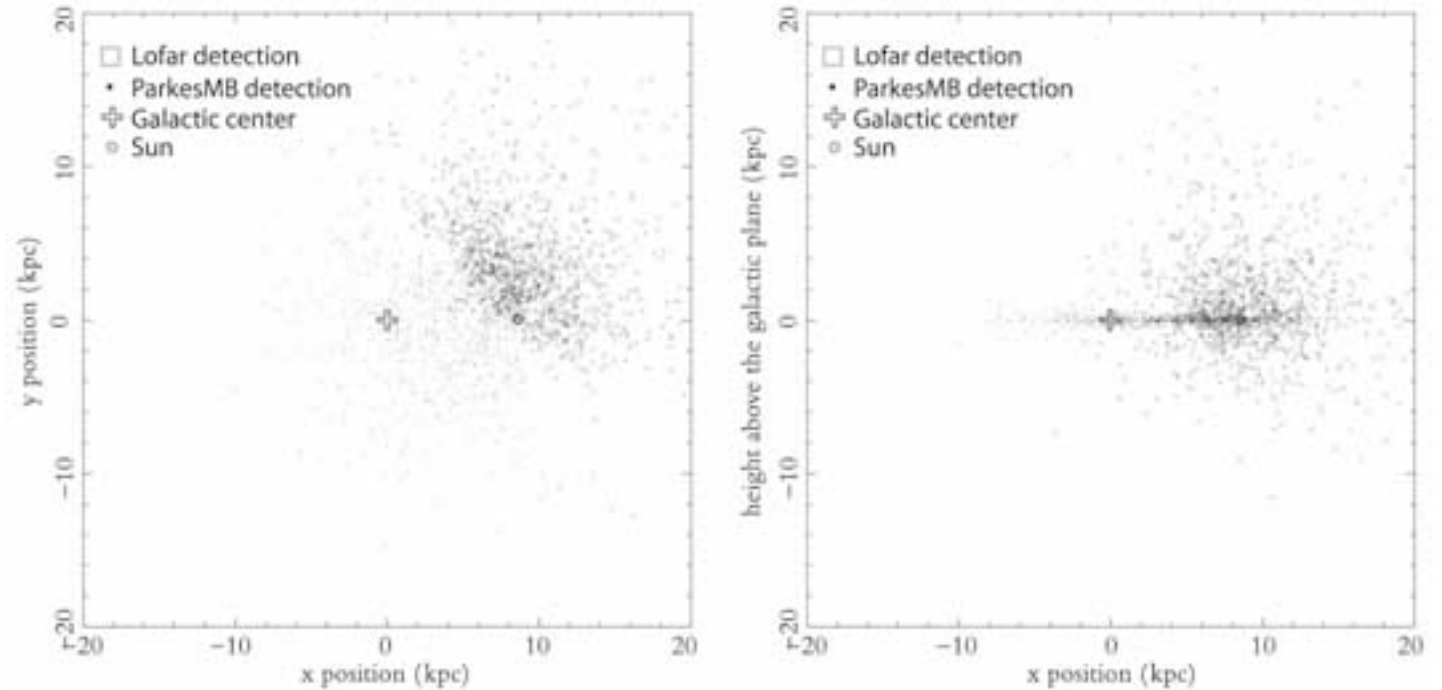

**Fig. 3.** Pulsars discovered with the Parkes Multibeam and Lofar surveys, for 1-hour Lofar pointings. Left) projected on the Galactic plane. Right) projected on the plane through the Galactic centre and sun, perpendicular to the disk.

Hartman et al. (1997). We choose initial positions and velocities and calculate the orbits through the Galaxy (van Leeuwen & Verbunt 2004). We evolve magnetic fields and periods, determine whether the pulsar is still above the death-line and estimate the luminosity at 400MHz, which we scale to the observing frequency as $\nu^{-1.5}$. We determine the sky background for the simulated position on the sky using the sky map of Haslam et al. (1982) and the abovementioned power law with spectral index −2.6. We use the Taylor & Cordes (1993) model to estimate the dispersion measure and, using eq. 1, the scatter broadening. We then determine which of these pulsars is detected by any of the 7 simulated pulsar surveys specified in table 1. These are modelled using their sky coverage and sensitivity function. By comparing the simulated pulsar sample with the real one, Hartman et al. (1997) have determined the most probable model parameters using the first four surveys mentioned. We use this best model with one modification: the number of pulsars born per unit area drops off exponentially with galactocentric radius, so the initial radius distribution is:

$$p(R)dR = \frac{R}{R_w} \exp(-\frac{R}{R_w})dR \qquad (3)$$

with the scale length $R_w$ as in Hartman et al. (1997). This distribution does not peak as strongly towards the galactic centre as eq. 4 in Hartman et al. (1997) does; for the local population there is no significant change, but surveys that reach the galactic centre, like the Parkes Multibeam survey, are better reproduced. We use this best model to determine the yield of future surveys like Lofar. As a test, we also predict the number of pulsars the Parkes Multibeam survey can find: our model predicts roughly 1000 pulsars, which compares quite well with the actual number of 1050 (Hobbs & Manchester 2004).

We have evaluated surveys with pointings of 5, 15, 30, 45, 60, 90 and 120 minutes duration (see fig. 1). With 60-minute pointings of the whole sky available over the Netherlands, Lofar could detect 1500 new pulsars in a total survey time of 60 days with 8 beams on the sky. If pulsar luminosity is the limiting factor in the detections, a larger time per pointing t decreases the minimum detectable flux $S_{min}$ as $t^{-\frac{1}{2}}$, increasing the distance d out to which pulsars can be detected as $t^{\frac{1}{4}}$. At the scale of the Galaxy pulsars are located in a disk; if one could observe pulsars throughout the Galaxy, the number of detectable pulsars is $N \sim d^2 \sim t^{\frac{1}{2}}$. Locally, pulsars are distributed roughly isotropically; then $N \sim d^3 \sim t^{\frac{3}{4}}$. We find that for Lofar surveys, N is less dependent on t than $N \sim t^{\frac{1}{2}}$ (fig. 1) because scatter broadening is the limiting factor, not luminosity (cf. Cordes 2002).

Compared to the Parkes Multibeam survey, the pulsars found with Lofar are different in several ways. The Parkes' focus on the Galactic centre is evident (fig. 3). The contours of the population detected by Lofar are determined by scatter broadening, and therefore follow lines of equal dispersion measure (fig. 3), to which the scatter broadening is closely related (eq. 1).

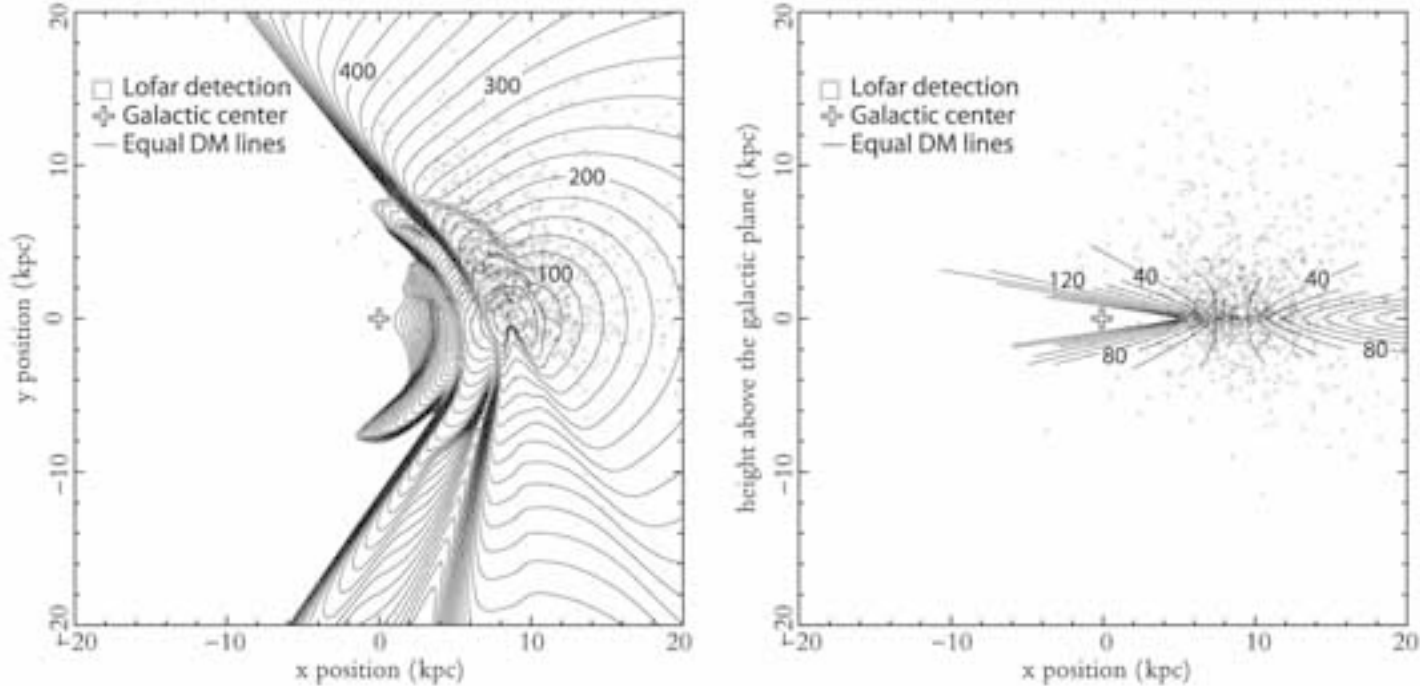

**Fig. 4.** Given sufficient computational power, dispersion smearing is not problematic, but scatter broadening is the limiting factor for a low-frequency survey like Lofar. The amount of scatter broadening follows from the dispersion measure DM through eq. 1. Here we show the 1500 pulsars discovered with a Galactic Lofar survey, for 1-hour pointings, overlaid on the lines of equal DM (and hence scatter broadening) in the plane of projection. The steps in dispersion measure are 20 pc/cm$^3$. Left) projected on the Galactic plane. Right) projected on the plane through the Galactic centre and sun, perpendicular to the disk. Clearly Lofar probes the local population very well. (Dispersion measure modelled after Taylor & Cordes 1993.)

Because of its higher sensitivity, Lofar detects more low-luminosity nearby pulsars (fig. 3a). A lower limit to pulsar luminosity of 1mJy kpc$^2$ (at 400MHz) has been suggested (Lyne et al. 1998); with a spectral index of $-1.8$, this compares to a luminosity at 140MHz of 7mJy kpc$^2$. If this lower limit exists, Lofar can see all pulsars that are beamed towards us, up to 1.7kpc. If it does not, then such a survey will certainly illuminate how many low-luminosity pulsars exist: a number of great importance if one wants to understand the neutron-star birthrate.

Furthermore, Lofar finds more pulsars out of the Galactic disk (fig. 3b). As pulsars are born in the plane (z=0), their observed z-distribution discloses their birth velocity, which is much debated (Hartman 1997; Arzoumanian et al. 2002).

Our Galactic pulsar survey will also be sensitive to millisecond pulsars, but here the scatter broadening (eq. 1) plays a more significant role. For DMs higher than 25 pc/cm$^3$, the sensitivity for the fastest millisecond second pulsars is reduced: there the scatter broadening is more than 0.1ms. In the Galactic plane, this means the fastest millisecond pulsars are visible up to 1kpc from the earth (see fig. 3). As many of the currently known Galactic-disk millisecond pulsars are within this range of 1kpc in which scatter broadening does not hinder detection, Lofar could probe the local population to a much lower flux limit, investigating the current lack of sub-millisecond pulsars.

| Name | d | DM | $r_c$ | $\Gamma$ | N |
|------|------|--------|--------|------|---|
|      | kpc | pc/cm$^3$ | arcmin |      |   |
| M15 | 10.3 | 67.3 | 0.07 | 665 | 8 |
| M92 | 8.2 | 29[+] | 0.23 | 106 |   |
| M5 | 7.5 | 30.0 | 0.42 | 79 | 2 |
| Pal 2 | 27.6 | 97[+] | 0.24 | 209 |   |
| M 3 | 10.4 | 17[+] | 0.55 | 66 | 3 |
| M 13 | 7.7 | 30.4 | 0.78 | 39 | 2 |
| NGC 6934 | 15.7 | 53[+] | 0.25 | 23 |   |
| M 71 | 4.0 | 74[+] | 0.63 | 3 |   |
| NGC 6229 | 30.4 | 26[+] | 0.13 | 22 |   |
| NGC 4147 | 19.3 | 17[+] | 0.10 | 7 |   |

**Table 2.** Name, distance (d), dispersion measure (DM) observed or [+]modelled after Taylor & Cordes (1993), core radius ($r_c$), collision number ($\Gamma$) and number of detected pulsars (N) for the ten highest-ranking candidates for a Lofar globular cluster survey. Data from Harris (1996).

*Galactic globular cluster surveys*

In a Galactic disk survey, one trades field of view for sensitivity by adding the signal of the stations incoherently in stead of coherently. For large areas, this trade-off increases the number of detectable pulsars because it allows for more time per pointing. If there were specific, smaller regions on the sky with higher densities of radio pulsars, we could target these with smaller field of view, but significantly higher sensitivity.

Globular clusters fit the description well; they are compact and form regions on the sky with high stellar densities. Furthermore these high densities cause globular clusters to contain more binaries and binary-products than are found in the disk. This makes globular clusters very good candidates for specific millisecond pulsar searches. To estimate which clusters are most promising for a Lofar search, we evaluated several of their properties: location on the sky, dispersion measure (DM), distance (d), angle on the sky (a), and the probable number of radio pulsars present.

Only globular clusters with declinations higher than 0° were considered. The next concern is the scatter broadening which, as mentioned above, makes far-away millisecond pulsars hard or impossible to detect, especially the fastest ones. Most globular cluster pulsars have periods of around 2–5 ms (see fig. 5). At 140MHz, these are detectable up to a DM of about 50 pc/cm$^3$. Observing at 200MHz extends this limit to 80 pc/cm$^3$. For the longer period pulsars also present in these clusters, these limits are less of a problem. Therefore we only consider globular clusters with DMs smaller than 100 pc/cm$^3$. We then rate the candidate clusters by the expected number of detectable pulsars. This number scales as d$^{-2}$, but also as a$^{-1}$ as a larger angle on the sky increases the number of pointings needed to cover the cluster, decreasing the time per pointing (assuming unchanging total time) and consequently the minimum detectable flux. Assuming $\frac{d\log L}{d\log N} = -1$ (Anderson 1992), the maximum flux depends on the number of radio pulsars present, which scales as the collision number $\Gamma \equiv \rho_c^{3/2} r_c^2$, where $\rho_c$ and $r_c$ are the central density and the core radius, respectively (Verbunt 2003). The most promising sources are listed in table 2. Highest-ranking cluster M15 is very compact and fits well into a coherent compact-core beam (see fig. 6). We can now use the full gain of the coherent addition, which makes Lofar at least twice as sensitive as the Arecibo survey that found the other milliseconds pulsars in M15 (Anderson 1992), in the same integration time, at 140MHz, assuming a millisecond spectral index of −1.8. The weakest pulsar in M15 is 0.97mJy; in a 10-hour pointing with the compact core at 140MHz the minimum detectable flux (eq. 2 for a zenith distance of 40°) is 0.18mJy. Thus far, the luminosities of the pulsars in M15 follow $\frac{d\log L}{d\log N} = -1$. If this relation extends to lower luminosities, observing up to a minimum flux of 0.18mJy should yield over 10 new millisecond pulsars. In many of the best candidate globular clusters listed in table 2, the first pulsars have yet to be discovered.

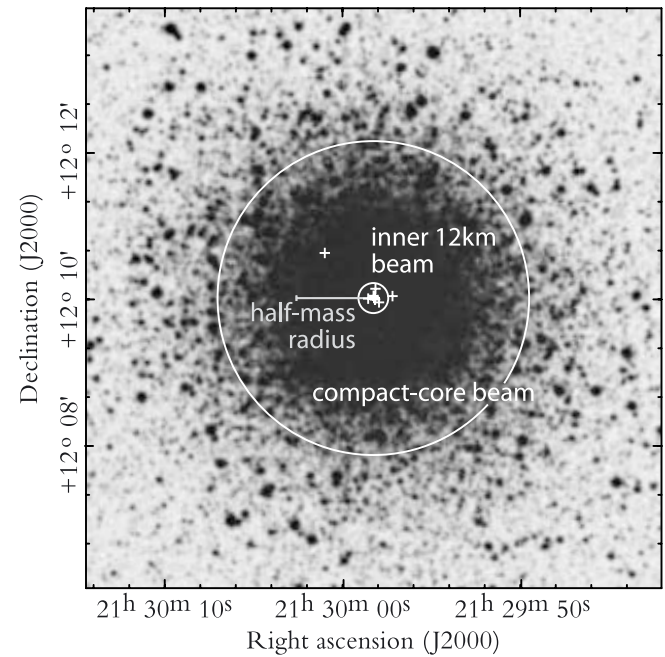

**Fig. 6.** Two different beam setups for a millisecond pulsar survey of M15, the globular cluster with the highest success probability in table 2. The cluster core-radius (0.07 arcmin) falls well within the middle 12-km baseline beam (0.35 arcmin). The crosses mark the 8 pulsars currently known (Anderson 1992).

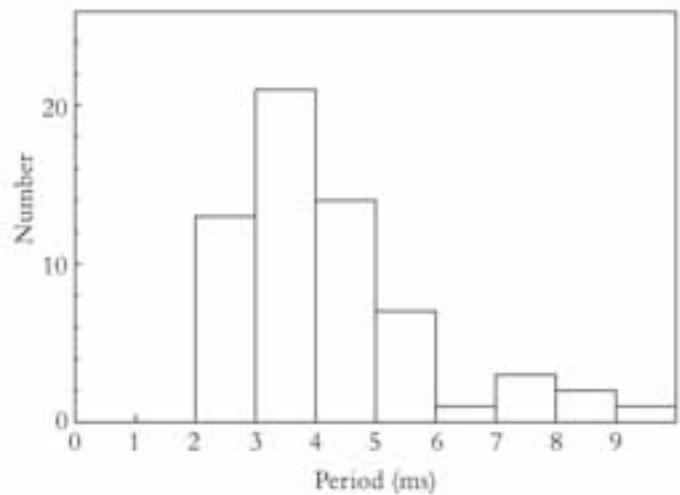

**Fig. 5.** Periods of the 63 millisecond pulsars in globular clusters with periods less than 10 seconds. There are another 17 pulsars, with longer periods. (Freire 2004).

*Pulsars in other galaxies*

The next obviously dense regions on the sky are galaxies. Their distance is a problem, but compared to globular clusters, galaxies have several advantages for a Lofar pulsar survey. They are distributed equally over both hemispheres, and if visible face-on and located in the part of the sky that is pointed away from our Galactic disk, the scatter broadening is relatively low. The main difference, however, is in the type of pulsars they host. Globular clusters contain old, spun-up pulsars, while spiral and irregular galaxies will also host young, bright, Crab-like pulsars.

In a relatively close galaxy like M33, a 10 hour pointing with the compact core at 140MHz could detect all pulsars brighter than 57Jy kpc² (see eq. 2). In our own Galaxy, ten pulsars of such brightness are known. We could observe the brightest pulsars in our Galaxy (like B1302−64, 130Jy kpc²) to distances of about 1.2Mpc.

Although the sample of pulsars with known giant pulses is very small (only two young and two recycled pulsars), young pulsars appear to have relatively brighter giant pulses. If the ratio of the flux of the giant pulses to the normal pulsar flux is more than roughly $10^5$ (McLaughlin & Cordes 2003), the pulsar in question can be discovered through its giant pulses before the normal signal becomes apparent. The flux-ratio for the Crab, for example, is higher than this number, and it was indeed first detected through its giant pulses. The distance d out to which one can detect single giant pulses in a 1-hour pointing (McLaughlin & Cordes 2003) depends on the system noise flux $S_{sys}$, the giant-pulse flux $S_{GP}$ and the bandwidth dν:

$$d \approx 0.85 \text{ Mpc} \left(\frac{5 \text{ Jy}}{S_{sys}}\right)^{\frac{1}{2}} \left(\frac{S_{GP}}{10^5 \text{ Jy}}\right)^{\frac{1}{2}} \left(\frac{d\nu}{10 \text{ MHz}}\right)^{\frac{1}{4}} \qquad (4)$$

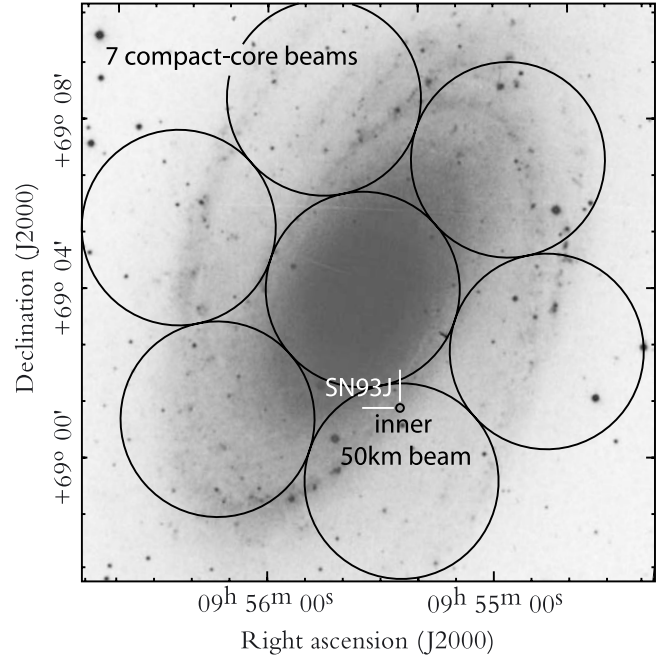

**Fig. 7.** Coherently formed beams from the compact core and the inner 12-km of Lofar, projected on the core of M81, the second best candidate galaxy in table 3 and on supernova 1993J (Ripero et al. 1993).

The above standard values are for a 1-hour Arecibo 430MHz giant-pulse survey for Crab-like pulsars. For Lofar, the system temperature is composed of 120K antenna, 100K ground and 800K sky noise; as shown section 2, the gain in the coherent compact-core setup is approximately 21K/Jy, so the Lofar $S_{sys}$=47Jy. The bandwidth dν is 32MHz. If we assume a giant pulse spectral index of −3 (Sallmen et al. 1999; Voûte 2001; Popov 2004), d=1.9Mpc; if we assume −4.5 (McLaughlin & Cordes 2003), d=4.5Mpc.

Like for the globular clusters, we have ranked target galaxies for their distance, size on the sky and expected number of sources (here assumed to scale with their mass). In table 3, we list the 10 highest-ranked galaxies that are in the Lofar field of view, have Galactic latitudes and inclination angles larger than 30° (indicating little scatter broadening) and that are within 4.5Mpc from the earth.

The two highest-ranked galaxies are large compared to lower-ranked ones, but their core can still be captured with a single compact-core beam (see fig. 7), or the whole galaxy by a pattern of 7 beams. Some of the smaller galaxies could be mapped using one inner 12km beam, which is 1.5 times more sensitive than the compact-core beam.

| Name | d | r | i | gb | M | type |
|------|------|--------|------|--------|-------|------|
| | Mpc | arcmin | deg | deg | Msol | |
| M33 | 0.7 | 28.2 | 56. | −31.33 | 10.10 | Sc |
| M81 | 3.5 | 11.0 | 60. | 40.91 | 10.73 | Sb |
| M94 | 4.3 | 6.1 | 33. | 76.02 | 10.78 | Sb |
| NGC 4236 | 2.2 | 9.8 | 73. | 47.35 | 9.82 | Sc |
| NGC 4395 | 3.6 | 6.9 | 38. | 81.55 | 10.07 | S |
| NGC 4244 | 3.1 | 7.9 | 90. | 77.16 | 9.99 | Sc |
| UGC 7321 | 3.8 | 2.5 | 90. | 81.04 | 9.65 | Sc |
| NGC 3077 | 2.1 | 2.7 | 43. | 41.66 | 9.12 | I |
| UGC 5666 | 2.7 | 6.5 | 55. | 43.62 | 9.59 | dSm |
| UGCA 205 | 1.3 | 2.2 | 38. | 39.89 | 8.44 | I |

**Table 3.** Name, distance (d), radius on the sky (r), inclination angle (i), Galactic latitude (gb), mass (M) and Hubble type for the ten highest-ranking candidates for a Lofar extragalactic pulsar survey. Data taken from Tully (1988) and Freedman et al. (2001).

The second ranked galaxy, M81, was also the host of SN93J, the second-closest supernova of modern times. Using the inner 50km of Lofar, we can form a 10" beam that has twice the sensitivity of the compact-core beam, and target such radar pulsar birth-grounds specifically. Because of the distance (3.5 Mpc) a 10-hour observation only measures down to a luminosity of 600Jy kpc$^2$, on order of magnitude higher than the highest pulsar luminosity known. known. Furthermore, the Galactic dispersion measure in the direction of M81 is roughly 40 pc/cm$^3$ and if we assume a similar contribution of M81 itself, the scatter broadening from the young pulsar possibly formed in SN93J would be about 10ms (see eq. 1), roughly twice the initial period of that other young pulsar, the Crab. Periodicity searches therefore appear to have a low chance of success.

Of all signs of life seen from any pulsar, Crab-like giant pulses are the brightest. A giant pulse search is also much less affected by scatter broadening. According to eq. 4, a 10-hour observation of SN93J with the inner 50km of Lofar (having a gain of 50 K/Jy), assuming a Crab-like giant pulse luminosity function $\frac{dN}{dS} = S^{-3.5}$ (McLaughlin & Cordes 2003) and spectral index of $-3.0$ (Voûte 2001), can detect giant pulses from SN93J even if it is 2 times weaker than the Crab; if the giant-pulse spectral index is $-4.5$ (McLaughlin & Cordes 2003) a pulsar up to 10 times weaker than the Crab-pulsar would be detectable. Due to its large gain, Lofar could then discover a pulsar only 14 years old; the implications would be tremendous.

## 4. Discussion

### Steep spectrum sources

The generally steep spectrum of pulsars typically turns over at lower frequency, between 100 and 250MHz (Malofeev et al. 1994). There is, however, a small fraction of pulsars for which no such spectral break has been observed, to frequencies as low as 50MHz. Sources with a spectral index steeper than the spectral index of the sky background of $-2.6$ (e.g. PSR B0943+10, spectral index $-4.0$ Ramachandran & Deshpande 1994) will be more more easily detectable by a telescope like Lofar, producing quantitative input for radio emission models.

### Location

The locations considered for building Lofar are sites in Western Australia, Southwestern United States and the Netherlands. Thus far we have described the results of pulsars searches from the Dutch site (1500 new pulsars at 60 minutes per pointing). We now describe the impact of different sites on the galactic survey. The Australian site, at a latitude of $-26°$ has the largest sky coverage of the three sites; in the same total survey time of 60 days (thus, a smaller time per pointing) Lofar finds 2200 new pulsars. Opposed to the Dutch and US sites, the Australian site points towards the centre of the galaxy. As the scatter broadening outside the galactic plane is much less (cf. fig. 3) Lofar is most sensitive outside the plane. The Parkes Multibeam survey already scanned the galactic plane so the two surveys are still complementary. The sky coverage of the US site, located at a latitude of 34°, falls in between that of the two other sites, as does its galactic plane coverage. A 60-day survey now finds 1700 new pulsars. As Lofar is mainly sensitive to the local population, all three surveys find the same kind of pulsars.

For globular cluster sources, the southern locations could also be more productive. Most globular clusters are positioned towards the galactic centre. Although the scatter broadening only allows observations of relatively nearby clusters, the Australian site can target more candidates, at more sensitive zenith distances.

For extragalactic searches the location is less important, as the sources are distributed more evenly on the sky.

## 5. Conclusions

Because of its large area and wide beam on the sky Lofar can probe the local population of pulsars to a very deep luminosity limit. Assuming the Dutch site, a 60-day Galactic survey at 140MHz would find 1500 new pulsars, doubling the known population in size. If we add all antennae coherently the sensitivity increases even further; with this setup, millisecond pulsars in nearby globular clusters can be detected to much lower flux limits than previously possible. Assuming the pulsar population in other galaxies is similar to that in ours, we can detect periodicities or giant pulses from extragalactic pulsars up to several Mpcs away.

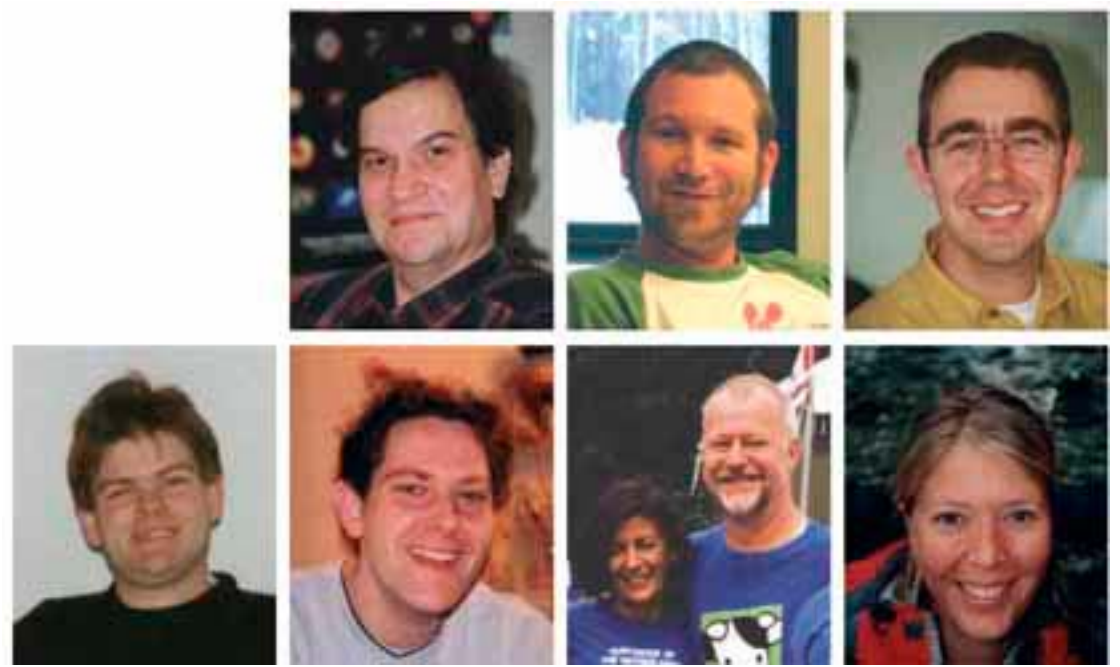

Joeri van Leeuwen was born on January 11, 1975 in Waardenburg, The Netherlands. After the VWO (Rosmalen, 1987-1993) he studied physics and astronomy at Utrecht University (1993-1998) and rowed for the Dutch national team (1995-2000). A masters-thesis project with Prof. Frank Verbunt was the prelude to the PhD work of which the book before you gives an overview. The results were also presented at talks in Amsterdam, Groningen, Melbourne, Los Angeles, San Francisco, Sydney and Utrecht.